\documentclass[12pt]{article}
\pdfoutput=1 
\usepackage[top=30truemm,bottom=30truemm,left=25truemm,right=25truemm]{geometry}

\usepackage{graphicx} 
\usepackage{color}

\usepackage{authblk}
\usepackage{amsthm}
\usepackage{amsmath,amssymb,amsfonts,bm}
\usepackage{cite}
\usepackage[]{hyperref}
\hypersetup{colorlinks,linkcolor={blue},citecolor={blue},urlcolor={blue}}  
\usepackage[titletoc]{appendix}
\usepackage{url}
\usepackage{graphicx,color}
\usepackage{bbold}
\usepackage{braket}
\usepackage{qcircuit}
\usepackage{here}
\usepackage{mathrsfs}

\usepackage{cancel,tabularx,moreverb,fancybox,float,braket,slashbox,accents,ascmac}

\usepackage{latexsym}
\usepackage{tcolorbox}

\newcommand{\pa}[1]{\left(#1 \right)}

\newcommand{\bb}[1]{\mathbb{#1}}

\newcommand{\ca}[1]{\mathcal{#1}}

\newcommand{\abs}[1]{\left|#1\right|}

\newcommand{\kket}[1]{\ket{#1}\rangle}

\newcommand{\ti}[1]{\tilde{#1}}

\newcommand{\fr}{\frac}

\def\be{\begin{equation}}
\def\ee{\end{equation}}
\def\ba{\begin{eqnarray}}
\def\ea{\end{eqnarray}}

\def\m{{\mu}}

 \def\n{{\nu}}
 \def\ep{{\epsilon}}

 \def\a{{\alpha}}
 \def\ba{{\bar{\alpha}}}

\def\dd{{\mathrm{d}}}

\def\CM{{\mathcal{M}}}

\newtheoremstyle{factstyle} % style name
    {0}%                   % upper space
    {}%                  % lower space
    {}%              % text font
    {}%                  % first line indent
    {\bfseries}%         % title font
    {}%                     % title comma
    {0mm}%                   % space after title
    {}%                      % title style

\theoremstyle{factstyle}

\newtheorem{innercustomfact}{Fact}
\newenvironment{customfact}[1]
  {\renewcommand\theinnercustomfact{#1}\innercustomfact}
  {\endinnercustomfact}

  \makeatletter
    
    \@addtoreset{equation}{section}
  \makeatother
\AtBeginDocument{%
  {\footnotesize\global\footnotesep=0.8\baselineskip}%
}

\begin{document}

\begin{titlepage}
\thispagestyle{empty}

\begin{flushright}
CALT-TH 2022-034
\\
RIKEN-iTHEMS-Report-22
\\
YITP-22-108

\end{flushright}

\begin{center}
\noindent{\bf \Large AdS/BCFT from Conformal Bootstrap:}\\
\vspace{0.3cm}
{\bf \Large Construction of Gravity with Branes and Particles }\\
\vspace{0.9cm}

{\bf \normalsize Yuya Kusuki$^{a,b}$ and Zixia Wei$^{c,d}$}
\vspace{0.6cm}\\

${}^{a}${\it
Walter Burke Institute for Theoretical Physics, \\[-1mm]
California Institute of Technology, Pasadena, CA 91125, USA
}\\[1.5mm]

${}^{b}${\it RIKEN Interdisciplinary Theoretical and Mathematical Sciences (iTHEMS), \\[-1mm]
Wako, Saitama 351-0198, Japan}\\[1.5mm]

${}^c${\it
Center for Gravitational Physics and Quantum Information,\\[-1mm]
Yukawa Institute for Theoretical Physics,
Kyoto University, Kyoto 606-8502, Japan
}\\[1.5mm]

${}^d${\it
Kavli Institute for Theoretical Physics, \\[-1mm]
University of California, Santa Barbara, CA 93106, USA
}

\vskip 1.5em
\end{center}

\begin{abstract}

We initiate a conformal bootstrap program to study AdS$_3$/BCFT$_2$ with heavy excitations. 
We start by solving the bootstrap equations associated with two-point functions of scalar/non-scalar primaries under the assumption that one-point functions vanish.  
These correspond to gravity with a brane and a non-spinning/spinning particle where the brane and the particle do not intersect with each other.
From the bootstrap equations, we obtain the energy spectrum and the modified black hole threshold. 
We then carefully analyze the gravity duals and find the results perfectly match the BCFT analysis. In particular, brane self-intersections, which are usually considered to be problematic, are nicely avoided by the black hole formation. 
Despite the assumption to solve the bootstrap equations, one-point functions of scalar primaries can be non-zero in general. We construct the holographic dual for a non-vanishing one-point function, in which the heavy particle can end on the brane, by holographically computing the R\'enyi entropy in AdS/BCFT.
As a bonus, we find a refined formula for the holographic R\'enyi entropy, which appears to be crucial to correctly reproduce the boundary entropy term. 
On the other hand, we explain why one-point functions of non-scalar primaries always vanish from the gravity dual. 
The non-sensitivity of the solution for the bootstrap equation to the boundary entropy helps us to construct gravity duals with negative tension branes. 
We also find a holographic dual of boundary primaries.

\end{abstract}

\end{titlepage}

\restoregeometry

\tableofcontents

%%%%%%%%%%%%%%%%%%%%%%%%%%%%%%%%%%%%%%%%%%%%%%%%%%%%%%%%%%%%%%%%%%%%%%%%%%%%%%%%%%%%%%%%%%%%%%
%%%%%%%%%%%%%%%%%%%%%%%%%%%%%%%%%%%%%%%%%%%%%%%%%%%%%%%%%%%%%%%%%%%%%%%%%%%%%%%%%%%%%%%%%%%%%%
\section{Introduction and Summary}
%%%%%%%%%%%%%%%%%%%%%%%%%%%%%%%%%%%%%%%%%%%%%%%%%%%%%%%%%%%%%%%%%%%%%%%%%%%%%%%%%%%%%%%%%%%%%%
%%%%%%%%%%%%%%%%%%%%%%%%%%%%%%%%%%%%%%%%%%%%%%%%%%%%%%%%%%%%%%%%%%%%%%%%%%%%%%%%%%%%%%%%%%%%%%

%%%%%%%%%%%%%%%%%%%%%%%%%%%%%%%%%%%%%%%%%%%%%%%%%%%%%%%%%%%%%%%%%%%%%%%%%%%%%%%%%%%%%%%%%%%%%%
\subsection{Introduction}
%%%%%%%%%%%%%%%%%%%%%%%%%%%%%%%%%%%%%%%%%%%%%%%%%%%%%%%%%%%%%%%%%%%%%%%%%%%%%%%%%%%%%%%%%%%%%%

The AdS/BCFT (anti-de-Sitter/boundary conformal field theory) correspondence  \cite{Karch:2000gx,Takayanagi2011,Fujita2011} plays an important role in recent developments of the black hole information loss paradox\cite{Penington2020,Almheiri2019,Almheiri2020}, which is one of the most important problems in quantum gravity. This is because that certain kinds of BCFT are dual to a setup where a region with dynamical gravity is coupled to a non-gravitational heat bath, and hence an evaporating black hole has a UV-complete description given by a BCFT. This setup is called the island model \cite{Penington2020,Almheiri2019,Almheiri2020} and profoundly related to the Karch-Randall type brane-world holography \cite{Karch:2000ct}. Recently, this novel correspondence between AdS/BCFT and the island model has brought many insights into quantum gravity to us. Efforts towards this direction include analyses of the island model from the higher dimensional gravity dual \cite{Sully2021,Chen:2020uac,Bousso:2020kmy,Akal2020,Akal2021,Geng2021,Ghosh:2021axl,Chu:2021gdb,Akal:2021foz,Omiya2021,Geng:2021hlu,Ageev:2021ipd,Geng2021a}, attempts to directly relate island models to BCFT \cite{Kusuki2021,Suzuki2022} and explorations of generalized setups \cite{Anous2022,Izumi2022,Kusuki2022,Geng2022,Lee2022,Miyaji2022,Hu2022,Suzuki2022a,Biswas2022}.

While the holographic dual of a CFT is an AdS spacetime according to AdS/CFT\cite{Maldacena:1997re}, the holographic dual of a BCFT can be obtained by adding a brane into the original AdS \cite{Takayanagi2011,Fujita2011}. The brane is a codimension-1 physical object where the AdS spacetime comes to an end. Therefore, such a brane is often called an end-of-the-world brane. 
This new ingredient enriches the structure of our gravitational theory.
This richness gives us various ideal laboratories where we can learn more about quantum gravity.
However, in the current situation, we know too little about the AdS/BCFT correspondence to fully apply it to quantum gravity. There are still many unclear aspects of AdS/BCFT that should be figured out. To this end, we are going to focus on AdS${}_3$/BCFT${}_2$, which is under good analytic control, and consider the following issues,
\begin{enumerate}
   \item Self-intersection \\
   When solving Einstein equations with the presence of a brane on the AdS side, one can get configurations where the brane is bent by gravitational effects and gets intersected with itself \cite{Cooper2019}. These configurations are considered to be unphysical in previous literature and used to give constraints by prohibiting these configurations \cite{Cooper2019,Geng:2021iyq}. We will show, from both the gravity side and the CFT side, that any excitation that was expected to cause a brane self-intersection will in fact form a black hole. Therefore, one does not need to worry about self-intersections, at least below the black hole threshold. 
   
   \item Holographic dual of boundary primary \\
 While a standard CFT has a set of primaries, a BCFT has two sets of primaries: one lives on the boundary of the BCFT, and the other lives in the bulk\footnote{In the context of holography, ``bulk'' and ``boundary" are often used to refer to the higher dimensional gravitational theory and the lower-dimensional non-gravitational theory, respectively. In this paper, we will not use this terminology since there are many different notions of bulk and boundary in AdS/BCFT. Instead, we will use ``bulk" for the bulk region of the BCFT, and ``boundary" for the boundary region of the BCFT. The AdS will be referred to as ``gravity''.}. Note that boundary primaries and bulk primaries are two distinct sets of operators interacting with each other, and a boundary primary cannot be obtained by simply moving a bulk primary to the vicinity of the boundary. The holographic dual of bulk primaries can be obtained by straightforwardly applying the AdS/CFT dictionary to the AdS/BCFT case. However, it is unknown how boundary primaries should be properly treated. We will present gravity duals for certain boundary primaries, and explain how they are distinct from bulk primaries.
   
   \item Spinning particle \\
   On the BCFT side, one can easily find the following fact using the rotational invariance,
\begin{screen}
    For a bulk primary $\phi_i$ with chiral conformal dimension $h_i$ and anti-chiral conformal dimension $\bar{h}_i$,
	\begin{equation}\label{eq:fact3}
    h_i \neq \bar{h}_i~~~ \Longrightarrow~~~ \braket{\phi_i}_{\text{disk}} = 0.
\end{equation}
    Here, $\braket{\cdots}_{\rm disk}$ is the expectation value evaluated in the BCFT defined on a disk. 
\end{screen}
   However, it is unclear how to reproduce this property from the gravity side. Such a non-scalar primary corresponds to a spinning particle in AdS. As we will see in the main text, it is straightforward to realize a non-vanishing one-point function for a scalar primary with $h_i = \bar{h}_i$ on the gravity side. From this stance, it looks puzzling why slightly changing the spin $h_i-\bar{h}_i$ from zero would drastically change the behavior of the one-point function. In fact, the knowledge about how to deal with spinning particles in gravity with branes is highly limited.
   Against this backdrop, we will present an explicit construction of spinning particles in AdS$_3$ and apply it to discuss non-scalar primaries in AdS/BCFT. 
   
   \item Negative tension brane \\ 
   In the negative tension case,
   there is a possibility that a particle's worldline is excluded by the end-of-the-world brane.
   To our current understanding of AdS/BCFT,
   we do not know how to deal with the particle behind the end-of-the-world brane.
   We will give a clear understanding of how this hidden particle should be treated and show that our prescription is consistent with CFT results.

\end{enumerate}

For our purpose, the conformal bootstrap turns out to be useful.
Even if we have very limited knowledge about the gravity side,
the conformal bootstrap can give robust answers or hints to our questions.
In fact, one can find various achievements in AdS/CFT by using the conformal bootstrap such as 
black hole thermodynamics \cite{Cardy1986, Hartman2014,Mukhametzhanov2019,Pal2020},
eigenstate thermalization hypothesis  \cite{Kraus2017, Hikida2018, RomeroBermudez2018, Brehm2018,Das2021},
existence/non-existence of pure gravity \cite{Hellerman2011, Friedan2013, Hartman:2019pcd, AfkhamiJeddi2019, Benjamin2019, Besken:2021eps},
binding energy of two particles with large angular momentum \cite{Fitzpatrick2013, Komargodski2013,Alday2015,Kaviraj2015,Kaviraj2015a, Alday2017a, SimmonsDuffin2017,Alday2017,Sleight2018,Fitzpatrick2014, Kusuki2019,Kusuki2019a, Collier2019},
and averaged theory as semiclassical gravity  \cite{Chandra2022,Kusuki2022}.
Recently, the conformal bootstrap is also solved in BCFTs numerically \cite{Collier2021} and analytically in the asymptotic regime \cite{Kusuki2021, Numasawa2022, Kusuki2022}.
These achievements strongly suggest the high usefulness of the conformal bootstrap in our context. 
In this paper, we will first address the questions by using the conformal bootstrap. Then we will properly construct the corresponding gravity duals using results from the conformal bootstrap as clues.

To approach the above problems,
we need knowledge about the details of the bulk configuration.
This is not the common case where only the information near the asymptotic boundary is necessary to evaluate
(i.e., the case where we can use the Fefferman-Graham expansion).
On this background, we have to introduce a new method that can simply tell us the details deep into the bulk,
and then one of our main results is to give a formulation of how to construct semiclassical gravity with branes and massive particles in a simple way.

We will in particular consider the bottom-up model proposed in \cite{Takayanagi2011,Fujita2011} as the gravity dual of BCFT. 
Since this construction is a naive and bottom-up one, it is not surprising that it may break down when we are trying to use it to tackle the issues listed above, and we may eventually end up with constructing a new effective model. However, interestingly, we find that these issues can be perfectly resolved in the framework of this naive bottom-up model, and the results on the gravity side are completely consistent with those from the BCFT side. The only thing we should do is treat heavy excitations and their interaction with the brane properly. We summarize our results below.

%%%%%%%%%%%%%%%%%%%%%%%%%%%%%%%%%%%%%%%%%%%%%%%%%%%%%%%%%%%%%%%%%%%%%%%%%%%%%%%%%%%%%%%%%%%%%%
\subsection{Summary}
%%%%%%%%%%%%%%%%%%%%%%%%%%%%%%%%%%%%%%%%%%%%%%%%%%%%%%%%%%%%%%%%%%%%%%%%%%%%%%%%%%%%%%%%%%%%%%

\paragraph{Matching between self-intersection bound \& black hole threshold (Section \ref{sec:intersection})}~\par
A heavy bulk primary excitation on the BCFT side corresponds to a heavy particle on the gravity side. 
This heavy particle deforms the end-of-the-world brane through gravitational interaction.
Such a deformation sometimes leads to problematic configurations.
If we consider an excitation induced by a heavy scalar primary with $h_i=\bar{h}_i > \fr{c-1}{32}$,
we can find that the brane in the corresponding geometry intersects with itself.
In \cite{Geng:2021iyq}, it has been proposed that to avoid the self-intersection,
the holographic BCFT cannot have primaries with conformal dimension $h_i \in (\fr{c-1}{32},\fr{c-1}{24})$.
The upper bound comes from the fact that bulk primaries with $h_i > \fr{c-1}{24}$ lead to a black hole formation.
In Section \ref{sec:two-CFT}, we show that in the BCFT, this black hole threshold $h_i = \fr{c-1}{24}$ is modified and turns out to be $h_i = \fr{c-1}{32}$.
On the gravity side, this modification is caused by the non-trivial graviton interaction between the heavy particle and the end-of-the-world brane.
In the CFT language, this interaction can be rephrased by the interaction between these bulk primaries and their mirror images (which are defined by the method of images).
Unlike higher-dimensional AdS${}_d$ ($d\geq4$), the gravitational interactions in AdS${}_3$ create a deficit angle, which can be detected even at infinite separation. Due to this, the effect of the gravitational interactions is a little bit non-trivial in AdS${}_3$.
Interestingly, this non-trivial gravitational effect is nicely encoded in the fusion matrix on the BCFT side.
In \cite{Kusuki2019a, Collier2019}, it has been shown that the spectrum of two-particle states at large spin is encoded in the pole structure of the fusion matrix.
In this paper, we show that the spectrum of the conical singularities in AdS${}_3$ with the presence of the end-of-the-world brane comes from the same pole structure of the fusion matrix in a similar way.
In other words, one can find the spectrum of the Virasoro mean field in the BCFT.
From this spectrum structure, one can find that the black hole threshold for the bulk scalar primary is given by $h_i = \fr{c-1}{32}$.
It implies that we can always avoid the self-intersection because of the black hole formation.

\paragraph{Gravity interpretation of the bulk-boundary OPE (Section \ref{sec:bulk-two})}~\par
With the bootstrap equation,
one can show that the spectrum in the bulk-boundary OPE is completely the same as the quantum Regge trajectory \cite{Kusuki2019a, Collier2019}.
In a similar way to \cite{Kusuki2019a, Collier2019},
we can give a natural gravitational interpretation.
The BCFT setup with two bulk primary insertions corresponds to a one-particle state on the gravity side.
Since we have an end-of-the-world brane in our setup, this particle interacts with the brane via gravitational force. 
We can explicitly calculate the binding energy of this one-particle state from the quantum Regge trajectory.
One can find that the spectrum of this binding energy completely reflects the attractive nature of gravity at the quantum level.

\paragraph{Holographic dual of boundary primaries (Section \ref{subsec:BPdual})}~\par

Boundary primaries are primary operators that live only on the boundary of a BCFT, and they are sometimes called the boundary condition changing operators. A straightforward yet important question to ask is how boundary primaries, which are distinct from bulk primaries, should be treated on the gravity side. In Section \ref{subsec:BPdual}, we propose the gravity description of a boundary primary correlation function $\braket{\phi_I \phi_I}$,
which is given by a geometry with a particle and a brane.
Roughly speaking, this particle corresponds to a bulk primary $\phi_i$ with the bulk-boundary OPE $\phi_i \sim \phi_I$. 
We will show that our proposal nicely reproduces a boundary primary two-point function.
An alternative proposal has been given in \cite{Miyaji2022, Biswas2022}, which relates a boundary primary correlation to a defect on the end-of-the-world brane.
There is also another proposal in \cite{Kusuki2022a} in terms of a generalization of the Ryu-Takayanagi formula.

\paragraph{Holographic R\'enyi entropy in AdS/BCFT (Section \ref{sec:Renyi})}~\par

In the setups summarized so far, $\braket{\phi_i} = 0$, i.e. vanishing of one-point functions, is explicitly or implicitly assumed. However, for scalar primaries, the one-point functions do not necessarily vanish. Therefore, we would like to know whether a non-vanishing one-point function can be realized in AdS/BCFT, and if the answer is yes, how to do so. To answer this question, we present the first-ever holographic computation of the $n$-th R\'enyi entropy in AdS/BCFT, since the geometry arising in this procedure is expected to correspond to the holographic one-point function. In short, we find that the geometry corresponding to a non-vanishing one-point function is a conical defect ending on the brane. 

Moreover, we find it is worth focusing more on the R\'enyi entropy itself. Consider a Euclidean disk given by $x^2+\tau^2 \leq R^2$ and look at the quantum state realized at $\tau = 0$. If we divide it into $A = (-R,0)$ and $A^C = (0,R)$, it is straightforward to find on the BCFT side that the $n$-th R\'enyi entropy is given by 
\begin{align}
    S^{(n)}(\rho_A) = \frac{c}{12} \left(1+\frac{1}{n}\right) \log\frac{R}{\ep} + \log g,
\end{align}
where $g$ is the $g$-function defined as the disk partition function $g\equiv \braket{\mathbb{I}}_{\text{disk}}$, and $\epsilon$ is a UV-cutoff corresponding to the lattice distance. Here, $\log g$ is often called the boundary entropy and denoted as $S_{\rm bdy}$. 
We apply the replica method to recover this result from the gravity side. This is more challenging than computing the holographic R\'enyi entropy in standard AdS/CFT, since it is sufficient to recover the logarithmic divergent term using the geometry near the asymptotic boundary, and hence the structure deep into the bulk is often ignored \cite{Hung2011}. However, in the AdS/BCFT case, we would like to recover both the logarithmic divergent part and the boundary entropy part, and this requires us to know the details deep into the bulk. 

In this procedure, we get a bonus. We find the usual Ryu-Takayanagi formula \cite{Ryu:2006bv,Ryu:2006ef} for computing the von-Neumann entropy and Dong's formula \cite{Dong2016} for computing the R\'enyi entropy needs to be refined to give the correct boundary entropy contribution. In the standard RT formula and Dong's formula, the bulk dual of the entanglement entropy or the $n$-th modular R\'enyi entropy are given by the length of a geodesic or a cosmic string extending from the asymptotic boundary. However, the length of the geodesic or the cosmic string is divergent at the asymptotic boundary, so one needs to specify a UV-cutoff. One cutoff which is often imposed is a universal cutoff at $z = \epsilon$, where $z$ is the bulk direction of the Poincar\'e coordinate. See Figure \ref{fig:stand_cut}. While this cutoff regime works well to recover the logarithmic divergent part, it cannot correctly recover the boundary entropy part for $n>1$. As a resolution, we treat the entangling surface $\partial A$ as a tiny hole with radius $\epsilon$, and impose a conformal boundary condition on the hole after \cite{Ohmori2015}.\footnote{
One may wonder if this prescription still works in a CFT with a non-zero chiral central charge (CCC), where we cannot impose the Cardy boundary condition.
The answer is no. It is known that, if the CCC is non-zero, we cannot define the Schmidt decomposition or the reduced density matrix \cite{Ohmori2015, Fan2022}, and hence the entanglement entropy cannot be defined properly either.} Corresponding to the boundary associated with the hole, we introduce an end-of-the-world brane on the gravity side. Accordingly, the geodesic or the cosmic string does not extend from the asymptotic boundary but from the end-of-the-world brane, and it ends on another brane. See Figure \ref{fig:brane_cut}.  In this description, we can correctly recover the expected boundary entropy term. The point is that the cosmic string, as a massive object, nontrivially interacts with the brane introduced by the hole at the entangling surface, and the effect from this nontrivial interaction should be taken into account. This prescription also unifies the holographic R\'enyi entropy in AdS/BCFT and the standard AdS/CFT. 

\paragraph{Non-scarlar primaries in BCFT and spinning particles in AdS (Section \ref{sec:spinning})}~\par

We explore the AdS/BCFT correspondence with heavy non-scalar primary insertions on the BCFT side. On the AdS side, correspondingly, we have geometries including branes and spinning particles in them. While it is in general difficult to solve the interaction between heavy objects, thanks to the nature of the 3D Einstein gravity, the metric must be locally AdS, and this allows us to cut and paste the global AdS$_3$ to get the answer. This is exactly how we treat the interaction between the brane and the non-spinning particle above. However, compared to the conical defect geometry which realizes a non-spinning particle, it is less well-studied how to realize a spinning particle starting from global AdS$_3$. 

In order to construct the gravity dual for BCFT with non-scalar primaries, we present a method to construct a spinning defect from global AdS$_3$. While a non-spinning conical defect can be realized by cutting the AdS$_3$ and pasting it along the same time slice, we find that a spinning defect can be realized by cutting the AdS$_3$ and performing a ``twisted identification", i.e. pasting it between different time slices. This procedure perfectly reproduces analytically continued Kerr-BTZ solutions.

Using this construction, we give answers to the following two questions. First, the one-point function of a non-scalar primary must be zero, unlike a scalar primary. This means that there should be a mechanism that prohibits a spinning particle to end on a brane on the gravity side. The question is how it works. Using the construction described above, we can show that if a spinning particle could end on a brane, then it would cause a mismatch on the brane so that it would cause a divergence in the gravitational action. Therefore, a spinning particle ending on a brane is forbidden. In other words, we succeed to give a nice explanation for the fact (\ref{eq:fact3}) from the gravity side.
Second, when considering the gravity dual of two-point functions of non-scalar primaries, the existence of a spinning particle will also cause self-intersections of the brane in a much more complicated way. The question is whether this kind of self-intersections is avoided by black hole formation. We specify the parameter region where self-intersections occur on the gravity side and find that the threshold is exactly the same as the black hole threshold predicted by the conformal bootstrap. Therefore, we succeed to show that the brane self-intersections will not occur when a spinning particle is introduced.

\paragraph{Negative tension brane (Section \ref{sec:negative})}~\par

Our question related to the negative brane case is ``what happens if a brane cross a conical singularity (See Figure \ref{fig:negative})?''.
It seems to drastically change the solution to the Einstein equation when the brane attaches a particle's worldline.
This happens when the brane tension is negative.
For this reason,
it was conjectured in \cite{Bianchi2022} that the boundary primary spectrum may be changed in the negative tension case.
And the authors in \cite{Bianchi2022}  proposed that this ``transition'' can be found by the conformal bootstrap.
In fact, as we will see later,
the solution to the bootstrap equation is not sensitive to the brane tension.
Based on this hint,
we clarify how to deal with the conical singularity behind the brane on the gravity side.
We show that the hidden conical singularity appears as a defect localized on the brane (see Figure \ref{fig:corner}).
This solution can be found if we include a generalized Hayward corner term in the gravity action.
We also give a natural explanation of why we need to include such a term.
Finally, we show that the mass of this geometry with the brane defect precisely matches the CFT prediction.

%%%%%%%%%%%%%%%%%%%%%%%%%%%%%%%%%%%%%%%%%%%%%%%%%%%%%%%%%%%%%%%%%%%%%%%%%%%%%%%%%%%%%%%%%%%%%%
%%%%%%%%%%%%%%%%%%%%%%%%%%%%%%%%%%%%%%%%%%%%%%%%%%%%%%%%%%%%%%%%%%%%%%%%%%%%%%%%%%%%%%%%%%%%%%
\section{Review and Notation}
%%%%%%%%%%%%%%%%%%%%%%%%%%%%%%%%%%%%%%%%%%%%%%%%%%%%%%%%%%%%%%%%%%%%%%%%%%%%%%%%%%%%%%%%%%%%%%
%%%%%%%%%%%%%%%%%%%%%%%%%%%%%%%%%%%%%%%%%%%%%%%%%%%%%%%%%%%%%%%%%%%%%%%%%%%%%%%%%%%%%%%%%%%%%%

In this section, we give a brief review of the boundary conformal field theory and the AdS/BCFT correspondence. We will also fix the notations and terminologies used throughout this paper. 

%%%%%%%%%%%%%%%%%%%%%%%%%%%%%%%%%%%%%%%%%%%%%%%%%%%%%%%%%%%%%%%%%%%%%%%%%%%%%%%%%%%%%%%%%%%%%%
\subsection{Boundary Conformal Field Theory}
%%%%%%%%%%%%%%%%%%%%%%%%%%%%%%%%%%%%%%%%%%%%%%%%%%%%%%%%%%%%%%%%%%%%%%%%%%%%%%%%%%%%%%%%%%%%%%

A boundary conformal field theory (BCFT) is a CFT which is defined on a manifold with boundaries and possesses boundary conditions that maximally preserve the conformal symmetries. In 2D BCFTs, the conformal symmetries can be maximally preserved by the so-called Cardy boundary condition \cite{Cardy1991}, 
\begin{equation}\label{eq:T=T}
\left. \left(T(z) - \bar{T}(\bar{z})\right)\right|_{\text{bdy}} = 0.
\end{equation}
Thanks to this constraint, we can apply the method of images (also called the doubling trick) to relate Ward identity in a BCFT to that in a standard CFT with no boundaries. Therefore, for a 2D BCFT, many ingredients can be fixed merely from the Virasoro algebra.  

Besides the original ingredients inherited from the parent CFT, the presence of the boundaries introduces some new ingredients to the BCFT. The first new ingredient is the boundary condition imposed on the boundary. Let us use $a,b,,c,\cdots$ to denote the boundary conditions. One crucial ingredient related to the boundary condition is the $g$-function. More precisely, the $g$-function of a boundary condition $a$ is defined as the partition function evaluated on a unit disk with $a$ imposed on the boundary: 
\begin{align}
    g^a \equiv \braket{\mathbb{I}}_{{\rm disk}}^a. 
\end{align}
An equivalent quantity, the boundary entropy, is defined as follows, 
\begin{align}
    S_{\rm bdy} \equiv \log g^a. 
\end{align}
The second new ingredient introduced by the boundary is the primary operator living on the boundary. Let us call them boundary primaries. Since one can in principle impose different boundary conditions on the two sides of a boundary primary insertion, boundary primaries are also called the boundary condition changing operators. Let us use capital letters $I,J,K,\cdots$ to label boundary primaries. A boundary primary which is labeled by $I$ and changes the boundary condition $a$ on the left side to the boundary condition $b$ on the right side may be denoted as $\phi_{I}^{ab}$. We choose the normalization such that the boundary two-point functions are given on an upper half plane (UHP) as
\begin{align}\label{eq:metric}
        \braket{\phi^{ab}_I (0) \phi^{bc}_J (x)}_{\rm UHP} = \delta_{IJ} \sqrt{g^{a}  g^{c}   }_I x^{-2h_I},
\end{align}
where $h_I$ is the conformal dimension of $\phi_I^{ab}$. Here, we use $(z,\bar{z}) = (x+i\tau,x-i\tau)$ to parameterize the upper half-plane. The physical region is given by $\tau\geq 0$, and the boundary is given by $\tau = 0$. Associated with boundary primaries, we can consider the operator product expansion (OPE) between them:
\begin{align}
        \phi^{ab}_I (0) \phi^{bc}_J (x) \sim \sum_K C^{abc}_{IJK} x^{h_K-h_I-h_J} \phi^{ac}_K(x)+\cdots. 
\end{align}
The OPE coefficients $C^{abc}_{IJK}$ appearing here are called the boundary-boundary-boundary OPE coefficients. Besides the boundary primaries, there are the primary operators inherited from the original parent CFT live in the bulk of the BCFT, and we call them the bulk primaries. We use lowercase letters $i,j,k,\cdots$ to label bulk primaries. The OPE coefficients appearing in the OPE between bulk primaries 
\begin{align}
    \phi_i (0) \phi_j (z) \sim \sum_k C_{ijk} z^{h_k-h_i-h_j}   \bar{z}^{\bar{h}_k-\bar{h}_i-\bar{h}_j}  \phi_k(z)+\cdots.
\end{align}
are called the bulk-bulk-bulk OPE coefficients. Besides all the ingredients introduced above, there is also an OPE which expands bulk primaries in terms of boundary primaries: 
\begin{align}
        \phi_i (z) \sim \sum_I (2 \Im z)^{h_I-h_i-\bar{h}_i} C^a_{iI} \phi^{aa}_I(\Re z) + \cdots. 
\end{align}
The OPE coefficients $C^{a}_{iI}$ appearing here are called the bulk-boundary OPE coefficients.

\begin{figure}[H]
 \begin{center}
  \includegraphics[width=8.0cm,clip]{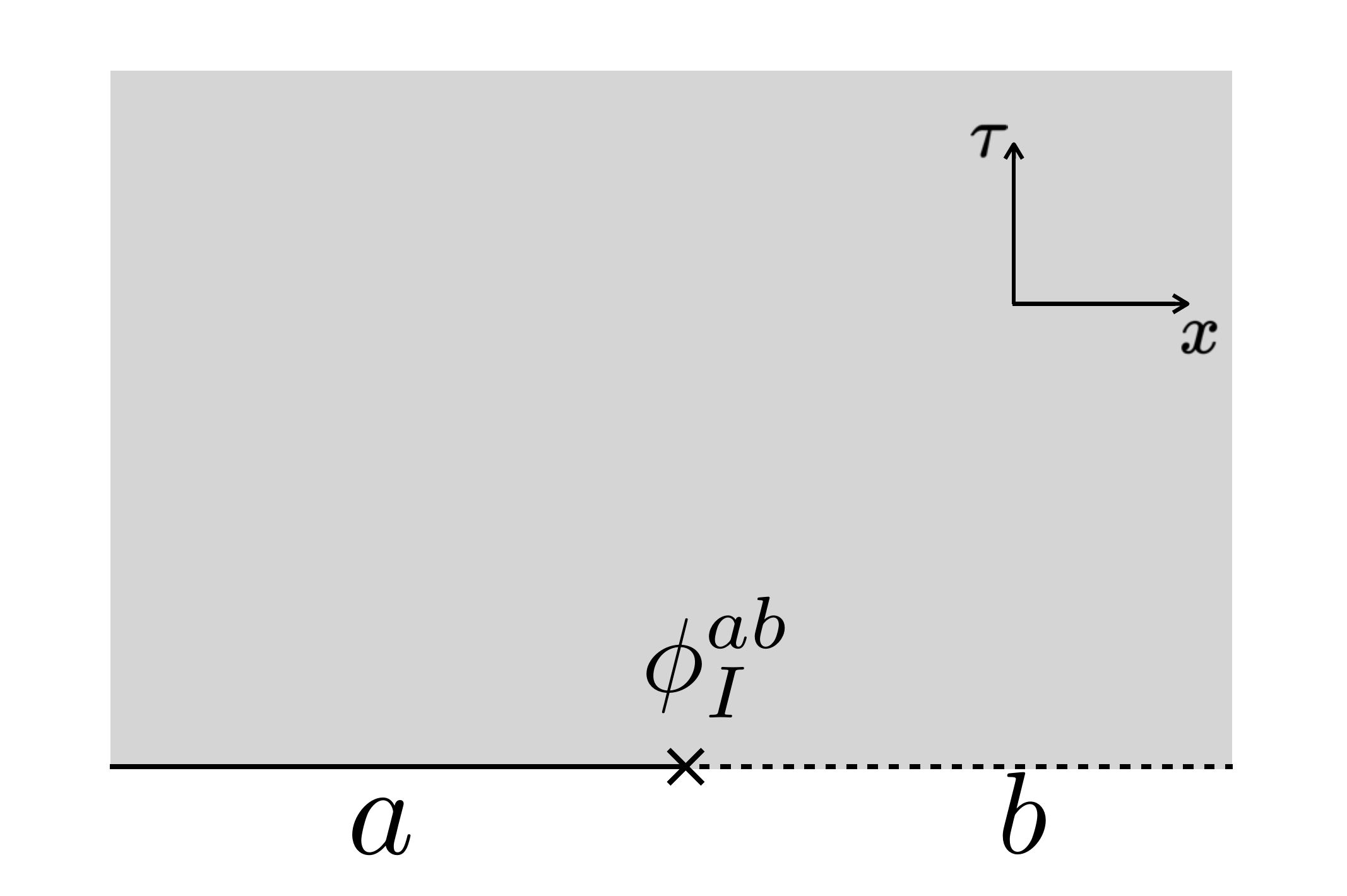}
 \end{center}
 \caption{A boundary primary $\phi^{ab}_I$ defined on an upper half plane. The superscript $ab$ means that the boundary conditions on the two sides of $\phi^{ab}_I$ are $a$ and $b$, respectively. The subscript $I$ labels the primary field. }
 \label{fig:bdyop}
\end{figure}

Boundary conditions can be mapped to states (called boundary states) in a similar way to the state/operator correspondence.
The Cardy boundary condition (\ref{eq:T=T}) can be re-expressed in terms of the boundary state as
\begin{equation}
L_n \ket{B^a} = \bar{L}_{-n} \ket{B^a},
\end{equation}
where $L_n$ is the Virasoro generator and we label the boundary condition by the superscript $a$.
It is known that the solution to this equation can be expanded in the basis of Ishibashi states,
\begin{equation}\label{eq:Ishibashi}
|{j}\rangle\rangle \equiv \sum_{N} \ket{j;N} \otimes   U \overline{\ket{j;N}},
\end{equation}
Here, the state $\ket{j;N}$ is in the Verma module labeled by $j$, and the state itself is labeled by $N$. The operator $U$ appearing in front of the anti-holomorphic part is an anti-unitary operator. By definition,
the coefficients of the boundary state when expanding in the Ishibashi state basis are given by the bulk-boundary OPE coefficients. Therefore, the boundary state can be written in the following way. 
\begin{equation}
\ket{B^a} = g^a \sum_j C^a_{j \mathbb{I}} \kket{j}.
\end{equation} 
In diagonal RCFTs, the boundary state consistent with the modular bootstrap equation is known as the Cardy state,
\begin{equation}\label{eq:Cardy_state}
\ket{\ti{a}} = \sum_j \fr{S_{aj}}{\sqrt{S_{\mathbb{I}j}}} \ket{j}\rangle,
\end{equation}
where $S_{aj}$ is the modular-S matrix.
If one can figure out the coefficients appearing in this expansion,
one can find
\begin{equation}
g^a = \braket{\mathbb{I}}_a =  \fr{S_{a\mathbb{I}}}{\sqrt{S_{\mathbb{I}\mathbb{I}}}},
\end{equation}
and\footnote{
We will use the same character $\mathbb{I}$ for the boundary vacuum and the bulk vacuum.
We distinguish them by the rule that for the bulk-boundary OPE coefficients (e.g., $C_{\mathbb{I} \mathbb{I}}^a$), the left index describes the bulk primary and the right index describes the boundary primary. 
}
\begin{equation}
C^a_{i \mathbb{I}} =\fr{\braket{i|\ti{a}}}{\braket{0|\ti{a}}  } = \fr{S_{ai} \sqrt{S_{\mathbb{I}\mathbb{I}}}}{S_{a\mathbb{I}}\sqrt{S_{\mathbb{I}i}}}.
\end{equation}
Besides these results from the modular bootstrap,
other OPE coefficients can be fixed by the conformal bootstrap equation for the bulk and boundary operators.

%%%%%%%%%%%%%%%%%%%%%%%%%%%%%%%%%%%%%%%%%%%%%%%%%%%%%%%%%%%%%%%%%%%%%%%%%%%%%%%%%%%%%%%%%%%%%%
\subsection{Holographic Dual of BCFT}\label{sec:review_adsbcft}
%%%%%%%%%%%%%%%%%%%%%%%%%%%%%%%%%%%%%%%%%%%%%%%%%%%%%%%%%%%%%%%%%%%%%%%%%%%%%%%%%%%%%%%%%%%%%%

We will then explain how the gravity dual is constructed for a BCFT. Here, we will focus on the minimal bottom-up construction\footnote{See, for example, \cite{DHoker2007,DHoker2007a,Fujita2011,Aharony2011,Chiodaroli2012,Chiodaroli2012a,Berdichevsky2013,Uhlemann:2021nhu,VanRaamsdonk:2021duo,Uhlemann:2021itz,Coccia:2021lpp,Martinec:2022ofs,Karch:2022rvr} for various top-down constructions for gravity duals of BCFTs.} introduced in \cite{Takayanagi2011,Fujita2011}. For a holographic BCFT defined on a manifold $\Sigma$ with a nontrivial boundary $\partial \Sigma$, its gravity dual $\CM$ is given by solving the Einstein equation induced from the following action. 
\begin{align}\label{eq:gravaction_brane}
    I_{\rm grav}[\CM] = -\frac{1}{16\pi G_N}\int_{\CM} \sqrt{g}(R+2) - \frac{1}{8\pi G_N} \int_Q \sqrt{h} (K-T).
\end{align}
The first term is the standard Einstein-Hilbert term where $G_N$ is the Newton constant, $g_{\m\n}$ is the metric of $\CM$, and $R$ is the scalar curvature. Here we have set the cosmological constant to $-1$. In this case, the AdS radius turns out to be $1$. The second term is the Gibbons-Hawking term on the end-of-the-world brane $Q$, where $K$ is its extrinsic curvature, $T$ is its tension, and $h_{ab}$ is the induced metric on it. We have omitted the Gibbons-Hawking term on the asymptotic boundary $\Sigma$ and the counter term for simplicity. 

The end-of-the-world brane ends on the boundary of $\Sigma$, i.e. $\partial Q = \partial \Sigma$. The boundary of the gravity dual is given by $\partial \CM = \Sigma \cup Q$. The boundary condition imposed on the brane $Q$ is of the Neumann type, 
\begin{align}
    K_{ab} - K h_{ab} -T h_{ab} = 0,
\end{align}
which fixes a brane profile for a given setup as we will see below.
This makes a distinction from the asymptotic boundary on which the Dirichlet boundary condition is imposed. 

To be concrete, let us present some simple examples of the gravity duals. 

\paragraph{Gravity dual of the BCFT on a disk}~\par

\begin{figure}[H]
    \centering
    \includegraphics[width=12cm]{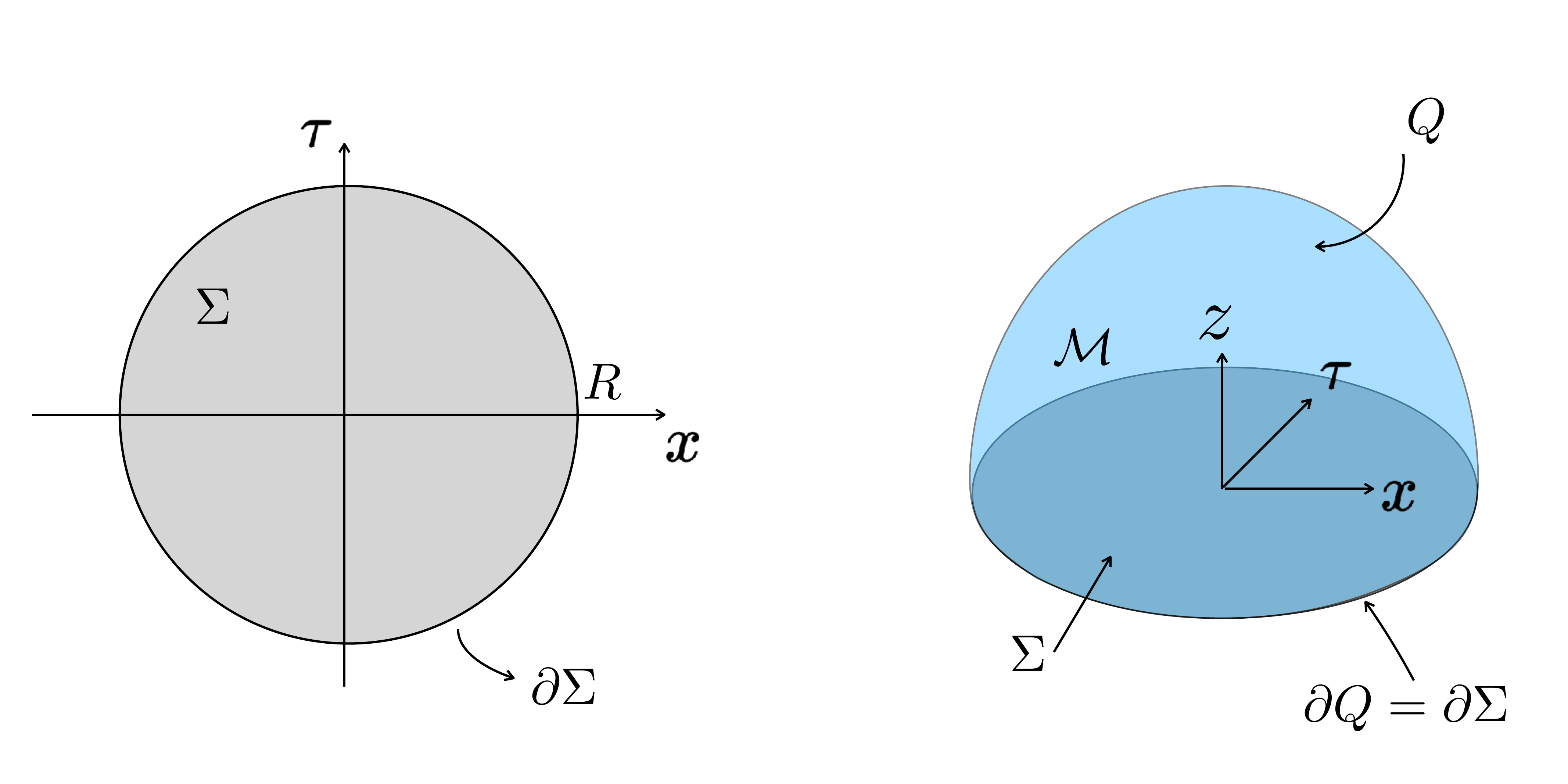}
    \caption{BCFT defined on a disk (left) and its gravity dual (right).}
    \label{fig:disk_adsbcft}
\end{figure}

Let us first consider the case where $\Sigma$ is a disk with radius $R$. We can use $w=x+i\tau$ and $\bar{w} = x-i\tau$ to parameterize it. The disk is given by $x^2 + \tau^2 \leq R^2$. In this case, it is straightforward to find that the gravity dual $\CM$ is a region surrounded by $Q$ and $\Sigma$, where the metric is 
\begin{align}
    ds^2_{\CM} = \frac{dw d\bar{w} + dz^2}{z^2},
\end{align}
and the brane profile is 
\begin{align}
    w\bar{w} + (z-R \sinh \left[{\rm arctanh}(T)\right])^2 = \left(R \cosh \left[{\rm arctanh}(T)\right]\right)^2,
\end{align}
which intersects with the asymptotic boundary $z=0$ at
\begin{align}
    |w| = R.
\end{align}
See Figure \ref{fig:disk_adsbcft} for a sketch of $\CM$. 

Since the $g$-function is defined as the partition function evaluated on a unit disk, and the boundary entropy $S_{\rm bdy}$ is given by $S_{\rm bdy} \equiv \log g$, by computing the partition function with saddle point approximation on the gravity side, one can figure out that the boundary entropy is given by \cite{Takayanagi2011}
\begin{align}
    S_{\rm bdy} =\frac{1}{4G_N} {\rm arctanh}(T). 
\end{align}
This result will be extensively used later.

\paragraph{Gravity dual of the BCFT on a half plane}~\par

\begin{figure}[H]
    \centering
    \includegraphics[width=12cm]{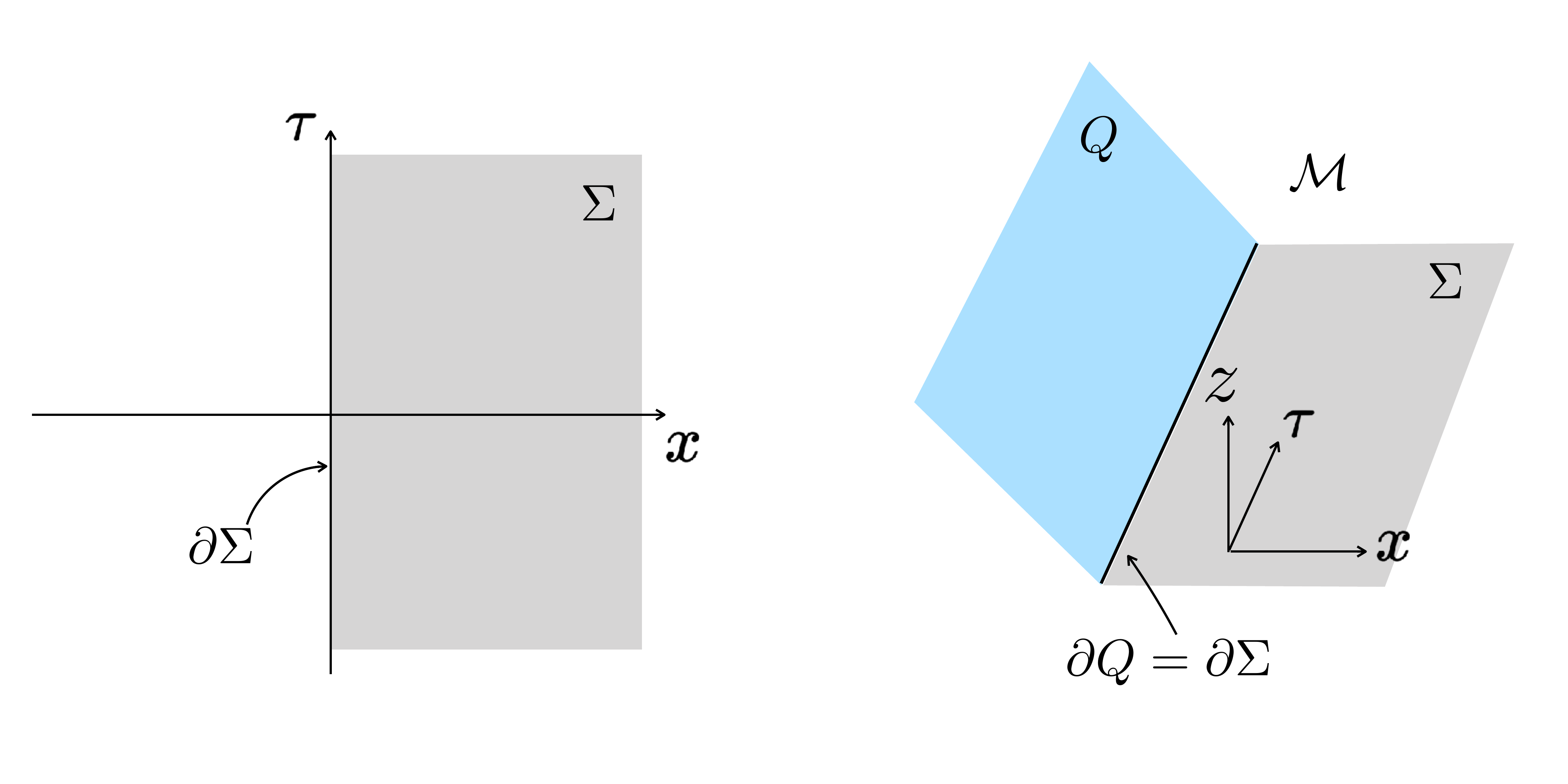}
    \caption{BCFT defined on a half plane (left) and its gravity dual (right).}
    \label{fig:HP_adsbcft}
\end{figure}

Let us then consider a half-plane parameterized by $(x,\tau)$ and given by $x\geq0$. The gravity dual is a Poincar\'e AdS with metric 
\begin{align}
    ds^2 = \frac{dz^2 + dx^2 + d\tau^2}{z^2}. 
\end{align}
The end-of-the-world brane $Q$ locates at 
\begin{align}
    T z + \sqrt{1-T^2}~x = 0. 
\end{align}
The physical region is given by 
\begin{align}
    T z + \sqrt{1-T^2}~x \geq 0. 
\end{align}
See Figure \ref{fig:HP_adsbcft}. Since everything is extremely simple in this setup, when considering a BCFT with a complicated shape, it is convenient to map it into a half plane, and then compute physical quantities or construct the complicated gravity dual with the help of the one associated with the half-plane \cite{Ugajin2013,Shimaji:2018czt,Caputa:2019avh,Akal2020,Akal2021,Akal:2021dqt,Akal:2022qei}.

\paragraph{Gravity dual of the BCFT on an infinite strip}~\par

Last but not least, let us present the gravity dual of a BCFT defined on an infinite strip. This can be realized by introducing an end-of-the-world brane in global AdS$_3$. This setup will be the main starting point of many setups discussed in this paper.  

\begin{figure}[H]
    \centering
    \includegraphics[width=12cm]{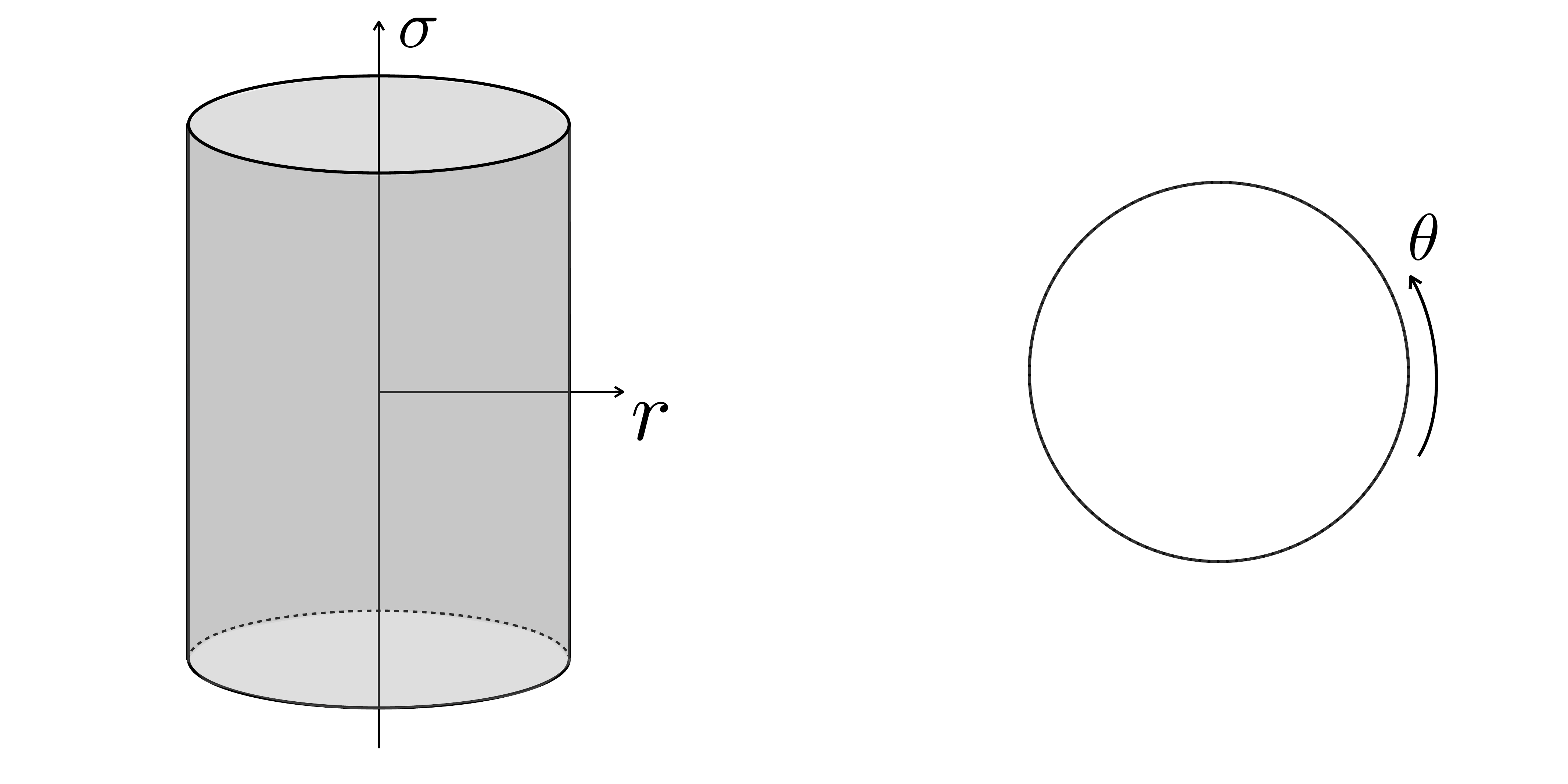}
    \caption{Coordinates in global AdS$_3$ (left) and its time slice (right).}
    \label{fig:global_coord}
\end{figure}

Let us first have a look on a Euclidean global AdS$_3$ with no brane in it. The metric is given by
\begin{align}\label{eq:globalAdS}
    ds^2 = (r^2+1)d\sigma^2 + \frac{dr^2}{r^2+1} + r^2 d\theta^2, 
\end{align}
with $\theta\in[0,2\pi)$. See Figure \ref{fig:global_coord} for a sketch. We will refer to the $\sigma,r, \theta$ direction as the (Euclidean) time direction, the radial direction, and the angular direction, respectively. The global AdS$_3$ corresponds to a CFT defined on a cylinder. The perimeter of the cylinder depends on the holographic renormalization regime one takes to get to gravity dual. The holographic energy stress tensor evaluated in this coordinate is 
\begin{align}
    T_{\sigma\sigma}^{\rm (hol)} = -\frac{1}{16\pi G_N},
\end{align}
and the ADM mass turns out to be 
\begin{align}
    M_\sigma = \int_0^{2\pi} d\theta~T_{\sigma\sigma}^{\rm (hol)} = -\frac{1}{8G_N}.
\end{align}
By identifying $1/4G_N$ with $c/6$, $T_{\sigma\sigma}^{\rm (hol)}$ and $M_{\sigma}$ match the energy density and the Casimir energy of the vacuum state in a CFT with central charge $c$ defined on a cylinder with perimeter $2\pi$, parameterized by $(\sigma,\theta)$.\footnote{Note that the standard convention in CFT differs from the holographic computation here by a coefficient as $T^{\rm (CFT)} = 2\pi T^{\rm (hol)}$. } 

Note that if we want to identify the dual CFT as one defined on a cylinder with perimeter $2L$, it would be convenient to perform the coordinate transformation $\tau = (L/\pi)\sigma$ and $x = (L/\pi) \theta$ with which the metric turns out to be 
\begin{align}
    ds^2 = (r^2+1)\left(\frac{\pi}{L}\right)^2 d\tau^2 + \frac{dr^2}{r^2+1} + r^2 \left(\frac{\pi}{L}\right)^2 dx^2 .
\end{align}
The holographic energy stress tensor and the ADM mass then turn out to be 
\begin{align}
    T_{\tau\tau}^{\rm (hol)} = -\frac{1}{16\pi G_N} \left(\frac{\pi}{L}\right)^2,
\end{align}
and 
\begin{align}
    M_{\tau} =  \int_0^{2L} dx~ T_{\tau\tau}^{\rm (hol)} =  -\frac{1}{8G_N}\left(\frac{\pi}{L}\right).
\end{align}
These match the CFT results defined on a cylinder with perimeter $2L$, parameterized by $(\tau,x)$.

\begin{figure}[H]
    \centering
    \includegraphics[width=12cm]{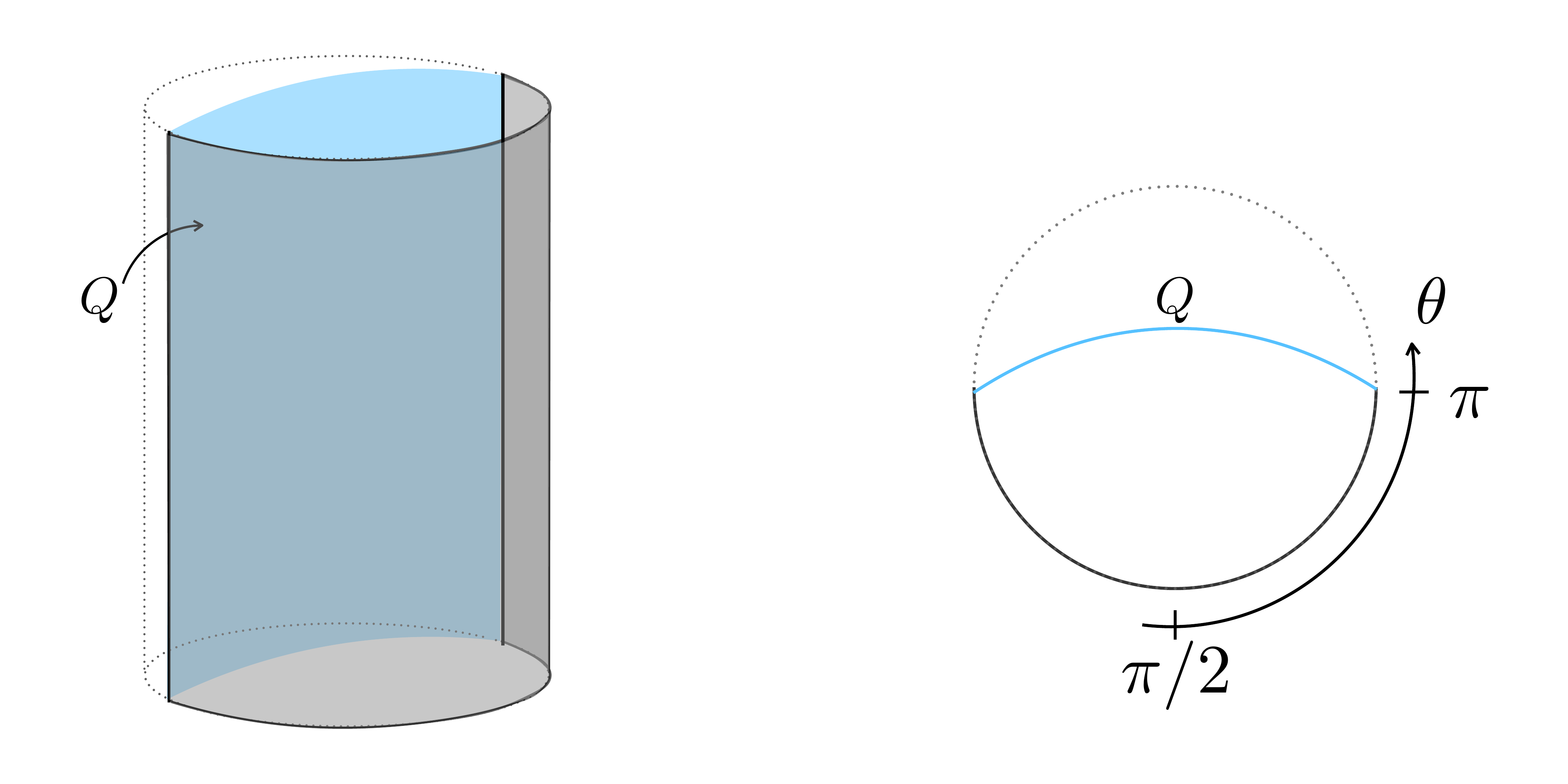}
    \caption{Global AdS$_3$ with an end-of-the-world brane (left) and its time slice (right).}
    \label{fig:onlybrane}
\end{figure}

Introducing an end-of-the-world brane with tension $T>0$ in the global AdS results in the following brane profile:
\begin{align}
    \sin \theta = -\frac{T}{\sqrt{1-T^2}} \frac{1}{r}, \ \ \ \ \ \ \ \ \ (\pi \leq \theta \leq 2\pi)
\end{align}
which intersects the asymptotic boundary at $\theta = \pi, 2\pi\sim0$. In this case, the bulk region is still covered by $\theta \in [0, 2\pi)$, while only the $\theta \in [0,\pi]$ region remains on the asymptotic boundary. See Figure \ref{fig:onlybrane}. This corresponds to a CFT defined on a strip parameterized by $(\sigma,\theta)$ with length $\pi$. Performing the coordinate transformation given above to the $(\tau,r,x)$ coordinate makes it convenient to consider the corresponding BCFT as an infinite strip with length $L$ parameterized by $(\tau,x)$. In this case, the holographic energy stress tensor is again evaluated as 
\begin{align}
    T_{\tau\tau}^{\rm (hol)} = -\frac{1}{16\pi G_N} \left(\frac{\pi}{L}\right)^2. 
\end{align}
The ADM mass then turns out to be 
\begin{align}\label{eq:M_onlyb}
    M_{\tau} =  \int_0^{L} dx~ T_{\tau\tau}^{\rm (hol)} =  -\frac{1}{16G_N}\left(\frac{\pi}{L}\right).
\end{align}
These again match the energy stress tensor and the Casimir energy evaluated from the BCFT side. 

%%%%%%%%%%%%%%%%%%%%%%%%%%%%%%%%%%%%%%%%%%%%%%%%%%%%%%%%%%%%%%%%%%%%%%%%%%%%%%%%%%%%%%%%%%%%%%
%%%%%%%%%%%%%%%%%%%%%%%%%%%%%%%%%%%%%%%%%%%%%%%%%%%%%%%%%%%%%%%%%%%%%%%%%%%%%%%%%%%%%%%%%%%%%%
\section{Self-intersection and Black Hole Threshold}\label{sec:intersection}
%%%%%%%%%%%%%%%%%%%%%%%%%%%%%%%%%%%%%%%%%%%%%%%%%%%%%%%%%%%%%%%%%%%%%%%%%%%%%%%%%%%%%%%%%%%%%%
%%%%%%%%%%%%%%%%%%%%%%%%%%%%%%%%%%%%%%%%%%%%%%%%%%%%%%%%%%%%%%%%%%%%%%%%%%%%%%%%%%%%%%%%%%%%%%

\begin{figure}[H]
 \begin{center}
  \includegraphics[width=10.0cm,clip]{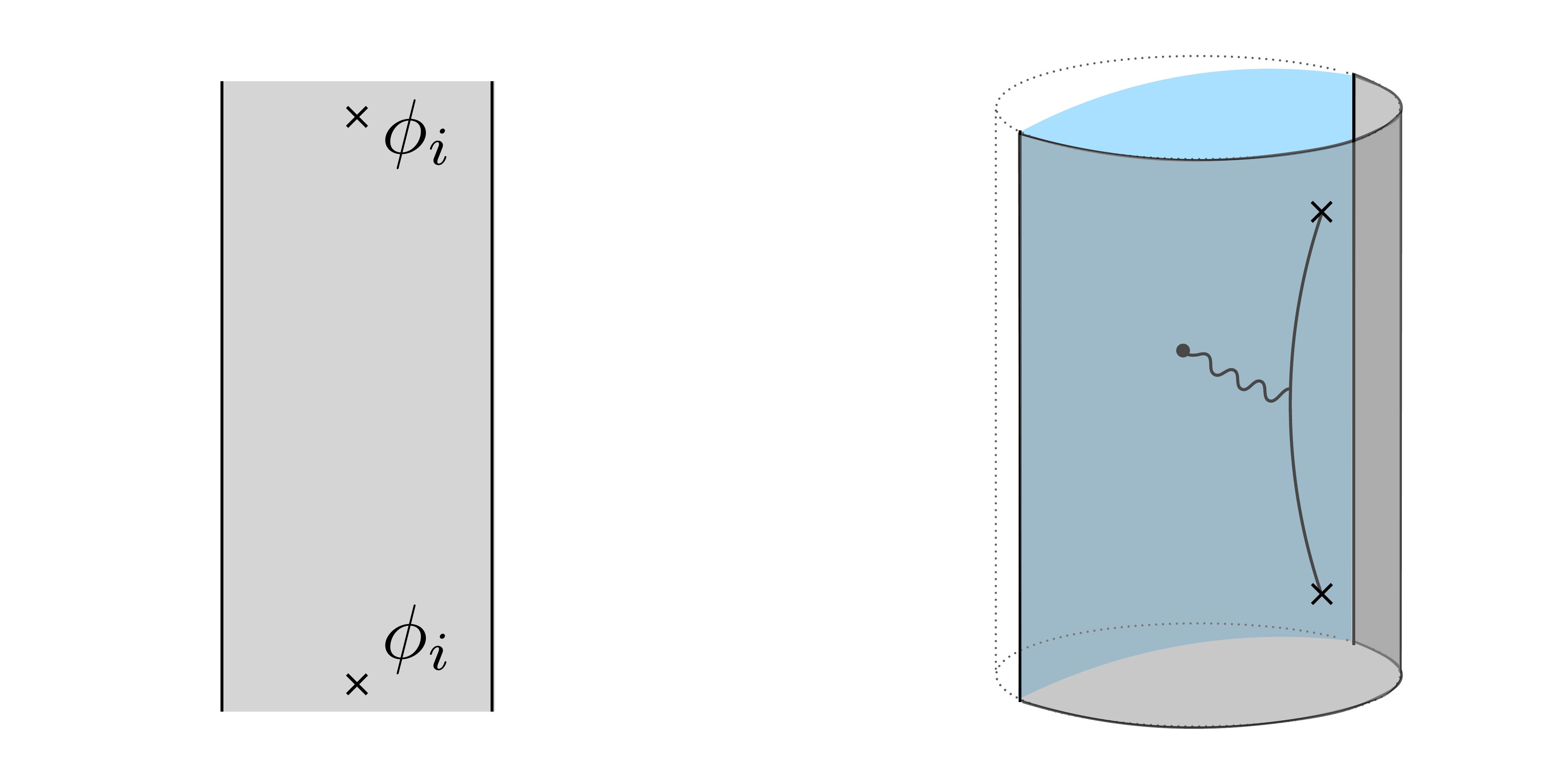}
 \end{center}
 \caption{Two point function of bulk primaries defined on an infinite strip (left) and its gravity dual (right). For convenience, we depict the primary insertions at finite points in these pictures. The primary insertions introduce a massive particle (solid black curve) in the gravity dual, which interacts with the brane (shown in blue). The interaction between the massive particle and the brane is schematically depicted as a wave line. 
 }
 \label{fig:setup}
\end{figure}

If a conical defect is introduced in an AdS spacetime with an end-of-the-world brane, the brane will be bent due to the gravitational interaction. When the deficit angle exceeds a certain threshold, the brane will intersect itself \cite{Cooper2019, Geng:2021iyq, Kawamoto2022, Bianchi2022}. Configurations with brane self-intersections indeed are solutions to the Einstein equation. The question is whether these configurations are physical or not, and if not, in what sense. 

A conical defect can be caused by a massive point particle (with no angular momentum), and a massive particle can be introduced in AdS$_3$ by considering a bulk-primary two-point function on an infinite strip $\braket{\phi_i(-i \infty) \phi_i(i \infty) }_{\text{strip}}$, where $\phi_i$ is a scalar parimary whose conformal dimension $h_i = \bar{h}_i$ is at $\mathcal{O}(c)$ (see Figure \ref{fig:setup}).
Its holographic dual has an end-of-the-world brane and a massive particle. 

\begin{figure}[H]
    \centering
    \includegraphics[width=4cm]{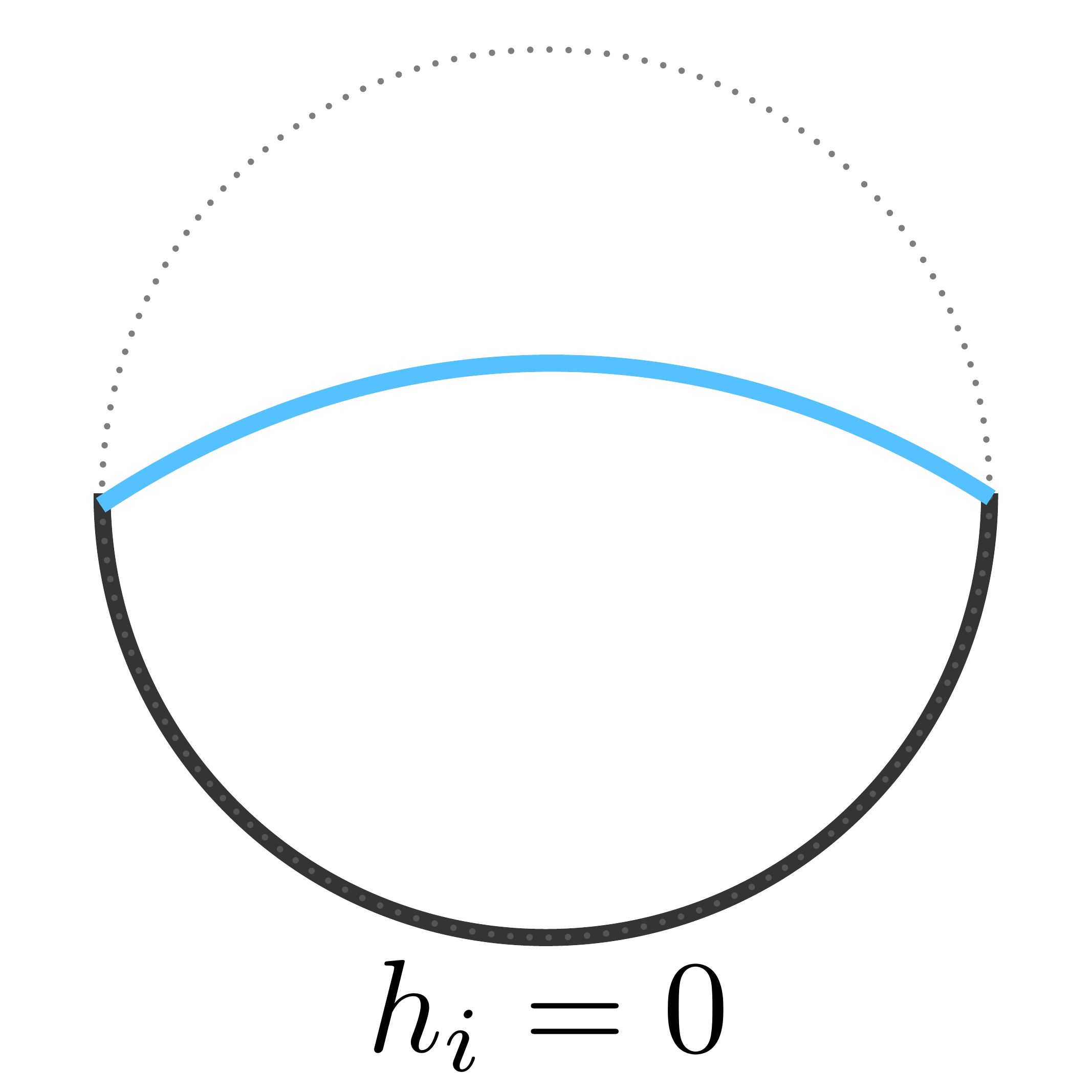}
    \includegraphics[width=4cm]{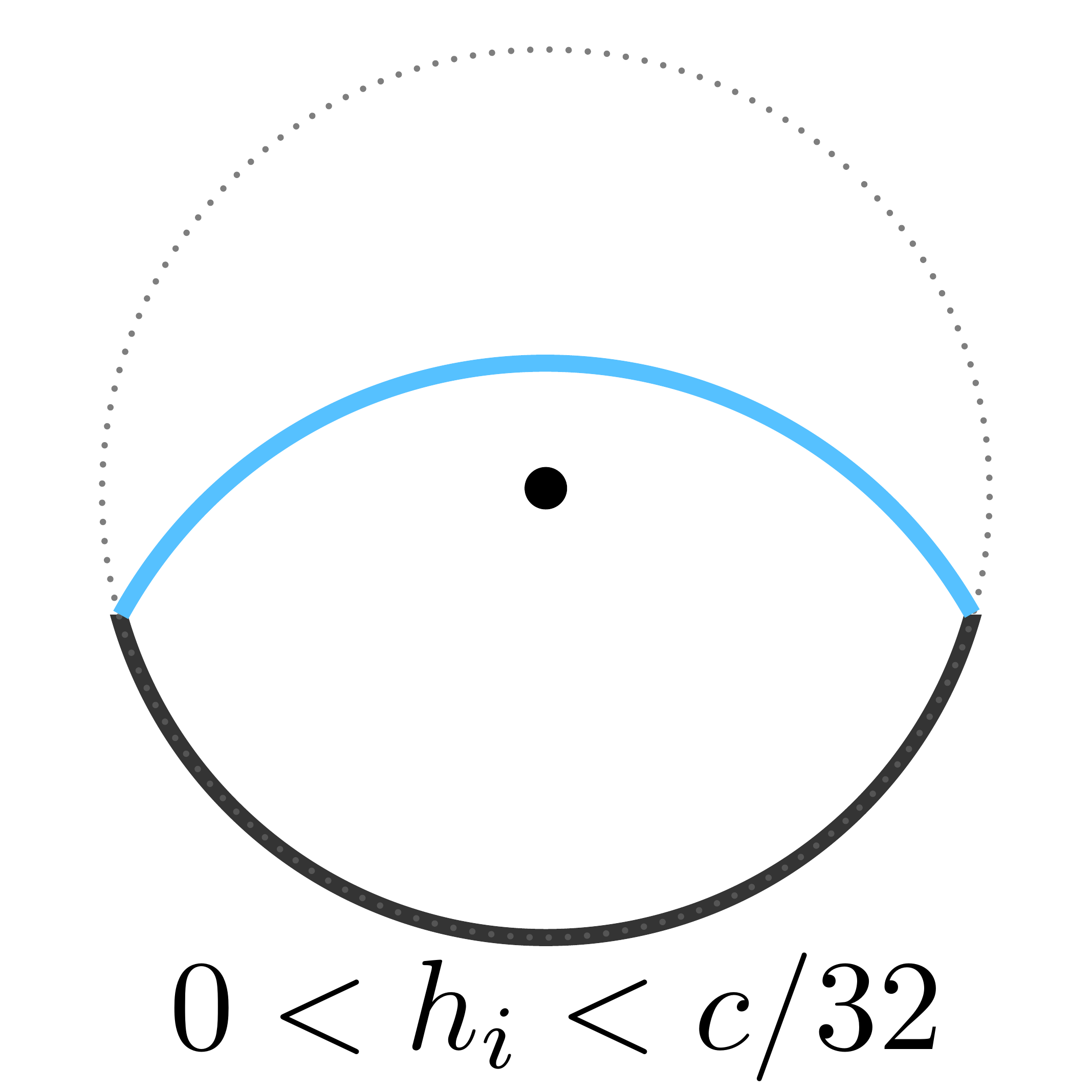}
    \includegraphics[width=4cm]{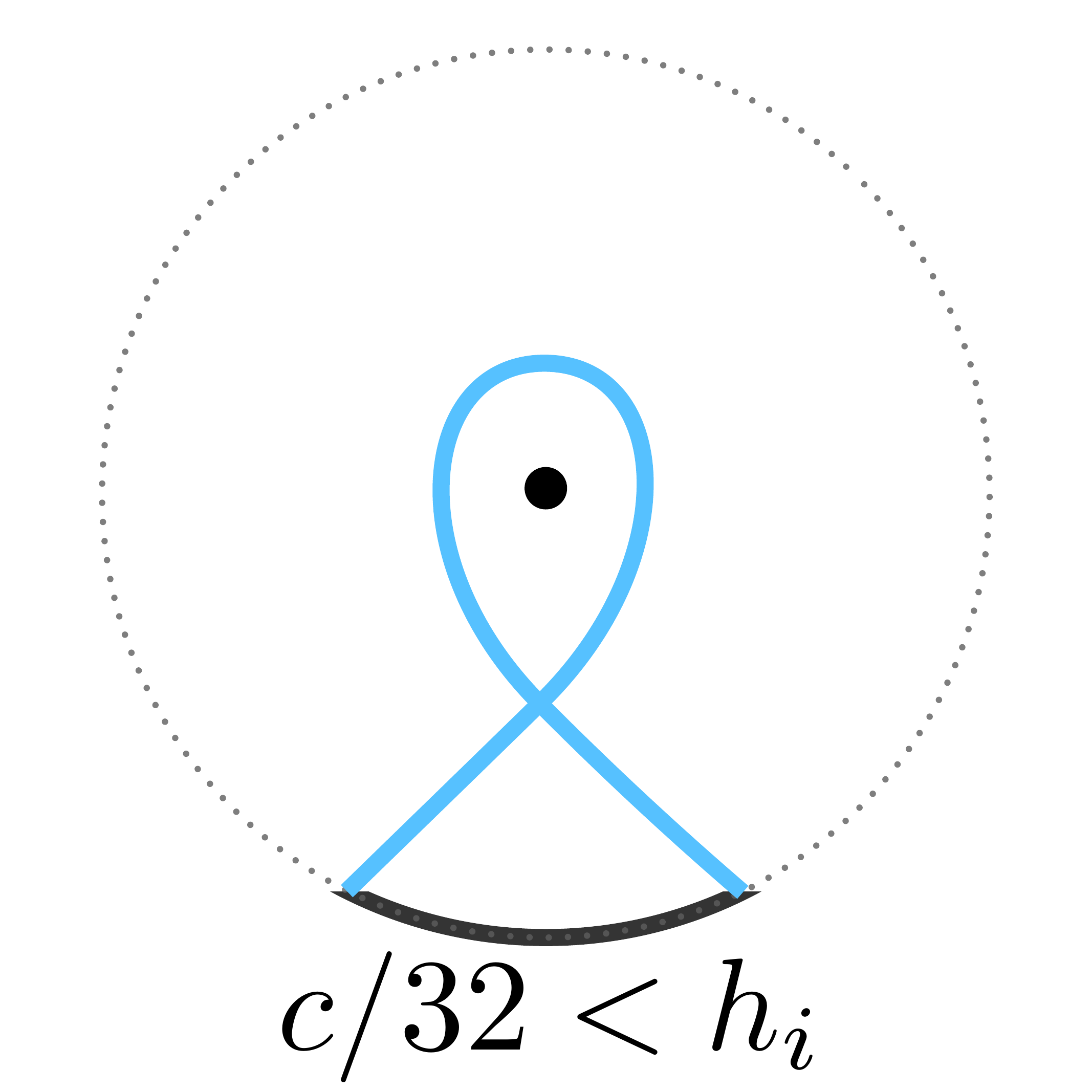}
    \caption{Time slices of global AdS$_3$ with a brane (shown in blue) and a conical defect (the black dot at the center). The conical defect is created by inserting scalar primaries with conformal dimension $h_i=\bar{h}_i$ at two infinities in the BCFT dual. When $h_i=\bar{h}_i$ exceeds $c/32$, the deficit angle around the conical defect will exceed $\pi$, and the brane will intersect itself.}
    \label{fig:self}
\end{figure}

As we will review in Section \ref{sec:bulk-two}, a scalar primary with scaling dimension $\Delta_i=h_i+\bar{h}_i$ produces a conical defect with deficit angle $\delta\theta= 2\pi \pa{1-\sqrt{1-\fr{24 h_i}{c}}}$. On the other hand, for the brane to satisfy the Einstein equation, the visual angle when looking at the defect from the two edges of the brane must be $\pi$. Therefore, when $\delta \theta > \pi$, the brane will intersect with itself. This happens at $h_i > \frac{c}{32}$ (see Figure \ref{fig:self}). 
In the following, we refer to the physical condition $h_i<\fr{c}{32}$ as the self-intersection bound. 

If one considers brane self-intersections to be unphysical, one may exclude the parameter regions where self-intersections occur. Therefore, one naive resolution to the self-intersection described above is to exclude all scalar primaries with $h_i \in \pa{\fr{c}{32},\fr{c}{24} }$.\footnote{This argument is based on the philosophy in \cite{Geng:2021iyq}. However, the authors of \cite{Geng:2021iyq} related their gravitational setups to boundary primaries in BCFT, while we are considering bulk primaries in BCFT. As emphasized throughout our paper, boundary primaries and bulk primaries are two distinct sets of objects and should be treated in a different way.} Here, the value $\frac{c}{24}$ comes from the usual BTZ black hole threshold. A scalar primary with $h_i > \frac{c}{24}$ will directly form a black hole on the gravity side where self-intersections are avoided.\footnote{
There is an analog of the Hawking-Page transition in AdS/BCFT \cite{Fujita2011, Takayanagi2011}.
} However, there is no reason to expect the above ``resolution" should work on the BCFT side. 

In fact, one cannot apply the black hole threshold $\frac{c}{24}$ naively to gravity with a brane
because both the brane and the particle are massive objects and the effect induced by the gravitational interaction between them should also be accounted. This interaction is expected to change the black hole threshold. On the BCFT side, correspondingly, the presence of the bulk-boundary OPE will change the black hole threshold. With the doubling trick \cite{Cardy2004}, we can translate it into the interaction with its mirror image. 
This interaction makes the binding energy of the one-particle state in BCFT (or the two-particle state formed from the original primaries and their mirror images) non-trivial.

In this section, we will first give the correct black hole threshold from the BCFT side by using the conformal bootstrap and show that it completely matches the self-interaction bound $\frac{c}{32}$. Then we will explain in the gravity dual, how the interaction between the massive particle and the brane changes the black hole threshold. 
All these results imply that the self-intersection can always be avoided by the black hole formation, and need not be treated in a particular way.
Note that the CFT analysis is similar to that presented in \cite{Kusuki2022}, but here we will present a more general statement and alternative interpretations. 

%%%%%%%%%%%%%%%%%%%%%%%%%%%%%%%%%%%%%%%%%%%%%%%%%%%%%%%%%%%%%%%%%%%%%%%%%%%%%%%%%%%%%%%%%%%%%%
\subsection{Bootstrapping a Two-point Function on a Strip}\label{sec:two-CFT}
%%%%%%%%%%%%%%%%%%%%%%%%%%%%%%%%%%%%%%%%%%%%%%%%%%%%%%%%%%%%%%%%%%%%%%%%%%%%%%%%%%%%%%%%%%%%%%

\newsavebox{\boxba}
\sbox{\boxba}{\includegraphics[width=90pt]{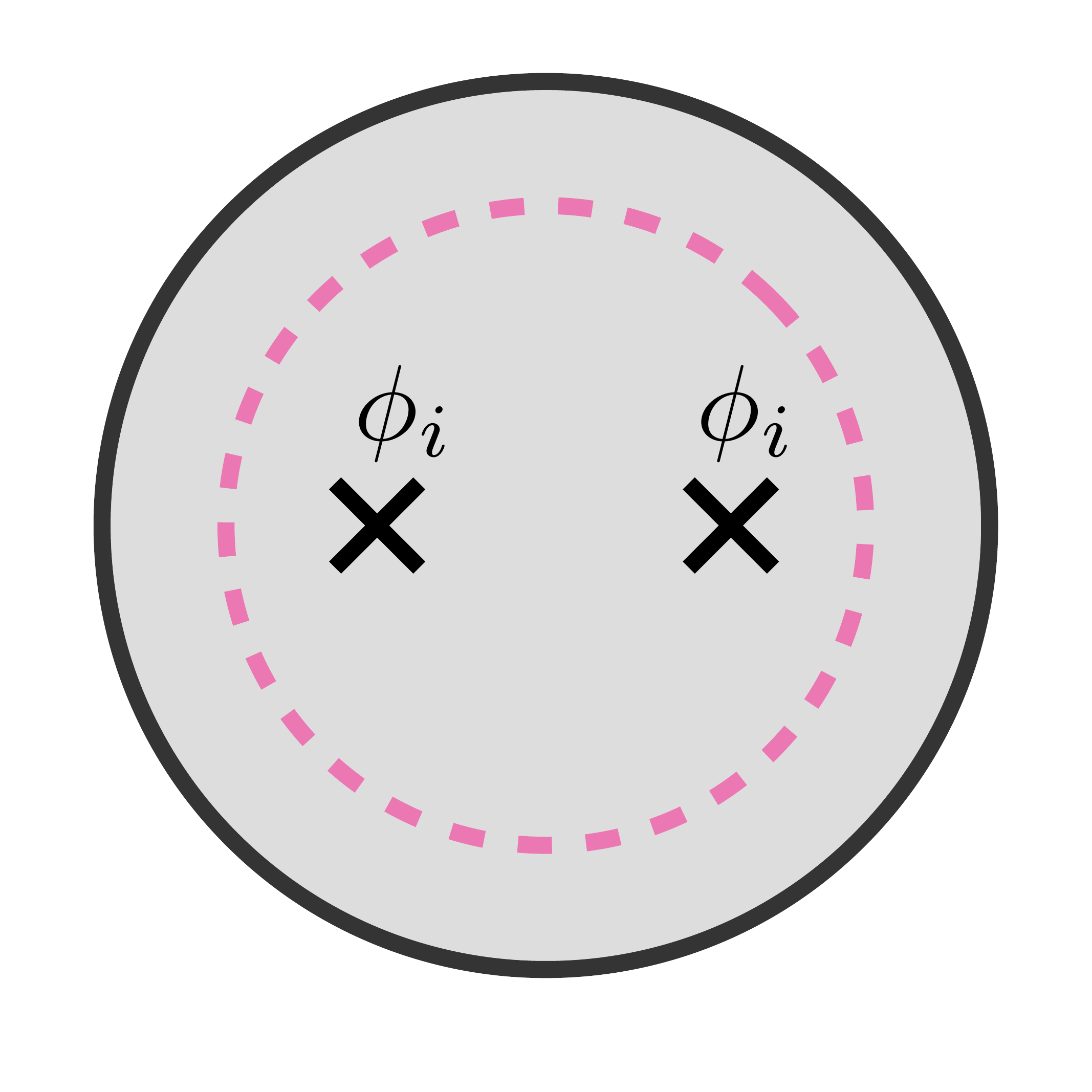}}
\newlength{\boxbaw}
\settowidth{\boxbaw}{\usebox{\boxba}} 

\newsavebox{\boxbb}
\sbox{\boxbb}{\includegraphics[width=90pt]{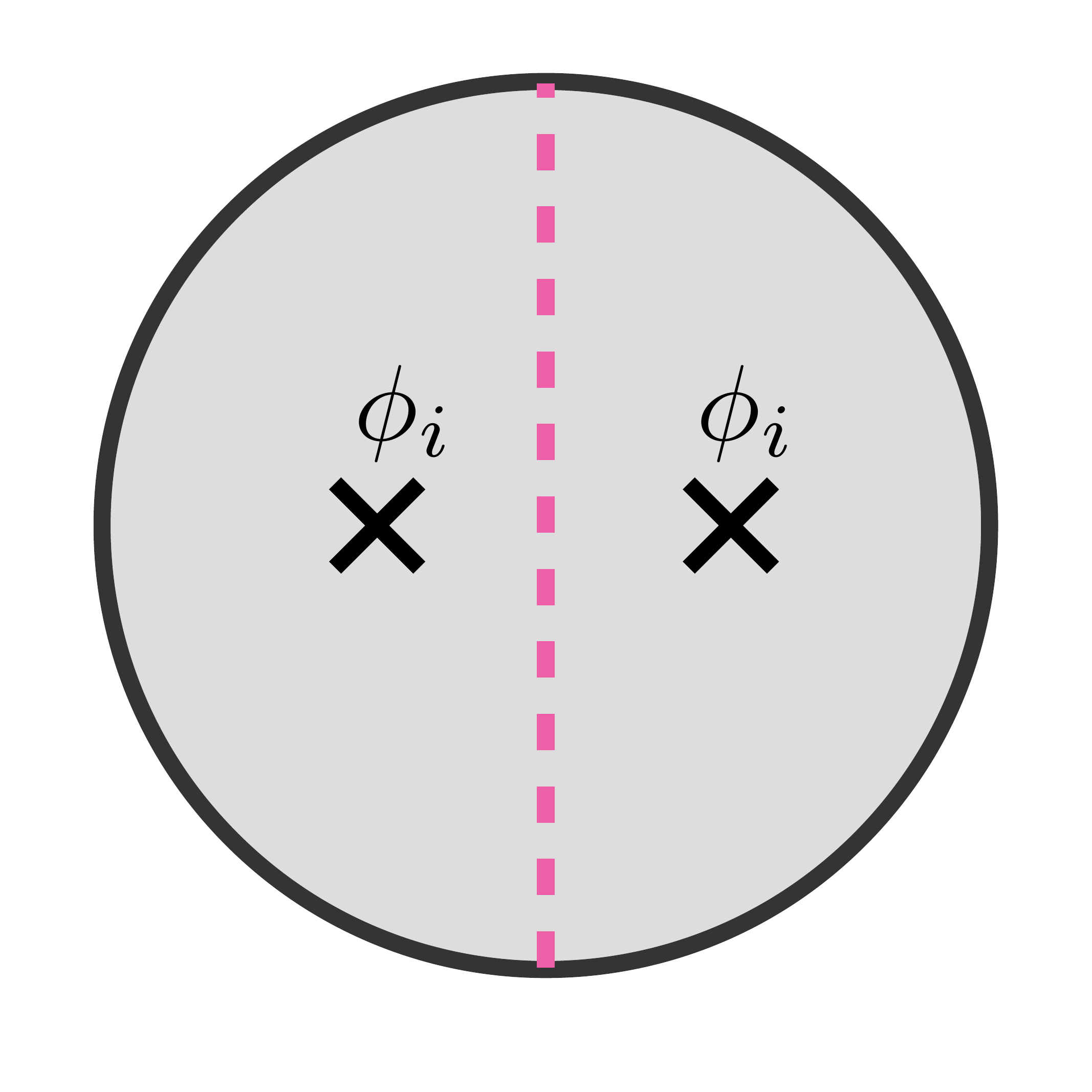}}
\newlength{\boxbbw}
\settowidth{\boxbbw}{\usebox{\boxbb}} 

Let us consider a bulk primary two-point function on a strip (see Figure \ref{fig:setup}),
\begin{equation}
\braket{\phi_i(x_1)\phi_i(x_2)}_{\text{strip}}.
\end{equation}
where $x_1$ and $x_2$ are largely separated from each other.
This correlation function can be mapped to that  on an UHP by the following conformal transformations,
\begin{equation}
\begin{aligned}
z_\text{{UHP}} &=e^{\fr{\pi}{L} z_{\text{strip}}}, \\
\end{aligned}
\end{equation}
where $L$ is the width of the strip.
If we define the conformal map,
\begin{equation}
f(z) =e^{\fr{\pi}{L} z}
\end{equation}
the two-point function on a strip can be re-expressed as
\begin{equation}
\abs{
\pa{\left. \fr{d f(x)}{d x} \right|_{x=x_1}}^{h_i}
\pa{\left. \fr{d f(x)}{d x} \right|_{x=x_2}}^{h_i}
}^2
\braket{\phi_i(z_1)\phi_i(z_2)}_{\text{UHP}},
\end{equation}
where $z_k=f(x_k)$.

To compute a bulk primary two-point function on an UHP, we can insert the identity operator $\Sigma_p |p\rangle \langle p|$ at two different places (depicted as the red dashed line) shown below,
\begin{equation}
\parbox{\boxbaw}{\usebox{\boxba}} = \parbox{\boxbbw}{\usebox{\boxbb}},
\end{equation}
where the disk represents Riemann hemisphere with a boundary.
We will call the expansion on the left-hand side the bulk channel, and that on the right-hand side the boundary channel. By explicitly writing down the equivalence between these two expansions, we can get the following bootstrap equation: 
\begin{align}\label{eq:bootstrap}
    \sum_{p} C_{iip}C^a_{p\mathbb{I}} \mathcal{F}^{ii}_{ii}(h_p|z) 
    =
    \sum_{P}\left(C^a_{iP}\right)^2 \ca{F}^{ii}_{ii}(h_P|1-z).
\end{align}
Here, the cross ratio $z$ is defined as 
\begin{align}
    z \equiv \frac{(z_1-z_2)(z_2^*-z_1^*)}{(z_1-z_2^*)(z_2-z_1^*)}. 
\end{align}
$C_{iip}$ is an OPE coefficient when expanding the product of two bulk primaries with bulk primaries, i.e.  bulk-bulk-bulk OPE coefficient. On the other hand, $C^{a}_{p\mathbb{I}}$ and $C^{a}_{iP}$ are bulk-boundary coefficients. $\ca{F}^{ii}_{ii} (h_p|z)$ is the Virasoro conformal block where the four external operators are labeled by $i$ and the one internal operator is labeled by $p$. 
The sum on the left-hand side runs over primaries in the closed string sector and that on the right-hand side runs over primaries in the open string sector.

To solve the bootstrap equation \eqref{eq:bootstrap}, we need to impose some constraints as inputs. To this end, we would like to assume the following condition: for any primary $\phi_p$ that appears in the OPE between two bulk primaries $\phi_i \times \phi_i$
\footnote{
More precisely, we assume this condition at the semiclassical level. That is, we assume for this to be suppressed in the large $c$ limit.
}
\begin{equation}\label{eq:one}
\braket{\phi_p}_{\text{disk}}=0,
\end{equation}
except for the identity $\mathbb{I}$. This assumption is often considered in the AdS/BCFT construction to model setups where there is no direct coupling between the brane and the matter fields \cite{Fujita2011,Kawamoto2022}. Another context where such an assumption is often considered is discussions about the BCFT dual of the island model \cite{Suzuki2022, Kusuki2022}. However, we would like to note that the assumption (\ref{eq:one}) is not a necessity to consider the AdS/BCFT correspondence. Gravity setups with $\braket{\phi_i}_{\text{disk}} \neq 0$ will be discussed in Section \ref{sec:Renyi}. With the assumption \eqref{eq:one}\footnote{
One can also realize this approximation by considering non-scalar primaries $h_i \neq \bar{h}_i$,
where the boundary channel cannot be dominated by the vacuum.
}, we have 
\begin{align}
    C^{a}_{p\mathbb{I}} = 0 ~~ {\rm if}~~ p\neq\mathbb{I}. 
\end{align}
Substituting this into \eqref{eq:bootstrap}, the left hand side can be greatly simplified to $\mathcal{F}^{ii}_{ii}(0|z)$. Note that while $C_{i \mathbb{I}}$ turns out to be zero, a generic bulk-boundary OPE coefficient $C_{iP}$ is non-zero because the matters on the boundary can interact with the matters in the bulk.

Let us then introduce some new notations to rewrite the right-hand side of \eqref{eq:bootstrap}. First of all, we would like to introduce the following notations:
\begin{equation}\label{eq:Liouville_notation}
    c=1+6Q^2, \ \ \ \ \ Q=b+\fr{1}{b}, \ \ \ \ \ h_i=\a_i(Q-\a_i).
\end{equation}
These notations are often used in the context of the Liouville CFT, but they are also convenient to discuss the conformal bootstrap. Here, we call $\a_i$ the Liouville momentum.
Next, we would like to introduce the density of primary states in the open string sector as 
\begin{equation}
\rho^{aa} (\a)=\sum_p D_p \delta(\a-\a_p) ,
\end{equation}
where $D_p$ is the degeneracy of primary states whose Liouville momenta are given by $\alpha_p$. 
Here, the superscript $aa$ means that the boundary conditions imposed on both of the boundaries of the open string are labeled by $a$. With these new notations, 
\eqref{eq:bootstrap} can be rewritten as 
\begin{equation}\label{eq:two}
\ca{F}^{ii}_{ii}(0|z) 
=
\int d \alpha_P \rho^{aa}( \alpha_P ) \overline{\pa{C^a_{iP}}^2} \ca{F}^{ii}_{ii}(h_P|1-z).
\end{equation}
The overline in $\overline{\pa{C^a_{iP}}^2}$ stands for an average over the states with Liouville momentum $\alpha_P$.

Since the left-hand side is a single conformal block, it is completely determined by the Virasoro symmetry.
Accordingly, the OPE coefficients appearing on the right-hand side should also have a universal form.
To determine these OPE coefficients,
it is convenient to re-express the left-hand side in terms of the dual basis.
The integral transformation from the s-channel block to the t-channel block is called the fusion transformation,
which has been developed in \cite{Teschner2001,Ponsot1999,Teschner2003,Ponsot2001}.
With the fusion transformation, the vacuum block is given by
\footnote{Note that the fusion transformation presented here includes a discrete sum of residues and an integral. On the other hand, the expression of fusion transformation found in literature is often given solely by an integral.
However, the discrete sum part is also necessary: it comes from the residues due to the poles of the fusion matrix \cite{Teschner2001}. See appendix A of \cite{Kusuki2019a} for a detailed discussion about how this discrete sum part arises.
}
\begin{equation}\label{eq:fusion}
\begin{aligned}
\ca{F}^{AA}_{BB}(0|z)
&=
 \sum_{\substack{\a_{n,m}<\fr{Q}{2} \\ n,m \in \bb{Z}_{\geq0}}}\ \text{Res}  \biggl( \biggr.   -2\pi i 
  {\mathbf{F}}_{0, \a_t} 
   \left[
    \begin{array}{cc}
    \a_A   & \a_A  \\
     \a_B  &   \a_B\\
    \end{array}
  \right] 
  \ca{F}^{AB}_{AB}(h_{\a_t}|1-z);\a_t=\a_{n,m}  \biggl. \biggr)     \\
&+
\int_{\fr{Q}{2}+0}^{\fr{Q}{2}+i \infty} \dd \a_t {\mathbf{F}}_{0, \a_t} 
   \left[
    \begin{array}{cc}
    \a_A   & \a_A  \\
     \a_B  &   \a_B\\
    \end{array}
  \right]
  \ca{F}^{AB}_{AB}(h_{\a_t}|1-z).
\end{aligned}
\end{equation}
Here, the sum of the first term runs over all
\begin{align}
    \a_{n,m}\equiv\a_A+\a_B+mb+nb^{-1} < \fr{Q}{2}~~~~~~(n,m \in \bb{Z}_{\geq 0}).
\end{align}
$ {\mathbf{F}}_{\a_s, \a_t} $ is the kernel of this integral transformation, and is often called the fusion matrix or the crossing matrix. The closed form of $ {\mathbf{F}}_{\a_s, \a_t} $ can be found in \cite{Teschner2001,Kusuki2019a}.

With the fusion transformation \eqref{eq:fusion}, both sides of (\ref{eq:two}) are expanded in terms of the same basis. Therefore, we can determine the OPE coefficients by simply comparing the coefficients of the Virasoro blocks on both sides. Since in our setups, the primaries are inserted at infinities, we are in particular interested in the lightest boundary primary appearing on the right-hand side.
From the explicit form of the fusion transformation (\ref{eq:fusion}), it is straightforward to find that the lowest conformal dimension is given by $h_{\a_P}=\a_P(Q-\a_P)$ with
\begin{equation}\label{eq:deficit}
\begin{aligned}
\a_P
&=\left\{
    \begin{array}{ll}
     2\a_i  ,& \text{if } \a_i<\fr{Q}{4}   ,\\
     \fr{c-1}{24}  ,& \text{otherwise }   . \\
    \end{array}
  \right.\\
\end{aligned}
\end{equation}
The behavior of $\alpha_P$ changes at $\alpha_i=\fr{Q}{4}$, which corresponds to the black hole threshold in the holographic theory. From (\ref{eq:Liouville_notation}), this happens at 
\begin{align}
    h_i = \fr{c-1}{32}.
\end{align}
Thus, we can say from the conformal bootstrap analysis that the brane self-intersection mentioned above does not happen due to the black hole formation. 

Besides the lightest boundary primary, we can also give the full spectrum of the boundary primaries, which turn out to be
\begin{equation}\label{eq:VMFT}
\rho^{aa}(\a_P)
\overline{C_{iP}^2} = 
 \text{Res}\pa{   -2\pi i 
  {\mathbf{F}}_{0, \a_P} 
   \left[
    \begin{array}{cc}
    \a_A   & \a_A  \\
     \a_B  &   \a_B\\
    \end{array}
  \right]
;\a_P=\a_{n,m}}.
\end{equation}
For a CFT, if its OPE coefficients are determined by the fusion matrix in the same way as \eqref{eq:VMFT}, it is called the Virasoro mean-field theory (VMFT) \cite{Collier2019}.
In this sense, we can say that the BCFT with the assumption (\ref{eq:one}) is another realization of VMFT.

The above result for the bulk-boundary OPE coefficients has a natural gravitational interpretation.
The bulk primaries correspond to one-particle states on the gravity side.
This particle gravitationally interacts with the end-of-the-world brane.
According to the AdS/CFT dictionary, the excitation energy of the one-particle state compared to the vacuum state is given by the difference between the expectation value of the Hamiltonian evaluated for the excited state on the strip with length $L$ and that evaluated for the ground state, which turns out to be
\begin{align}\label{eq:excitation_energy}
    E_P = 2\alpha_i(Q-2\alpha_i) \left(\frac{\pi}{L}\right).
\end{align}
In the following subsection, we will perform the computation on the gravity side, and see if it matches the result obtained from the BCFT side. 
Note that the binding energy $E_{bind} $ is given by	
\begin{equation}\label{eq:energy_one_particle}	
E_{bind} = E_P -E_i,	
\end{equation}	
where $E_i$ is the mass of that particle $\phi_i$, 	
\begin{align}	
    E_i \equiv h_i+\bar{h}_i.
\end{align}	
From this expression, 	
one can find that the binding energy is always negative.	
This reflects the attractive nature of gravity.	
Moreover, we have the trajectory $2\a_i+mb+nb^{-1}$,	
which can be interpreted as the gravitational nature at the quantum level.
\footnote{
One may wonder if this spectrum can be different from the spectrum of the closed string sector.
A similar but more concrete example can be found in the generalized free CFT with boundary \cite{Kastikainen2021}.
In fact, this is not problematic.
There is a possibility that the spectrum of the open string sector is not related to the spectrum of the closed sector at all.
One example is known as Friedan states \cite{Janik2001}, where one can find the continuous spectrum in the open string sector even though the spectrum of the closed string sector is discrete.
Note also that our spectrum has a continuous region above $\fr{c-1}{24}$,
which is not problematic for the same reason.
But we expect that this continuous spectrum is resolved by sub-leading corrections.
}

Before ending this subsection, let us give some comments on the relation between the results obtained here and the light-cone bootstrap \cite{Kusuki2019a, Collier2019}.
The statement in \cite{Kusuki2019a, Collier2019} is that two-particle states with large spin become black holes if the twist of the particles (assuming that two are identical) satisfies $\tau\geq\fr{c-1}{16}$, where the twist is defined by $\tau = 2 \min (h,\bar{h})$.
Since this bound looks very similar to our bound (or equivalently, the self-intersection bound) $h_i\geq \fr{c-1}{32}$,
it is natural to expect that the self-intersection bound has some connection to the bound in \cite{Kusuki2019a, Collier2019} (as already pointed out in \cite{Kusuki2021}).
Now we can give a complete understanding with respect to their relation.
To obtain the Regge trajectory $2\a_i+mb+nb^{-1}$, we need the vacuum block approximation on one side of the bootstrap equation.
The authors in \cite{Kusuki2019a, Collier2019} consider the bootstrap equation for four-point functions on a plane,
\begin{equation}
\sum_p C_{iip}^2 \ca{F}^{ii}_{ii}(p|z) \overline{\ca{F}^{ii}_{ii}(p|z)}
=
\sum_p C_{iip}^2 \ca{F}^{ii}_{ii}(p|1-z) \overline{\ca{F}^{ii}_{ii}(p|1-z)},
\end{equation}
and approximate this by the vacuum block in the limit $z \to 0$ with $\bar{z}$ fixed (called the light-cone limit),
\begin{equation}
\ca{F}^{ii}_{ii}(\mathbb{I}|z) \overline{\ca{F}^{ii}_{ii}(\mathbb{I}|z)}
=
\sum_p C_{iip}^2 \ca{F}^{ii}_{ii}(p|1-z) \overline{\ca{F}^{ii}_{ii}(p|1-z)}.
\end{equation}
This approximated bootstrap equation is essentially the same as our bootstrap equation (\ref{eq:two}).
As a result, the OPE coefficient is completely fixed by the fusion matrix (\ref{eq:fusion}) in both cases.
This is the reason why we found similar bounds in two different contexts.
Note that recently this OPE spectrum (called the quantum Regge trajectory) has been reproduced from the spinning string in the gravity\cite{Maxfield2022}.
Since we obtain the same trajectory from a simpler setup with BCFT, it is expected that one can get new insights into quantum gravity by considering the gravity dual of our quantum Regge trajectory in BCFT in a similar way to \cite{Maxfield2022}.

%%%%%%%%%%%%%%%%%%%%%%%%%%%%%%%%%%%%%%%%%%%%%%%%%%%%%%%%%%%%%%%%%%%%%%%%%%%%%%%%%%%%%%%%%%%%%%
\subsection{Summary of BCFT Results}\label{sec:CFT}
%%%%%%%%%%%%%%%%%%%%%%%%%%%%%%%%%%%%%%%%%%%%%%%%%%%%%%%%%%%%%%%%%%%%%%%%%%%%%%%%%%%%%%%%%%%%%%

In this subsection, we summarize the facts obtained from the BCFT side for later reference.

\begin{tcolorbox}
\begin{customfact}{1}[Relation between ADM mass and mass of scalar particle (\ref{eq:deficit})]~\par
\begin{equation}\label{eq:fact1}
\begin{aligned}
\a_P
&=\left\{
    \begin{array}{ll}
     2\a_i  ,& \text{if } \a_i<\fr{Q}{4}   ,\\
     \fr{c-1}{24}  ,& \text{otherwise },\\
    \end{array}
  \right.\\
\end{aligned}
\end{equation}
where the ADM mass is defined by $h_{\text{ADM}} = \a_P(Q-\a_P)$ and the mass of the particle is defined by $h_i =\a_i(Q-\a_i) $.
\end{customfact}
\end{tcolorbox}
\noindent
To derive this relation, we assume $h_i=\bar{h}_i$ for simplicity.
In fact, the generalization to non-scalar primaries is straightforward.
For a non-scalar primary excitation, the simplified bootstrap equation is
\begin{equation}
\ca{F}^{ii}_{\bar{i} \bar{i}}(0|z) 
=
\int d \alpha_P \rho^{aa}( \alpha_P ) \overline{\pa{C^a_{iP}}^2} \ca{F}^{\bar{i}i}_{\bar{i}i}(h_P|1-z),
\end{equation}
and then we obtain the generalized statement,
\begin{tcolorbox}
\begin{customfact}{1'}[Relation between ADM mass and mass of spinning particle]\label{fact1'}
\begin{equation}\label{eq:fact1'}
\begin{aligned}
\a_P
&=\left\{
    \begin{array}{ll}
     \a_i+\bar{\a}_i  ,& \text{if } \a_i+\bar{\a}_i<\fr{Q}{2}   ,\\
     \fr{c-1}{24}  ,& \text{otherwise }.\\
    \end{array}
  \right.\\
\end{aligned}
\end{equation}
\end{customfact}
\end{tcolorbox}
\noindent
This result tells us how to properly construct the geometry with non-scalar primaries as we will see later.

Another point that we would like to emphasize is that the boundary entropy does not appear in the conformal bootstrap equation (\ref{eq:two}).
It implies
\begin{tcolorbox}
\begin{customfact}{2}[Non-sensitivity to brane tension]~\par\label{fact2}
The equations (\ref{eq:fact1}) and (\ref{eq:fact1'}) hold not only if the brane tension is positive but also if the brane tension is negative.
\end{customfact}
\end{tcolorbox}
\noindent
It has been proposed in \cite{Bianchi2022} that the spectrum is sensitive to whether the tension is positive or negative.
This proposal is totally opposite to the above fact.
We will resolve this tension in this article.

For reference, we again show the following fact for non-scalar primaries,
\begin{tcolorbox}
\begin{customfact}{3}[No coupling between non-scalar primary and boundary]~\par \label{fact3}
    For a bulk primary $\phi_i$ with chiral conformal dimension $h_i$ and anti-chiral conformal dimension $\bar{h}_i$,
	\begin{equation}
    h_i \neq \bar{h}_i~~~ \Longrightarrow~~~ \braket{\phi_i}_{\text{disk}} = 0.
\end{equation}
    Here, $\braket{\cdots}_{\rm disk}$ is the expectation value evaluated in the BCFT defined on a disk. 
\end{customfact}
\end{tcolorbox}
\noindent
In the rest of this article,
we will construct the corresponding gravity dual based on the above findings.

%%%%%%%%%%%%%%%%%%%%%%%%%%%%%%%%%%%%%%%%%%%%%%%%%%%%%%%%%%%%%%%%%%%%%%%%%%%%%%%%%%%%%%%%%%%%%%
\subsection{Gravity Dual of Two-point Function}\label{sec:bulk-two}
%%%%%%%%%%%%%%%%%%%%%%%%%%%%%%%%%%%%%%%%%%%%%%%%%%%%%%%%%%%%%%%%%%%%%%%%%%%%%%%%%%%%%%%%%%%%%%

In the following, we present the gravity dual of the setup considered above, i.e., a BCFT defined on a strip with width $L$ and primary operators $\phi_i$ inserted at two infinity points. The gravity dual is constructed by finding out an on-shell configuration of a natural gravitational action, as we will explain later. As a consistency check, we will compute the binding energy from the gravity side, and see that it indeed matches the expectation from the CFT calculation presented above. 

Applying the standard AdS/CFT and AdS/BCFT dictionary \cite{Takayanagi2011, Fujita2011}, the gravitational action corresponding to the current setup is in the form
\begin{align}\label{eq:gravaction}
    I_{\rm grav} = -\frac{1}{16\pi G_N}\int_{\CM} \sqrt{g}(R+2) - \frac{1}{8\pi G_N} \int_Q \sqrt{h} (K-T) + m\int_\Gamma \sqrt{\gamma} .
\end{align}
The third term is the action of a point particle with invariant mass $m$.

The point particle creates a conical defect whose deficit angle is given by $8\pi G_N m$, and $\Gamma$ is its world line. Thanks to the property that any solution of the Einstein equation with a negative cosmological constant is locally AdS away from the matters (in this case, the brane and the point particle), we can treat the brane and point particle separately, and then combine them to get an on-shell configuration of the action \eqref{eq:gravaction}. 

Let us firstly review the setup with no brane and only a particle in it. This corresponds to a CFT defined on a cylinder with two primaries inserted on the two infinities. By doing so, we can see how the mass $m$ is related to $h_i$, the conformal dimension of $\phi_i$.

\paragraph{Solution with only the particle and no brane}~\par
A particle with mass $m$ creates a deficit angle \begin{align}
    \delta \theta = 8\pi G_N m
\end{align}
in the global AdS$_3$ (\ref{eq:globalAdS}). The solution can be written in the same metric as that of the global AdS$_3$
\begin{align}\label{eq:conicaldefect}
    ds^2 = (r^2+1)d\sigma^2 + \frac{dr^2}{r^2+1} + r^2 d\theta^2, 
\end{align}
but with a different parameter range $\theta\in[0, 2\pi\chi)$ where 
\begin{align}
    \chi \equiv 1-4G_N m. 
\end{align}
If we want to identify the dual CFT as one defined on a cylinder with perimeter $2L$, it is convenient to perform the coordinate transformation $\tau = (L/\pi\chi)\sigma$ and $x = (L/\pi\chi)\theta$. The holographic energy stress tensor and the ADM mass turn out to be 
\begin{align}
    T_{\tau\tau}^{\rm (hol)} = -\frac{\chi^2}{16\pi G_N} \left(\frac{\pi}{L}\right)^2,
\end{align}
and 
\begin{align}
    M_{\tau} =  \int_0^{2L} dx ~ T_{\tau\tau}^{\rm (hol)} =  -\frac{\chi^2}{8G_N}\left(\frac{\pi}{L}\right).
\end{align}
Applying the Brown-Henneaux relation $1/4G_N = c/6$ \cite{Brown1986}, these match the CFT results defined on a cylinder with perimeter $2L$ and two scalar operators inserted on the two infinities where the chiral dimension $h_i$ of the operators satisfies
\begin{align}\label{eq:chi_and_h}
    \frac{24h_i}{c} = 1-\chi^2.
\end{align}
In this way, we have figured out that inserting scalar operators with scaling weight $\Delta_i = 2h_i$ on the CFT side corresponds to introducing conical defects with deficit angle 
\begin{align}
    2\pi \left(1-\sqrt{1-\frac{24h_i}{c}}\right),
\end{align}
or in other words, introducing point particles with invariant mass\footnote{Note that this relation between the mass of the particle and the conformal dimension is valid for heavy operators with $h=\mathcal{O}(c)$. This does not contradict with the well-known relation $m^2 = 4h(h-1)$ for light scalar operators with $h=\mathcal{O}(1)$. In fact, one can explicitly check these two relations match each other at the leading order when $1\ll h \ll c$.}
\begin{align}
    m = \frac{c}{6} \left(1-\sqrt{1-\frac{24h_i}{c}}\right). 
\end{align}
In the following, we will work on the configuration with both the brane and the particle based on this relation. 

\paragraph{Solution with both the brane and the particle}~\par
Let us then introduce the brane into the game. We can construct a solution with both the brane and the particle by simply cutting and pasting the geometries with only one of them. Introducing the brane in the conical defect geometry \eqref{eq:conicaldefect}, we get the following brane profile: 
\begin{align}\label{eq:brane_conf_defect}
    \sin (\theta+2\pi(1-\chi)) = -\frac{T}{\sqrt{1-T^2}} \frac{1}{r} \ \ \ \ \ \ \ \ ((2\chi-1)\pi \leq \theta \leq 2\pi\chi),
\end{align}
which intersects the asymptotic boundary at $\theta = \pi-2\pi(1-\chi), 2\chi \pi \sim 0$. In this configuration, the bulk of the gravity dual is still covered by $\theta\in [0,2\pi\chi)$, i.e. there is still a conical defect with deficit angle $2\pi(1-\chi)$ in the bulk of the gravity dual. On the other hand, only the $\theta\in[0,\pi-2\pi(1-\chi)] = [0,\pi(2\chi-1)]$ part is remained on the asymptotic boundary. Note that \eqref{eq:brane_conf_defect} is defined on $\theta\in (\pi(2\chi-1),2\pi\chi)$, since it is an object living in the bulk of the gravity dual. Also note that the brane covers a region with angle $\pi$ again in this case, which is a necessary condition for the configuration to be a solution to the equation of motions. 

Now since we would like to consider the CFT dual being defined on a strip with length $L$ parameterized by $(\tau,x)$, it is convenient to perform the coordinate transformation $\tau = \frac{L}{\pi(2\chi-1)} \sigma$ and $x = \frac{L}{\pi(2\chi-1)} \theta$ with which the metric turns out to be 
\begin{align}
    ds^2 = (r^2+1)\left(\frac{\pi(2\chi-1)}{L}\right)^2 d\tau^2 + \frac{dr^2}{r^2+1} + r^2 \left(\frac{\pi(2\chi-1)}{L}\right)^2 dx^2 .
\end{align}
Accordingly, the holographic energy stress tensor and the ADM mass becomes
\begin{align}
    T_{\tau\tau}^{\rm (hol)} = -\frac{1}{16\pi G_N} \left(\frac{\pi(2\chi-1)}{L}\right)^2,
\end{align}
and 
\begin{align}\label{eq:M_BandP}
    M_{\tau} =  \int_0^{L} dx ~ T_{\tau\tau}^{\rm (hol)} =  -\frac{(2\chi-1)^2}{16G_N}\left(\frac{\pi}{L}\right).
\end{align}
From this computation, we can straightforwardly see that the black hole threshold $M_{\tau} = 0$ is given by $\chi = \frac{1}{2}$. Plugging this into \eqref{eq:chi_and_h}, we obtain $h_i = \frac{c}{32}$. Thus, the black hole threshold predicted directly from the gravity side is $h_i = \frac{c}{32}$. This matches the one predicted from the conformal bootstrap on the BCFT side if we replace $c$ with $c-1$. This is because we are performing semiclassical computations on the gravity side, and it is valid only in the $c\rightarrow\infty$ limit. The shift of $c$ to $c-1$ is expected to come from the one-loop effects \cite{Maxfield2019}. Thus, we can say that our computation on the gravity side matches that on the BCFT side.

Note that properly choosing the holographic renormalization regime \cite{Balasubramanian1999} is important to get the correct result. We need to perform the holographic renormalization such that the corresponding BCFT dual is a strip with length $L$.

\paragraph{The excitation energy} ~\par
The excitation energy of the one-particle state with the presence of the brane, compared to the ground state where there is only a brane and no particle, is given by the difference between \eqref{eq:M_BandP} and \eqref{eq:M_onlyb},
\begin{align}
    \Delta E = -\frac{(2\chi-1)^2}{16G_N}\left(\frac{\pi}{L}\right) + \frac{1}{16G_N}\left(\frac{\pi}{L}\right) = \frac{1}{4G_N} \chi(1-\chi) \left(\frac{\pi}{L}\right). 
\end{align}
Using the relation $1/4G_N = c/6$ and $\chi=\sqrt{1-24h_i/c}$, we have 
\begin{align}
    \Delta E = \frac{c}{6} \sqrt{1-\frac{24h_i}{c}} \left(1-\sqrt{1-\frac{24h_i}{c}}\right). 
\end{align}
This again matches exactly with the BCFT result $E_P = 2\alpha_i\left(Q-2\alpha_i\right)(\pi/L)$ with $c$ replaced by $c-1$.

\paragraph{Intuitive interpretation of the gravitational computation} ~\par
Let us then explain the intuitive interpretation of the gravitational configuration presented above. Inserting scalar primary operators with chiral dimension $h_i$ corresponds to introducing a point particle with invariant mass 
\begin{align}
    m = \frac{c}{6} \left( 1 - \sqrt{1-\frac{24h_i}{c}} \right),
\end{align}
which creates a conical defect with deficit angle 
\begin{align}
    \delta\theta = 8\pi G_N m,
\end{align}
in the gravity dual. In other words, we can say that the metric in the vicinity of the particle can be written as 
\begin{align}
    ds^2 \approx dr^2 + r^2 d\theta^2 + \cdots,
\end{align}
where 
\begin{align}
    \theta \in [0,2\pi\chi), ~~ \theta \sim \theta + 2\pi \chi,~~\chi = 1-4G_N m = \sqrt{1-\frac{24h_i}{c}},
\end{align}
and the particle sits at $r=0$. Note again that this invariant mass $m$ is not the ADM mass of the spacetime, but the mass parameter appearing in the particle's action. 

We have already seen in the previous computations that the energy stress tensor, and hence the ADM mass, induced by the particle are changed when there is a brane. When there is only the particle and no branes, the energy stress tensor is 
\begin{align}
    -\frac{\chi^2}{16\pi G_N} \left(\frac{\pi}{L}\right)^2. 
\end{align}
When both the brane and the particle are in the spacetime, the energy stress tensor is 
\begin{align}
    -\frac{(2\chi-1)^2}{16\pi G_N} \left(\frac{\pi}{L}\right)^2.
\end{align}
From the asymptotic boundary, assuming there were no brane, it would look as if there was a conical defect with deficit angle
\begin{align}
    2\pi (1-(2\chi-1)) = 2\times 2\pi(1-\chi) = 2\delta \theta, 
\end{align}
i.e. twice the correct deficit angle. 

Intuitively, this is because the point particle is attracted by the brane. When there is no brane, the visual angle from the two sides of the strip (on which the CFT is defined) is $\pi\chi$, which covers exactly half of the point particle. On the other hand, when there is a brane, the particle is attracted by the brane and goes ``farther" from the asymptotic boundary. As a result, the visual angle from the two sides of the strip shrinks to $\pi(2\chi-1)$, which covers only $\chi-1/2 < 1/2$ of the point particle. This is reflected on the energy stress tensor. If an observer on the asymptotic boundary was not aware of the existence of the brane, they would think there was a larger deficit angle in the bulk, which is an illusion. 

We can get another good intuition by considering the mirror image of the current setup, and gluing them together. In this gravitational setup, we have two conical defects with the same deficit angles $\delta \theta$. The two defects are separated by a brane. As a whole, the total deficit angle in the bulk is $2\delta\theta$, reflected by the energy stress tensor on the asymptotic boundary. 
%%%%%%%%%%%%%%%%%%%%%%%%%%%%%%%%%%%%%%%%%%%%%%%%%%%%%%%%%%%%%%%%%%%%%%%%%%%%%%%%%%%%%%%%%%%%%%
\subsection{Holographic Dual of Boundary Primaries}\label{subsec:BPdual}
%%%%%%%%%%%%%%%%%%%%%%%%%%%%%%%%%%%%%%%%%%%%%%%%%%%%%%%%%%%%%%%%%%%%%%%%%%%%%%%%%%%%%%%%%%%%%%

While we are discussing the gravity dual of a bulk two-point function, we can also consider it as a gravity dual of a boundary two-point function in the following sense. Since we are considering the limit $|x_1-x_2| \gg L, x_1, x_2$, the most dominant contribution comes from the bulk-boundary OPE with the lowest chiral dimension $h_P$ which satisfies
\begin{align}
    1-\sqrt{1-\frac{24h_P}{c}} = 2\left(1 - \sqrt{1-\frac{24h_i}{c}}\right),
\end{align}
where we use $h_i$ to denote the chiral dimension of the bulk operator. 
Accordingly, we can regard the gravity dual of a two-point function of bulk scalar primaries with chiral dimension $h_i$ as the gravity dual of a two-point function of boundary primaries with conformal dimension $h_P$. 

In this description, boundary primaries with chiral dimension $h_P$ create a conical defect with deficit angle 
\begin{align}
    \delta \theta = \pi\left(1-\sqrt{1-\frac{24h_P}{c}}\right),
\end{align}
in the gravity dual. After a similar analysis presented before, we get the holographic energy stress tensor 
\begin{align}
    T_{\tau\tau}^{\rm (hol)} = -\frac{1}{16\pi G_N} \left(\frac{\pi(2\chi_i-1)}{L}\right)^2 = -\frac{\chi_P^2}{16\pi G_N} \left(\frac{\pi}{L}\right)^2,
\end{align}
where 
\begin{align}
    \chi_P = \sqrt{1-\frac{24h_P}{c}},~~\chi_i = \sqrt{1-\frac{24h_i}{c}}.
\end{align}
This again matches the CFT result. Self-intersection of the brane is expected to happen when $\delta\theta \geq \pi$. However, this is achieved when $h_P \geq c/24$, where conical defects are replaced by black holes. Therefore, self-intersection never happens below the standard black hole threshold $h_P = c/24$.
\footnote{
In the black hole phase, the branes are disconnected and do not interact with each other \cite{Takayanagi2011,Fujita2011}.
Note that if the tension is negative, the branes can still interact \cite{Biswas2022}.
}

It would also be instructive if we consider a mirror image of the current setup, paste it with the original setup, and regard it as the gravity dual of a defect CFT. Then the counterpart of boundary primary is the defect primary, which lives only on a defect. In this description, defect scalar primaries with chiral dimension $h_P$ create two conical defects with deficit angle 
\begin{align}
    \pi\left(1-\sqrt{1-\frac{24h_P}{c}}\right)
\end{align}
on the two sides of the brane. The defect primaries introduce deficit angles
\begin{align}
    2\pi\left(1-\sqrt{1-\frac{24h_P}{c}}\right)    
\end{align}
to the spacetime as a whole. Note that this description matches the standard AdS/CFT dictionary when we take the zero tension limit $T\rightarrow0$ of the brane.

From the BCFT point of view, what we did here is an inverse process of the bulk-boundary OPE. It would be an interesting future direction to formulate it on the BCFT side and construct gravity duals for general boundary primaries. 

%%%%%%%%%%%%%%%%%%%%%%%%%%%%%%%%%%%%%%%%%%%%%%%%%%%%%%%%%%%%%%%%%%%%%%%%%%%%%%%%%%%%%%%%%%%%%%
%%%%%%%%%%%%%%%%%%%%%%%%%%%%%%%%%%%%%%%%%%%%%%%%%%%%%%%%%%%%%%%%%%%%%%%%%%%%%%%%%%%%%%%%%%%%%%
\section{Particle-Brane Intersection and R\'enyi Entropy}\label{sec:Renyi}
%%%%%%%%%%%%%%%%%%%%%%%%%%%%%%%%%%%%%%%%%%%%%%%%%%%%%%%%%%%%%%%%%%%%%%%%%%%%%%%%%%%%%%%%%%%%%%
%%%%%%%%%%%%%%%%%%%%%%%%%%%%%%%%%%%%%%%%%%%%%%%%%%%%%%%%%%%%%%%%%%%%%%%%%%%%%%%%%%%%%%%%%%%%%%

In the previous section, we explicitly impose the vanishing one-point function condition \eqref{eq:one} when solving the bootstrap equation. Since our gravity dual perfectly matches the results from the conformal bootstrap, it means that the vanishing one-point function condition must be implicitly imposed also on the gravity side. This brings two questions to us. Is the vanishing one-point function \eqref{eq:one} an unavoidable nature of the AdS/BCFT construction?
 If not, how can we realize a non-vanishing one-point function on the gravity side? 

To answer this question, we can consider a setup in which a non-vanishing one-point function of heavy operators is unavoidable. As such an example, in this section, we compute the modular R\'enyi entropy
\begin{align}
    \tilde{S}^{(n)}(\rho_A) = n^{2} \frac{\partial}{\partial n}\left(\frac{n-1}{n} S^{(n)}(\rho_A)\right),
\end{align}
holographically from the gravity side for a simple setup in AdS$_3$/BCFT$_2$. Here, $S^{(n)}(\rho_A)$ is the $n$-th R\'enyi entropy defined as 
\begin{align}
    S^{(n)}(\rho_A) = \frac{1}{1-n} \log {\rm Tr} \left(\rho_A\right)^n.
\end{align}
To be concrete, we consider a BCFT$_2$ state defined on an interval with length $2R$, prepared by a Euclidean path integral on a disk with radius $R$. We consider the state realized on the reflection symmetric time slice, divide it into exactly two halves $A$ and $\bar{A}$, and consider the modular R\'enyi entropy between them.

\subsection{The BCFT Results}
Let us firstly present the CFT result for comparison. The R\'enyi entropy can be evaluated in the following way \cite{Calabrese2004,Calabrese2009}: 
\begin{align}
    S^{(n)}(\rho_A) = \frac{1}{1-n} \log \frac{\braket{\sigma_n(0,0)}_{\rm disk}}{\braket{\mathbb{I}}^n_{\rm disk}}.
\end{align}
Here, $\braket{\mathbb{I}}_{\rm disk}$ is the disk partition function of the CFT under consideration. Let us call this original CFT the seed CFT $\mathcal{C}$. On the other hand, $\sigma_n(w,\bar{w})$ is the twist operator defined in the orbifold CFT $\ca{C}^{\otimes n}/\mathbb{Z}_n$. The central charge of $\ca{C}^{\otimes n}/\mathbb{Z}_n$ is $nc$, and the conformal dimension of $\sigma_n$ is $h_n = \bar{h}_n = \frac{c}{24}\left(n-\frac{1}{n}\right)$. 

To evaluate the one-point function $\langle\sigma_n(0,0)\rangle_{\rm disk}$, it is convenient to map it to an upper half plane ${\rm Re}(z)\geq0$ parameterized by $z$. The conformal map is given by
\begin{align}
    w = R \frac{z-i}{z+i}.
\end{align}
The one-point function on an upper half plane can be straightforwardly evaluated using the doubling trick as 
\begin{equation}\label{eq:UHP}
\braket{\sigma_n(z,\bar{z})}_{\text{UHP}} = \pa{g^a}^{n}C_{\sigma_n \mathbb{I}}^a (z-\bar{z})^{-2h_{n}},
\end{equation}
where we label the boundary by $a$. The $g$-function, $g^a$, is not defined in the orbifold CFT $\ca{C}^{\otimes n}/\mathbb{Z}_n$ but in the seed CFT $\ca{C}$.
The bulk-boundary OPE coefficient $C_{\sigma_n \mathbb{I}}^a$ can be evaluated by considering the relation between the twist operator formalism and the partition function on the replica manifold (see for example, \cite{Akal2020, Sully2021}),
\footnote{
More precisely, we have additional contributions from the boundaries associated with the twist operators\cite{Ohmori2015}.
However, we usually absorb these contributions into the UV-cutoff $\epsilon$,
therefore, we express the one-point function on a disk just by $\pa{g^a}^{1-n} \epsilon^{2h_{n}}$.
}
\begin{equation}\label{eq:Csigma}
C_{\sigma_n \mathbb{I}}^a = \pa{g^a}^{1-n} \epsilon^{2h_{n}}.
\end{equation}
Thus we have 
\begin{align}
    \braket{\sigma_n(0,0)}_{\rm disk} = \left(\frac{dz}{dw}\right)^{h_n}_{w=0} \left(\frac{d\bar{z}}{d\bar{w}}\right)^{\bar{h}_n}_{\bar{w}=0} \braket{\sigma_n(i,-i)}_{\rm UHP} = g^a \left(\frac{R}{\epsilon}\right)^{-2h_n}.
\end{align}
Accordingly,
\footnote{
It is worth noting that the boundary entropy term in the entanglement entropy comes not from the $g$-function in $\braket{O_i}_a = g^a C_{i\mathbb{I}}^a$ (with the standard normalization)
but from the bulk-boundary OPE coefficients (\ref{eq:Csigma}). We need to care about this point to give the gravity interpretation of the boundary entropy.
}

\begin{align}\label{eq:RenyidiskBCFT}
    S^{(n)}(\rho_A) = \frac{1}{1-n} \log \frac{\braket{\sigma_n(0,0)}_{\rm disk}}{\braket{\mathbb{I}}^n_{\rm disk}} = \frac{c}{12} \left(1+\frac{1}{n}\right) \log\frac{R}{\ep} + S_{\rm bdy}, 
\end{align}
where 
\begin{align}
    S_{\rm bdy} = \log g^a
\end{align}
is the boundary entropy. From this, we can easily find out the modular R\'enyi entropy 
\begin{align}\label{eq:modularBCFT}
    \tilde{S}^{(n)}(\rho_A) = \frac{c}{6} \frac{1}{n} \log \frac{R}{\epsilon} + S_{\rm bdy}. 
\end{align}
From now on, we would like to perform the computation holographically on the gravity side, and see if it reproduces the BCFT results. 

\subsection{Holographic Computation of R\'enyi Entropy}\label{sec:holRenyi}
The computation performed in this subsection follows the philosophy of \cite{Lewkowycz2013,Dong2016}, where holographic formulae for entanglement entropy and R\'enyi entropy are derived. 

Since the most crucial ingredient in the computation of R\'enyi entropy 
\begin{align}
    S^{(n)}(\rho_A) = \frac{1}{1-n} \log {\rm Tr} (\rho_A)^n, 
\end{align}
is the evaluation of ${\rm Tr} (\rho_A)^n$, let us firstly consider realizing it on the gravity side. The primitive version of the replica trick \cite{Calabrese2004} on the BCFT side tells us that
\begin{align}
    {\rm Tr} (\rho_A)^n = \frac{Z_{\rm BCFT}[{\Sigma_n}]}{(Z_{\rm BCFT}[{\Sigma_1}])^n}
\end{align}
where $\Sigma_1$ is the disk with radius $R$, i.e. the original geometry we are considering. On the other hand, $\Sigma_n$ is the $n$-replicated manifold of $\Sigma_1$ pasted along $A$. In this case, it turns out to be a disk-like surface with a conical defect at the origin. The angle around the conical defect is $2\pi n$, and the radius is $R$. Note that $\Sigma_n$ can be mapped to a disk with radius $R^{\frac{1}{n}}$ via the following conformal map
\begin{align}
    f(w) = w^{\frac{1}{n}}. 
\end{align}
Using the GKP-Witten relation \cite{Gubser1998,Witten1998} and saddle point approximation, we have
\begin{align}
    {\rm Tr} (\rho_A)^n = \frac{Z_{\rm BCFT}[{\Sigma_n}]}{(Z_{\rm BCFT}[{\Sigma_1}])^n} = \frac{e^{-I_{\rm grav}[\CM_n]}}{e^{-nI_{\rm grav}[\CM_1]}},
\end{align}
where $I_{\rm grav}[\CM]$ is the gravitational action evaluated on the manifold $\CM$. Here, $\CM_1$ ($\CM_{n}$) is the most dominant solution of the Einstein equation with the asymptotic boundary given by $\Sigma_1$ ($\Sigma_n$). The gravitational action $I_{\rm grav}[\CM]$ is again given by 
\begin{align}\label{eq:gravaction2}
    I_{\rm grav}[\CM] = -\frac{1}{16\pi G_N}\int_{\CM} \sqrt{g}(R+2) - \frac{1}{8\pi G_N} \int_Q \sqrt{h} (K-T).
\end{align}
where we have again omitted the Gibbons-Hawking term on the asymptotic boundary. 

As we have reviewed in Section \ref{sec:review_adsbcft}, solving the Einstein equation with the boundary given by the disk $\Sigma_1$, it is straightforward to find that $\CM_1$ is given by a region surrounded by the brane $Q$, where the metric is 
\begin{align}
    ds^2_{\CM_1} = \frac{dw d\bar{w} + dz^2}{z^2},
\end{align}
and the brane profile is 
\begin{align}
    w\bar{w} + (z-R \sinh \left[{\rm arctanh}(T)\right])^2 = \left(R \cosh \left[{\rm arctanh}(T)\right]\right)^2,
\end{align}
which intersects the asymptotic boundary $z=0$ at
\begin{align}
    |w| = R.
\end{align}
See Figure \ref{fig:disk_adsbcft} for a sketch of $\Sigma_1$ and $\CM_1$. When we want to compute physical quantities in this system, what we need to do is to just place a cutoff surface at $z=\epsilon$, perform the calculation, and then apply an appropriate renormalization procedure (if needed). 

By computing the partition function with saddle point approximation on the gravity side, one can figure out that the boundary entropy is given by \cite{Takayanagi2011}
\begin{align}
    S_{\rm bdy} =\frac{1}{4G_N} {\rm arctanh}(T). 
\end{align}
This expression will be used later. 

Thanks to the topological nature of the AdS$_3$ gravity, $\CM_n$ has the same geometry as $\CM_1$. All we need to do is to choose an appropriate coordinate and be careful when taking a UV-cutoff. Let us again use $(w,\bar{w})$ to parameterize $\Sigma_n$ and perform the following diffeomorphism: 
\begin{align}\label{eq:ndisk_to_1disk}
    \xi = w^{\frac{1}{n}},~
    \bar{\xi} = \bar{w}^{\frac{1}{n}}.
\end{align}
The metric of $\Sigma_n$ is 
\begin{align}
    ds^2_{\Sigma_n} = dw d\bar{w} = n^2 \xi^{n-1} \bar{\xi}^{n-1} d\xi d\bar{\xi}.
\end{align}
Comparing with the $\CM_1$ case, we find that $\CM_n$ is given by the region surrounded by $Q$, where the metric is 
\begin{align}
    ds_{\CM_n}^2 = \frac{d\xi d\bar{\xi} + d\eta^2}{\eta^2}
\end{align}
and the brane profile is 
\begin{align}
    \xi\bar{\xi} + (\eta-R^{\frac{1}{n}} \sinh \left[{\rm arctanh}(T)\right])^2 = \left(R^{\frac{1}{n}} \cosh \left[{\rm arctanh}(T)\right]\right)^2,
\end{align}
which intersects the asymptotic boundary at 
\begin{align}
    |\xi| = R^{\frac{1}{n}}. 
\end{align}
Now we can see that both $\Sigma_n$ and $\CM_n$ have a $\mathbb{Z}_n$ symmetry. Therefore, we can accordingly introduce a quotient manifold 
\begin{align}
    \hat{\CM}_n = \frac{\CM_n}{\mathbb{Z}_n}.
\end{align}
This quotient manifold $\hat{\CM}_n$ has a conical defect with deficit angle $2\pi(1-1/n)$ which locates at $|w| = |\xi| = 0$. Except for this, the local structure is exactly the same as that of $\CM_n$. Note that $\hat{\CM}_n$ is not an on-shell configuration of $I_{\rm grav}$. However, we can introduce a point particle with invariant mass 
\begin{align}
    m_n = \frac{1}{4G_N} \frac{n-1}{n}
\end{align}
to the original action to get a new action
\begin{align}
    \hat{I}^{(n)}_{\rm grav} = I_{\rm grav} + \frac{1}{4G_N}\frac{n-1}{n} \int_{\Gamma_n} \sqrt{\gamma},
\end{align}
where $\Gamma_n$ is the world line of the conical defect. 
Then one can find that $\hat{\CM}_n$ is an on-shell configuration of $\hat{I}^{(n)}_{\rm grav}$, and 
\begin{align}
    I_{\rm grav}[\CM_n] = n \hat{I}^{(n)}_{\rm grav}[\hat{\CM}_n].
\end{align}
Coming back to the R\'enyi entropy, we have 
\begin{align}
    S^{(n)}(\rho_A) = \frac{1}{1-n} \left(-I_{\rm grav}[\CM_n] +nI_{\rm grav}[\CM_1] \right) = \frac{n}{1-n} \left(-\hat{I}^{(n)}_{\rm grav}[\hat{\CM}_n] + I_{\rm grav}[\CM_1] \right). 
\end{align}
Then it turns out that it is easier to evaluate the modular R\'enyi entropy instead, 
\begin{align}
    \tilde{S}^{(n)}(\rho_A) = n^{2} \frac{\partial}{\partial n}\left(\frac{n-1}{n} S^{(n)}(\rho_A)\right) = n^{2} \frac{\partial}{\partial n}
    \hat{I}^{(n)}_{\rm grav}[\hat{\CM}_n]. 
\end{align}
Since $\hat{\CM}_n$ is an on-shell action of $\hat{I}^{(n)}_{\rm grav}$, the variation of $\hat{I}^{(n)}_{\rm grav}[\hat{\CM}_n]$ with respect to the field configuration is zero. Therefore, we have 
\begin{align}
    \tilde{S}^{(n)}(\rho_A) = n^{2} \frac{\partial}{\partial n}
    \left(\frac{n-1}{n}\right) \times \frac{1}{4G_N} \int_{\Gamma_n} \sqrt{\gamma} = \frac{1}{4G_N} \int_{\Gamma_n} \sqrt{\gamma}.
\end{align}
In one word, the $n$-th modular entropy is given by the length of the conical defect $\hat{\CM}_n$, or equivalently, the length of the fixed line of $\CM_n$ under $\mathbb{Z}_n$. 

Now we would like to compute the length of the line connecting 
\begin{align}
    (\xi,\bar{\xi},\eta) = (0,0,R^{\frac{1}{n}} \cosh \left[{\rm arctanh}(T)\right]+
    R^{\frac{1}{n}} \sinh \left[{\rm arctanh}(T)\right]),
\end{align}
and 
\begin{align}
    (\xi,\bar{\xi},\eta) = (0,0,\delta_n),
\end{align}
in $\CM_n$, where $\delta_n$ is a proper UV-cutoff. 

In the following, we will present two ways to take the UV-cutoff. First, we will perform the standard cutoff regime and see that it does not reproduce the boundary entropy term correctly. Then we will introduce a new cutoff regime and see it reproduce the correct answer. This new regime suggests the existence of a refined version of the Ryu-Takayanagi formula \cite{Ryu:2006bv,Ryu:2006ef} and Dong's formula \cite{Dong2016}. 

\paragraph{Standard cutoff gives the wrong answer}~\par

The standard way to take a UV-cutoff in the AdS/CFT correspondence is as follows. If we use $(w,\bar{w})$ to denote the physical coordinate of the BCFT setup, the metric of its gravity dual in the vicinity of the asymptotic boundary can be approximately written into the Poincar\'e form 
\begin{align}
    ds^2 = \frac{dz^2+dw d\bar{w}}{z^2} + \cdots .
\end{align}
In this case, the standard way to take a UV-cutoff is to make a cut at $z=\epsilon$. By extending the map \eqref{eq:ndisk_to_1disk} to the bulk of the gravity dual via the Banados map, the cutoff surface should be chosen to be at 
\begin{align}\label{eq:naivecut}
    \eta = \frac{\epsilon}{n |\xi|^{n-1}}.
\end{align}
If we simply plug $\xi|_{w=0} = 0$ in it, then the cutoff will diverge. We need smearing to get rid of this divergence. 
It is natural to replace the endpoint of the conical defect from $|w| = 0$ to $|w|=\epsilon$. By plugging $\xi|_{w=\epsilon} = \epsilon^{\frac{1}{n}}$ into the above expression, we get a cutoff at 
\begin{align}
    \eta = \frac{\epsilon^{\frac{1}{n}}}{n}.
\end{align}
With this cutoff, the length of the conical defect $\Gamma_n$ can be evaluated as 
\begin{align}
    \int_{\Gamma_n} \sqrt{\gamma} &= \log\frac{R^{\frac{1}{n}} \cosh \left[{\rm arctanh}(T)\right]+
    R^{\frac{1}{n}} \sinh \left[{\rm arctanh}(T)\right]}{\ep^{\frac{1}{n}}/n} \nonumber\\
    &= \frac{1}{n} \log\frac{R}{\epsilon} + {\rm arctanh}(T) + \log n .
\end{align}
This does not match the BCFT result \eqref{eq:modularBCFT} due to the existence of the last term. In the following, we will introduce another cutoff regime and see how it produces the correct result.

\paragraph{A new cutoff regime gives the correct answer}~\par

On the BCFT side, when applying the replica method, the replica manifold $\Sigma_n$ is singular at $|w|=|\xi|=0$. To compute the partition function on $\Sigma_n$, we may introduce a hole at $|w| = \epsilon$, i.e. $|\xi| = \epsilon^{\frac{1}{n}}$ corresponding to the entangling surface. The hole introduces a new boundary to the original replica manifold. Although the details of this boundary cannot be determined in general, we can treat it as a conformal boundary with zero boundary entropy \cite{Ohmori2015}. 

On the gravity side, according to AdS/BCFT, this new boundary introduces another end-of-the-world brane located at 
\begin{align}
    \xi \bar{\xi} + \eta^2 = \left(\epsilon^{\frac{1}{n}}\right)^2.
\end{align}
Therefore, $\Gamma_n$ now turns out to be a line connecting the two poles of the two end-of-the-world branes, and the cutoff should be taken as $\delta_n = \epsilon^{\frac{1}{n}}$. As a result, the length of $\Gamma_n$ is 
\begin{align}
    \int_{\Gamma_n} \sqrt{\gamma} &= \log\frac{R^{\frac{1}{n}} \cosh \left[{\rm arctanh}(T)\right]+
    R^{\frac{1}{n}} \sinh \left[{\rm arctanh}(T)\right]}{\ep^{\frac{1}{n}}} \nonumber\\
    &= \frac{1}{n} \log\frac{R}{\epsilon} + {\rm arctanh}(T) .
\end{align}
Accordingly, the holographic modular R\'enyi entropy turns out to be 
\begin{align}
    \tilde{S}^{(n)}(\rho_A) = \frac{1}{4G_N}\frac{1}{n} \log \frac{R}{\epsilon} + S_{\rm bdy} . 
\end{align} 
We can also obtain the $n$-th R\'enyi entropy via integration over $n$:
\begin{align}
    S^{(n)}(\rho_A) = \frac{n}{n-1} \int_1^{n} dn' \frac{\tilde{S}^{(n')}}{n'^2} = \frac{1}{8G_N}\left(1+\frac{1}{n}\right) \log \frac{R}{\epsilon} + S_{\rm bdy}. 
\end{align}
These match the BCFT result \eqref{eq:RenyidiskBCFT} and \eqref{eq:modularBCFT} with $c/6 = 1/(4G_N)$.

\subsection{Holographic one-point Functions}
In the last subsection, we present a holographic computation of modular R\'enyi entropies on a disk. We have seen that geometries in which a conical defect extends from the asymptotic boundary and ends on the end-of-the-world brane were introduced naturally, when we considered dividing $\CM_n$, the gravity dual of the replica manifold, into $\hat{\CM}_n = \CM_n/\mathbb{Z}_{n}$. This computation is equivalent to the computation of the one-point function of the twist operator, which is non-vanishing. 

This tells us that, conical defects ending on the end-of-the-world branes are not prohibited by the first principle. 
Based on this, let us reconsider what is necessary for constructing a minimal theory of AdS$_3$/BCFT$_2$. In the AdS$_3$/CFT$_2$ case, a minimal theory includes Einstein gravity and conical defects. Besides these ingredients, a minimal model of AdS$_3$/BCFT$_2$ also needs end-of-the-world branes. On the other hand, {\it whether a conical defect can end on an end-of-the-world brane or not} is not determined by the first principle, but determined by hand. Therefore, we have the following two minimal choices: 
\begin{itemize}
    \item We do not allow the conical defects to end on the brane. 
    \item We allow the conical defects to end on the brane. 
\end{itemize}
The difference between the two choices is explicitly reflected in the behavior of one-point functions. The first choice corresponds to the case in which all the one-point functions of primaries vanish. On the other hand, the second choice implies all the one-point functions of primaries survive. 

If we consider the situation where the one-point function is non-vanishing, then the results derived in Section \ref{sec:two-CFT} do not hold except for some special cases (e.g., a spinning particle geometry). Correspondingly, on the gravity side, since the conical defects are allowed to end on the brane, the configurations discussed in Section \ref{sec:bulk-two} never dominate. 

\subsection{A Refined Formula for Holographic R\'enyi Entropy }

The computation of the holographic R\'enyi entropy in Section \ref{sec:holRenyi} indicates that there is a refined version of the Ryu-Takayanagi formula \cite{Ryu:2006bv,Ryu:2006ef} to compute the holographic entanglement entropy and Dong's formula \cite{Dong2016} to compute the holographic modular R\'enyi entropy. To see this, let us firstly review Dong's formula in AdS$_3$/CFT$_2$. According to \cite{Dong2016}, the $n$-th modular R\'enyi entropy is given by
\begin{align}\label{eq:oldRT}
    \tilde{S}^{(n)}(\rho_A) = \min_{\gamma_n} \frac{{\rm length~of~}\gamma_n}{4G_N},
\end{align}
where $\gamma_n$ is a cosmic string connecting the edges of $A$ with invariant mass 
\begin{align}
    m_n = \frac{1}{4G_N} \frac{n-1}{n}. 
\end{align}
The $n\rightarrow1$ limit gives the standard Ryu-Takayanagi formula. 

\begin{figure}[H]
    \centering
    \includegraphics[width=8cm]{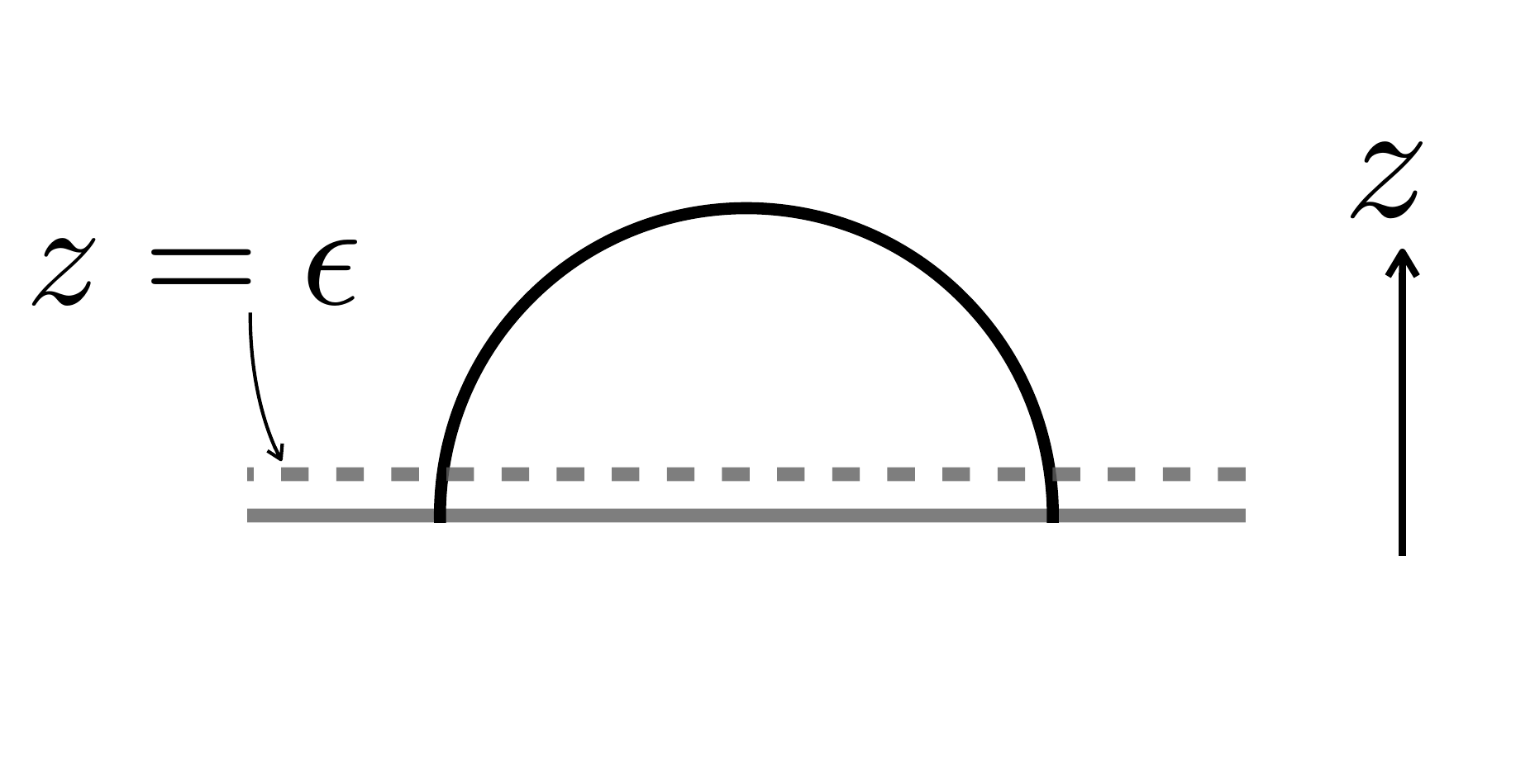}
    \caption{The standard way to make a cutoff by introducing the $z=\epsilon$ surface. }
    \label{fig:stand_cut}
\end{figure}

However, there is one insufficient point of \eqref{eq:oldRT}. Since the length of $\gamma_n$ is UV-divergent, one needs to specify a cutoff regime to get a finite answer. The standard way to impose a UV-cutoff is to introduce a constant cutoff surface in the gravity dual corresponding to the physical setup. See Figure \ref{fig:stand_cut} for a sketch. However, as we have seen in Section \ref{sec:holRenyi}, this kind of cutoff fails to reproduce the boundary entropy term appearing in the modular R\'enyi entropy. Moreover, this way to choose the cutoff is not covariant with respect to the coordinate transformation. 

\begin{figure}[H]
    \centering
    \includegraphics[width=8cm]{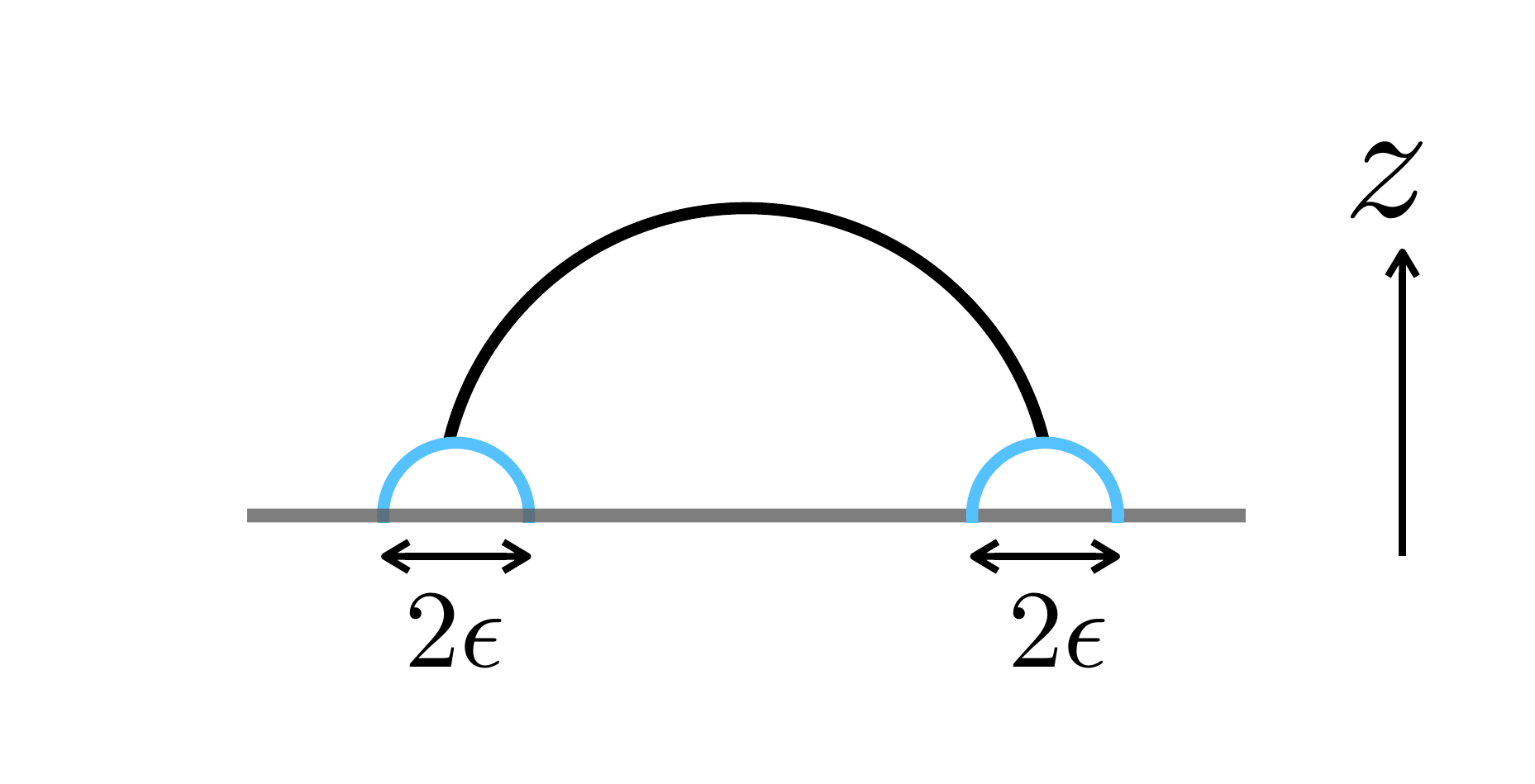}
    \caption{Cutoff by introducing end-of-world branes (blue) corresponding to the entangling surfaces. }
    \label{fig:brane_cut}
\end{figure}

In fact, how to choose a proper cutoff is a question not only in the holographic setup but also on the CFT side. On the CFT side, this question is answered in \cite{Ohmori2015}. The recipe given in \cite{Ohmori2015} is to firstly introduce disk-shaped holes at the edges of the subsystem $A$, and then impose conformal boundary conditions on the boundary induced by the holes. By applying this recipe on the CFT side, and using standard AdS/BCFT to map it to the gravity dual, each hole introduces an end-of-the-world brane. Accordingly, the $n$-th modular R\'enyi entropy is given by
\begin{align}\label{eq:newRT}
    \tilde{S}^{(n)}(\rho_A) = \min_{\Gamma_n} \frac{{\rm length~of~}\Gamma_n}{4G_N},
\end{align}
where $\Gamma_n$ is a cosmic string connecting the end-of-the-world branes with invariant mass 
\begin{align}
    m_n = \frac{1}{4G_N} \frac{n-1}{n}. 
\end{align}
As we have seen in Section \ref{sec:holRenyi}, this cutoff recipe reproduces the correct boundary entropy term. See Figure \ref{fig:brane_cut} for a sketch.

There are several advantages of the refined holographic formula, besides the fact that it can reproduce the correct boundary entropy term in the computation of modular R\'enyi entropies. First of all, this recipe for taking the UV-cutoff is covariant. All one needs to do is to solve the equation of motion with the new branes and cosmic strings, and then compute the length of the strings. This is directly related to the second advantage: the new recipe unifies the RT formula for AdS/BCFT \cite{Takayanagi2011}, which allows the RT surface to end on the end-of-the-world branes, and the RT formula for AdS/CFT, by letting the RT surface {\it always} end on some branes.

To see why our recipe is more than a fancy reformulation of the standard one, we also need to explain why it was not necessary for a bunch of previous works of computing holographic entanglement entropy and holographic R\'enyi entropy, and when it becomes necessary. First of all, if we only consider entanglement entropy, but do not consider R\'enyi entropy for general $n$, then the effect of choosing different UV-cutoff can be always absorbed by the redefinition of the cutoff parameter $\epsilon$. The effect only appears when considering R\'enyi entropy for general $n$, and letting setups with different $n$ share the same cutoff parameter $\epsilon$. However, holographic computations performed so far were not accurate enough to see this effect. For example, when the authors evaluated the on-shell gravity action to perform the holographic computation of R\'enyi entropy in \cite{Hung2011}, they did not perform the integral deep into the bulk, but just stopped at an intermediate point so that the integral was computable and sufficient to reproduce the leading area law term. As another example, in \cite{Dong2016}, the author computed only the $n$-th mutual information, which does not rely on how the UV-cutoff is taken. However, if one wants to compute the constant term as a function of $n$, then how to take the UV-cutoff becomes important. The boundary entropy term computed in Section \ref{sec:holRenyi} is such an example. Note that our cutoff recipe is essentially different from the standard one, in the sense that our cutoff surface takes part in the gravitational action, and should be determined dynamically by solving the equation of motions.

Another thing to note is that, although our computation and argument suggest the existence of such a refined version of the holographic R\'enyi entropy formula, we do not succeed to present it. This is because we have not given the equation of motion with the interaction occurring at the intersection between the branes and the cosmic strings. It would be an interesting future direction to specify the intersection term appearing in the action, and derive a holographic R\'enyi entropy formula in AdS/BCFT. 

%%%%%%%%%%%%%%%%%%%%%%%%%%%%%%%%%%%%%%%%%%%%%%%%%%%%%%%%%%%%%%%%%%%%%%%%%%%%%%%%%%%%%%%%%%%%%%
%%%%%%%%%%%%%%%%%%%%%%%%%%%%%%%%%%%%%%%%%%%%%%%%%%%%%%%%%%%%%%%%%%%%%%%%%%%%%%%%%%%%%%%%%%%%%%
\section{Non-scalar Primaries and Spinning Particles}\label{sec:spinning}
%%%%%%%%%%%%%%%%%%%%%%%%%%%%%%%%%%%%%%%%%%%%%%%%%%%%%%%%%%%%%%%%%%%%%%%%%%%%%%%%%%%%%%%%%%%%%%
%%%%%%%%%%%%%%%%%%%%%%%%%%%%%%%%%%%%%%%%%%%%%%%%%%%%%%%%%%%%%%%%%%%%%%%%%%%%%%%%%%%%%%%%%%%%%%

In this section, we present, to our knowledge, the first-ever analysis of AdS/BCFT setups with non-scalar primary insertions. At first glance, one may find some mysterious points by the following observations from the BCFT side. 

First, the vanishing nature of the one-point function of a non-scalar primary may seem to be puzzling on the gravity side. In the previous section, we realize the gravity dual of a BCFT setup with a non-vanishing one-point function by letting the massive particle, the holographic dual of scalar primaries, end on the end-of-the-world brane. Compared to this, non-scalar primaries' one-point functions always vanish on the BCFT side as 
\begin{align}
    \braket{\phi_{h,\bar{h}}(z,\bar{z})}_{\rm UPH} \propto \braket{\phi_{h}(z) \phi_{\bar{h}}(\bar{z})}_{\mathbb{R}^2} = 0,
\end{align}
where $\phi_{h,\bar{h}}$ is the non-scalar primary and $\phi_h$ ($\phi_{\bar{h}}$) is its chiral (anti-chiral) part. This indicates that there must be a mechanism which prohibits spinning particles, the holographic duals of non-scalar primaries, ending on the end-of-the-world brane. However, it is not obvious why this is the case on the gravity side at first glance\footnote{Or maybe obvious for some readers.}. While a massive with a zero spin can in principle end on the brane, it is hard to imagine how adding a tiny spin to the particle would kick it out of the brane. 

Second, self-intersection of the brane may seem to embrace a revival on the gravity side. In the previous sections, we show that self-intersections do not happen for two-point insertions of scalar primaries, since the parameter region $h=\bar{h}\geq c/32$ where self-intersections were expected in fact contains a black hole. This was firstly discussed from the conformal bootstrap, and then confirmed in its gravity dual. Compared to this, what happens in the non-scalar case is less clear. For example, when considering the chiral particle with $c/32<h<c/24$ and $\bar{h}=0$, it seems that self-intersections would happen in the chiral part if we could naively split the contributions into chiral and anti-chiral parts. While this does not happen, it is less trivial to answer whether the AdS/BCFT construction can realize holographic duals of two-point functions of non-scalar primaries in BCFT, and if the answer is yes, how to do so. 

To tackle these questions, we present a method which allows one to construct a geometry with a spinning conical defect from global AdS$_3$. We firstly show that the method works by identifying the resulting metric with a more well-known form and matching the ADM mass and the angular momentum of the configuration. Then we will focus on the setup where spinning primaries are inserted in a BCFT, by applying the above construction to configurations with an end-of-the-world brane. In particular, we will see how the one-point function of a spinning primary vanishes from the holographic point of view. Besides, we will also consider the configuration which corresponds to a two-point function of spinning primaries, and see that the self-intersection bound derived in the gravitational setup perfectly matches the black hole threshold derived from the BCFT side also in this case.

\subsection{Spinning Defect from Twisted Identification}

\begin{figure}[H]
    \centering
    \includegraphics[width=12cm]{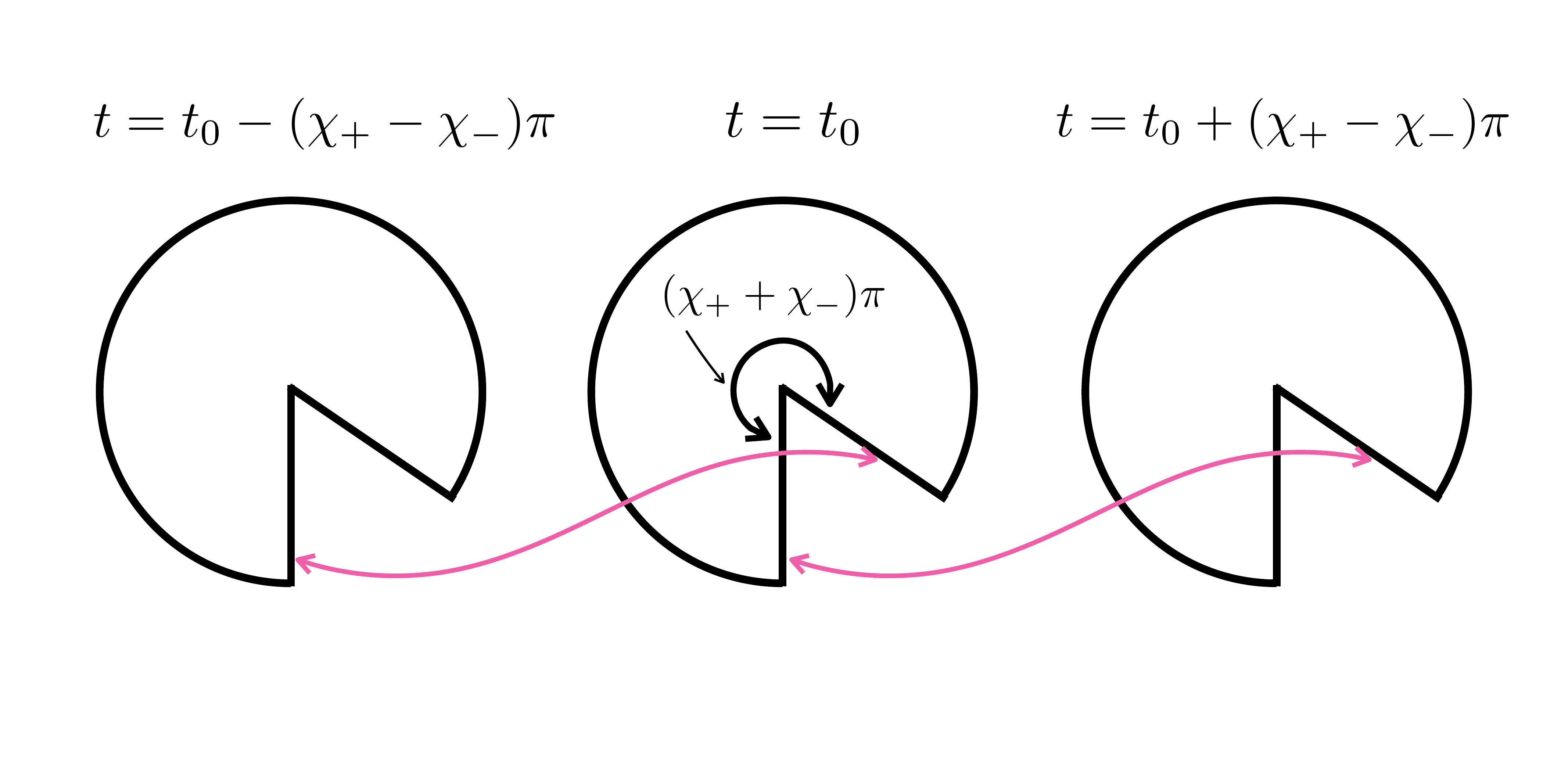}
    \caption{The twisted identification to create a spinning defect from global AdS$_3$. For each time slice $t=t_0$, we exclude a portion with angle $2\pi-(\chi_++\chi_-)\pi$, and then paste it with another two time slices $t=t_0-(\chi_+ - \chi_-)\pi$, $t=t_0+(\chi_+ - \chi_-)\pi$ in the way shown in the figure. Two lines connected by the pink arrows are identified with each other.}
    \label{fig:twisted_identif}
\end{figure}

In this section, we work on the Lorentzian signature, which turns out to be convenient when treating a spinning defect. The metric of the global AdS$_3$ may be written as 
\begin{align}
    ds^2 = -(r^2+1)dt^2 + \frac{dr^2}{r^2+1} + r^2 d\theta^2,
\end{align}
with $\theta\in[0,2\pi)$. As we have seen in Section \ref{sec:bulk-two}, a scalar primary with conformal dimension $h_i=\bar{h}_i$ introduces a conical defect that can be realized by restricting $\theta$ to $\theta \in [0, 2\chi\pi)$ and identify $\theta = 2 \chi \pi  \sim 0$, where $\chi$ is related to $h_i$ as 
\begin{align}
    \frac{24h_i}{c} = 1-\chi^2. 
\end{align}
From this expression, intuitively, it is plausible to believe that a non-scalar primary with chiral conformal dimension $h_i$ and antichiral conformal dimension $\bar{h}_i$ will introduce a spinning defect that can be realized by performing the following identification to lightcone coordinates in the global AdS$_3$,
\begin{align}\label{eq:identification_lightcone}
    (\theta-t,\theta+t) \sim (\theta-t+2\chi_-\pi, \theta+t+2\chi_+\pi), 
\end{align}
by considering the chiral part and the anti-chiral part separately. Here, we have
\begin{align}\label{eq:chi_pm}
    \frac{24h_{\pm}}{c} = 1-\chi_{\pm}^2,
\end{align}
where we denote $h_i (\bar{h}_i)$ by $h_+ (h_-)$ for convenience.
This can be alternatively written as 
\begin{align}\label{eq:identification_thetat}
    (\theta,t) \sim (\theta, t) + ((\chi_++\chi_-)\pi, (\chi_+-\chi_-)\pi). 
\end{align}
See Figure \ref{fig:twisted_identif} for a depiction.

In fact, this naive expectation is totally correct. We present a quantitative check in Appendix \ref{app:spinning} by performing a proper coordinate transformation and see that the spacetime obtained from the identification \eqref{eq:identification_lightcone} is exactly identical to the one obtained by performing an analytic continuation to the Kerr-BTZ black hole. Also, note that a similar construction has been discussed in \cite{Carlip1994}.

In this section, however, we will take the construction presented above as a fact, and use it to discuss holographic aspects associated with spinning particles.

\subsection{Vanishing One-point Function from Brane Mismatch}

In this subsection, we would like to explain why the holographic one-point function of a non-scalar primary is vanishing using the construction presented in the previous subsection. 

The answer is simple and straightforward. As depicted in Figure \ref{fig:twisted_identif}. If one wants to introduce an orientable spinning defect into a spacetime, then any codimension-1 surface which intersects the spinning defect will need to be pasted to another two codimension-1 surfaces on its two sides with respect to the direction of the spinning defect. As a result, the spinning defect cannot end at any finite point, otherwise a mismatch would happen at the vicinity of that point. If the spinning defect could end on an end-of-the-world brane, a mismatch of the brane would be unavoidable, at least at the vicinity of the intersection between the brane and the defect. In this way, the intersection between a brane and a spinning defect would cause a singular region in the spacetime and introduce a divergence to the gravitational action. Therefore, such a configuration is forbidden, which makes a great difference between a conical defect and a spinning defect. This difference leads to the fact that the one-point function of a non-scalar primary must be zero, while that of a scalar primary can be non-zero.\footnote{One may get confused why a spinning defect can ``end" on the asymptotic boundary. This is because the asymptotic boundary is not a part of the spacetime, which means that the spinning defect does not end anywhere.}

\subsection{No Self-intersection for Geometries with Spinning Excitations} ~\par 

Similar to the conical defect case, the spinning defect will also bend the brane, but in a more complicated way. Let us start from a global AdS$_3$ with a static end-of-the-world brane in it. The brane intersects the asymptotic boundary at $\theta = 0, \pi$. Therefore, if we introduce a spinning defect at the center of the geometry $r=0$, then we can see from \eqref{eq:identification_thetat} that the brane would intersect itself at 
\begin{align}
    (\chi_+ + \chi_-)\pi < \pi,
\end{align}
if no black hole was generated in the AdS$_3$. Translating it to the BCFT language via \eqref{eq:chi_pm}, we have 
\begin{align}\label{eq:spin_intsection_bound}
    \sqrt{1-\frac{24h_+}{c}} + \sqrt{1-\frac{24h_-}{c}} < 1. 
\end{align}
Let us compare this with Fact \ref{fact1'} obtained from the conformal bootstrap. According to Fact \ref{fact1'}, a non-scalar primary with conformal dimension $(h_+,h_-)$ will form a black hole in the gravity dual when 
\begin{align}
    \alpha_+ + \alpha_- > \frac{Q}{2},
\end{align}
where $\a_+(\a_-)$ is the Liouville momentum of the chiral (antichiral) part of the non-scalar primary.
Using the relation in \eqref{eq:Liouville_notation}, this gives 
\begin{align}
    \frac{\sqrt{\frac{c-1}{6}}-\sqrt{\frac{c-1}{6}-4h_+}}{2} + \frac{\sqrt{\frac{c-1}{6}}-\sqrt{\frac{c-1}{6}-4h_-}}{2} > \sqrt{\frac{c-1}{24}}.
\end{align}
Equivalently, we have 
\begin{align}
    \sqrt{1-\frac{24h_+}{c-1}} + \sqrt{1-\frac{24h_-}{c-1}} < 1.
\end{align}
This again perfectly matches \eqref{eq:spin_intsection_bound} by replacing $c-1$ with $c$. Therefore, self-intersection is nicely avoided by black hole formation also for the spinning particle case.

%%%%%%%%%%%%%%%%%%%%%%%%%%%%%%%%%%%%%%%%%%%%%%%%%%%%%%%%%%%%%%%%%%%%%%%%%%%%%%%%%%%%%%%%%%%%%%
%%%%%%%%%%%%%%%%%%%%%%%%%%%%%%%%%%%%%%%%%%%%%%%%%%%%%%%%%%%%%%%%%%%%%%%%%%%%%%%%%%%%%%%%%%%%%%
\section{Heavy Excitations and Negative Tension Branes}\label{sec:negative}
%%%%%%%%%%%%%%%%%%%%%%%%%%%%%%%%%%%%%%%%%%%%%%%%%%%%%%%%%%%%%%%%%%%%%%%%%%%%%%%%%%%%%%%%%%%%%%
%%%%%%%%%%%%%%%%%%%%%%%%%%%%%%%%%%%%%%%%%%%%%%%%%%%%%%%%%%%%%%%%%%%%%%%%%%%%%%%%%%%%%%%%%%%%%%

\begin{figure}[H]
 \begin{center}
  \includegraphics[width=14.0cm,clip]{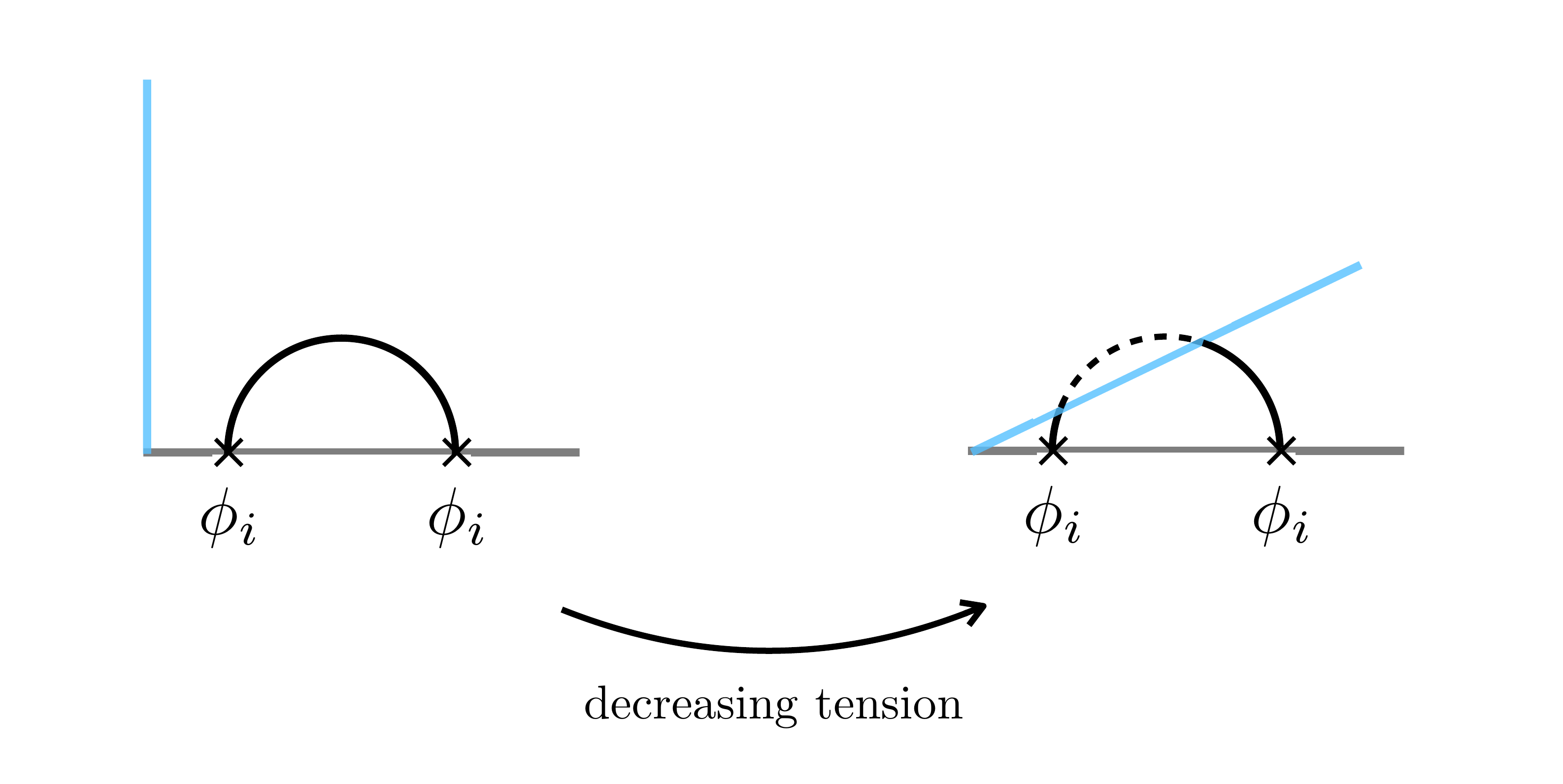}
 \end{center}
 \caption{
 The geodesics approximation of a two-point function on a half plane. The blue lines show the brane configurations. The left figure shows the zero tension case. By decreasing the brane tension to a negative value, there is a possibility that a part of the geodesics is excluded by the end-of-the-world brane.}
 \label{fig:negative}
\end{figure}

In this section, we will address the mysterious point associated with negative tension branes.
Let us first describe it in short.
Consider a two-point function on a half-plane.
There are cases where this correlation function is approximated by the bulk channel vacuum block (see the left of Figure \ref{fig:negative}).
For example, this happens if we assume the one-point functions for scalar primaries are vanishing or consider a non-scalar two-point function as investigated in previous sections.
On the gravity side, when the tension of the brane is negative, there is a possibility that the brane excludes a part of the geodesic corresponding to the two-point function (see the right of Figure \ref{fig:negative}).
It is not clear how to deal with geodesics behind the end-of-the-world brane.
A significant clue to this situation comes from Fact \ref{fact2},
which states that the solution to the bootstrap equation for the two-point function is not sensitive to the $g$-function or the boundary entropy.
Note that this negative tension problem has previously been pointed out in \cite{Bianchi2022} and the authors conjectured that the boundary primary spectrum would be changed if the brane tension is negative.
However, now we have solved the conformal bootstrap equation and found that the spectrum is not sensitive to the tension. Therefore, this possibility should be excluded.

\begin{figure}[H]
    \centering
    \includegraphics[width=4cm]{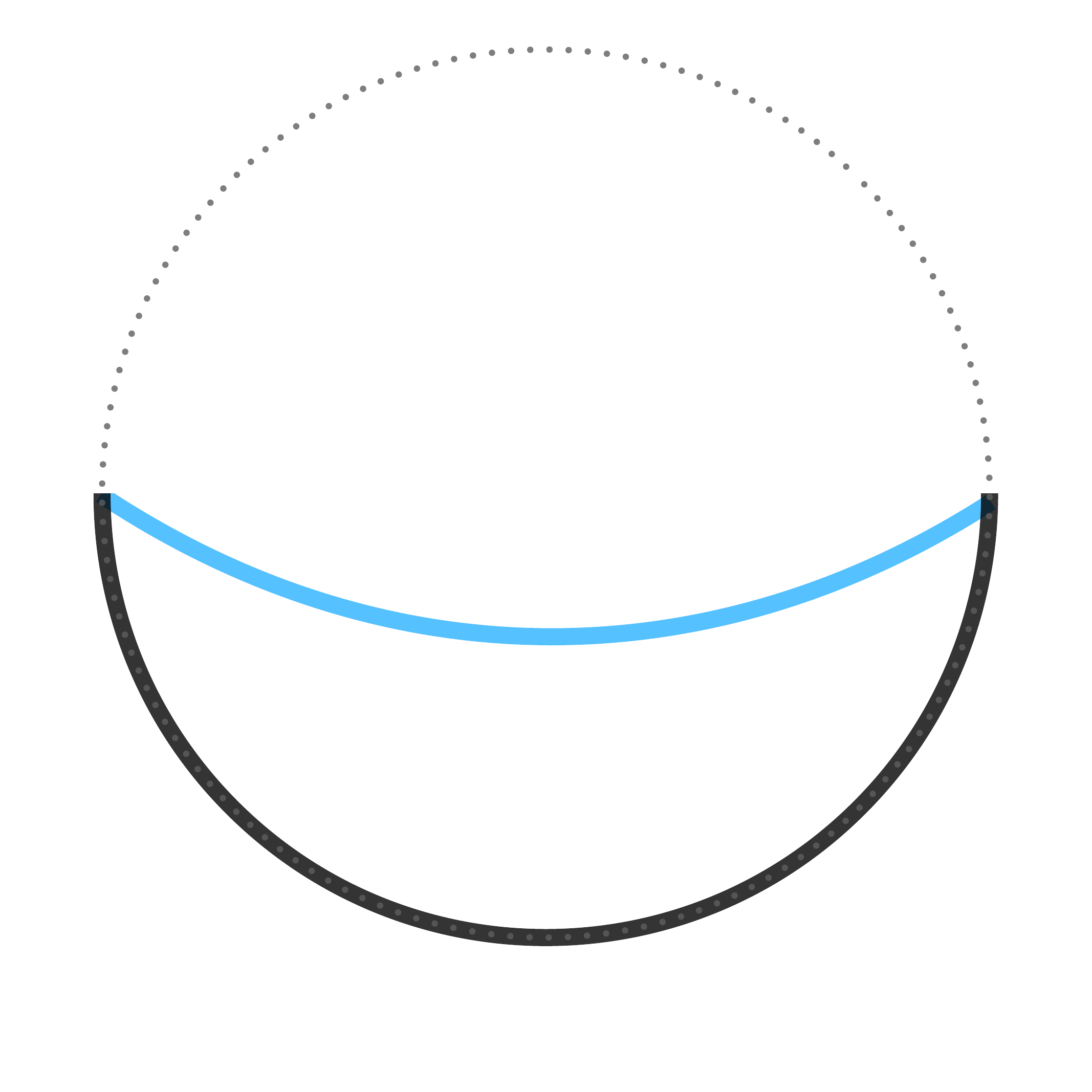}
    \includegraphics[width=4cm]{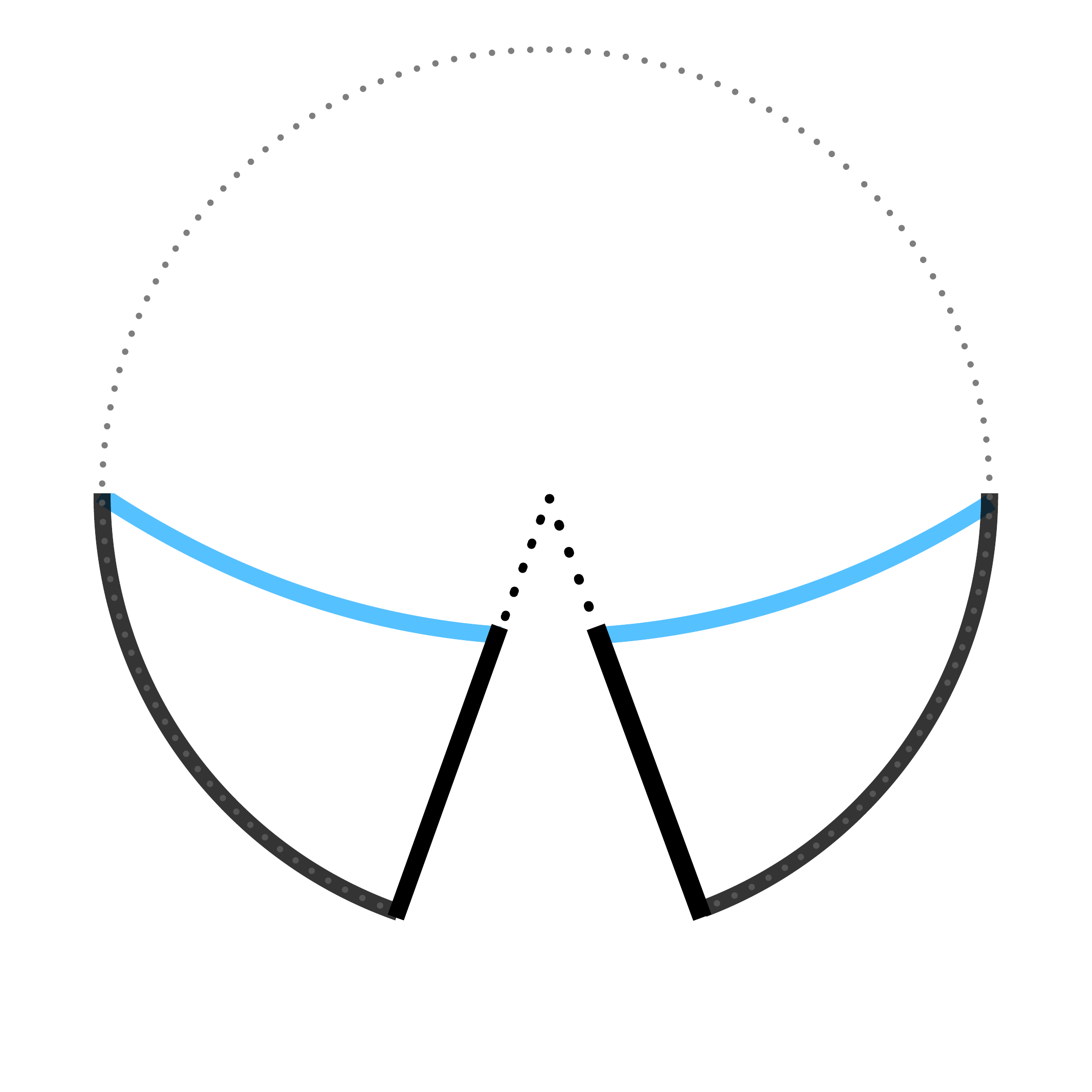}
    \includegraphics[width=4cm]{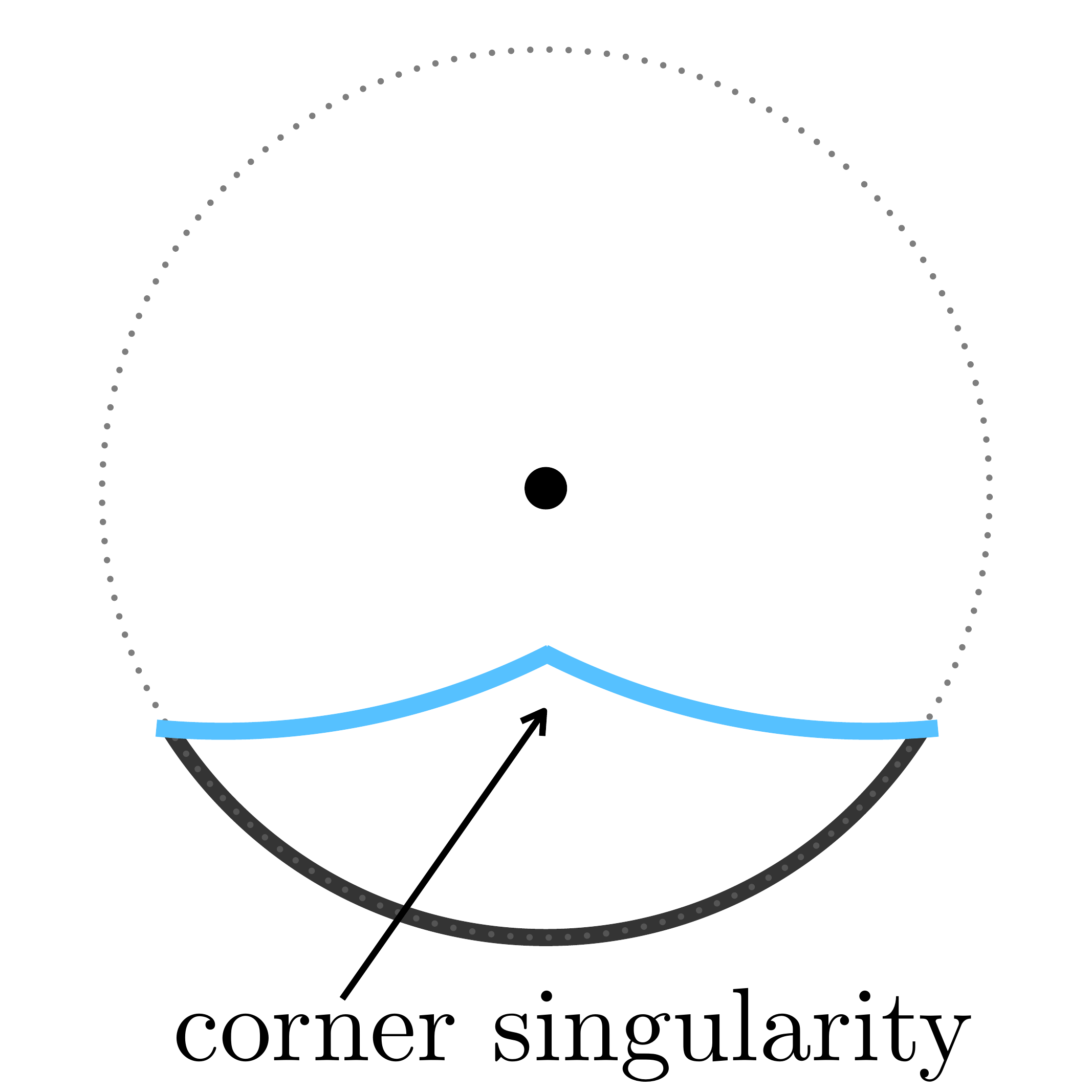}
    \caption{Cutting and pasting the spacetime with a negative tension brane (shown in blue). Starting from the vacuum configuration (left), excluding its portion in a symmetric way (middle) and then pasting the two boundaries, we get a configuration with a corner defect localized on the brane.}
    \label{fig:corner}
\end{figure}

To see what is really happening for the negative tension case, let us go back to the global coordinate case and repeat what we have done for the positive tension case in Section \ref{sec:bulk-two}.
Cutting the geometry to create a conical singularity at the ``center" of the global AdS$_3$ which is in fact beyond the end of the world, the brane gets a corner singularity (see Figure \ref{fig:corner}).
To treat this corner singularity,
we consider the following action including a generalized Hayward corner term \cite{Hayward1993},
\begin{equation}\label{eq:hayward}
\begin{aligned}
    I_{\rm grav} 
    &= -\frac{1}{16\pi G_N}\int_{\CM} \sqrt{g}(R+2) - \frac{1}{8\pi G_N} \int_Q \sqrt{h} (K-T)\\
    &+ m\int_\Gamma \sqrt{\gamma} 
    - \frac{1}{8\pi G_N} \int_{\ca{C}} \sqrt{\eta}(\Theta-T_{\mathcal{C}}),
\end{aligned}
\end{equation}
where $\eta_{\dot{a}\dot{b}}$ is the induced metric on the corner $\ca{C}$, $\Theta$ is the internal angle between two branes, and $T_{\mathcal{C}}$ is the tension of the defect on the corner. The last term is a generalized Hayward term in the sense that the standard Hayward term corresponds to the $T_{\mathcal{C}} = \pi$ case.  

Note that the defect tension $T_{\mathcal{C}}$ cannot be chosen freely. For this action to give solutions shown in Figure \ref{fig:corner}, $T_{\mathcal{C}}$ should be dynamically determined by the brane tension $T$ and the particle mass $m$. 
Even though the corner contribution non-trivially changes the solution to the Einstein equation,
the ADM mass is again given by (\ref{eq:M_BandP}) and the energy compared to the ground state is
\begin{align}\label{eq:DE}
    \Delta E = \frac{c}{6} \sqrt{1-\frac{24h_i}{c}} \left(1-\sqrt{1-\frac{24h_i}{c}}\right),
\end{align}
which precisely matches the CFT result.
It means that our bottom-up model is still consistent with the conformal bootstrap even if the brane tension is negative.

\begin{figure}[H]
 \begin{center}
  \includegraphics[width=9.0cm,clip]{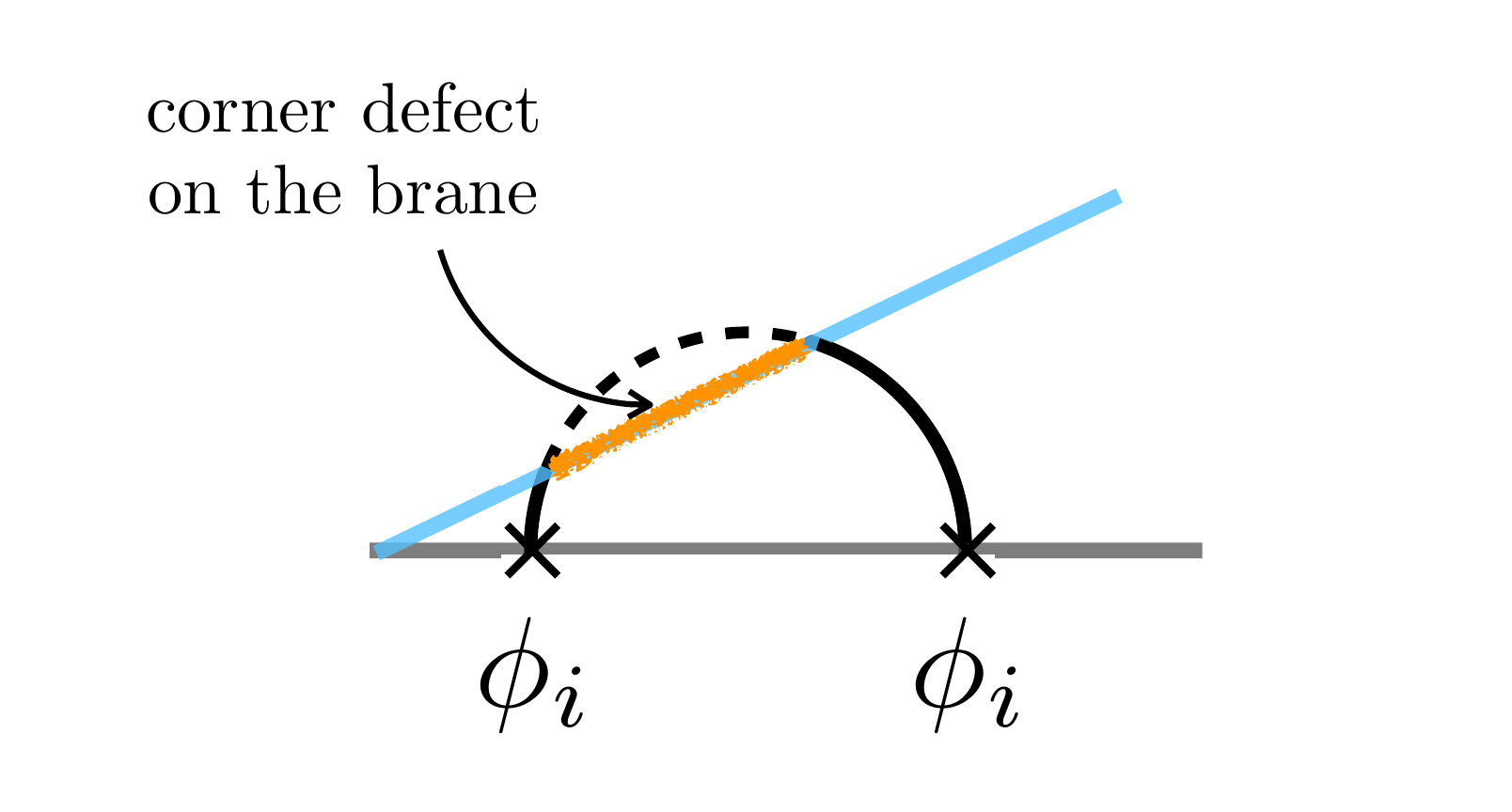}
 \end{center}
 \caption{
The holographic dual of a two-point function on a half plane with negative tension.
There is a corner defect line on the brane.
The tension of the defect is uniquely fixed by the brane tension and the mass of the particle.
}
 \label{fig:defect}
\end{figure}

In summary, we have seen that in the negative tension case, although the originally expected bulk channel seems to be problematic since it is cut by the brane in the middle as shown in Figure \ref{fig:negative}, what is really happening is that the massive particle nontrivially interacts with the brane and creates a defect on it as shown in Figure \ref{fig:defect}. 
Although the solution to the Einstein equation seems to be drastically changed, the formula (\ref{eq:DE}) is not changed, which is consistent with the conformal bootstrap.

\begin{figure}[H]
 \begin{center}
  \includegraphics[width=7.0cm,clip]{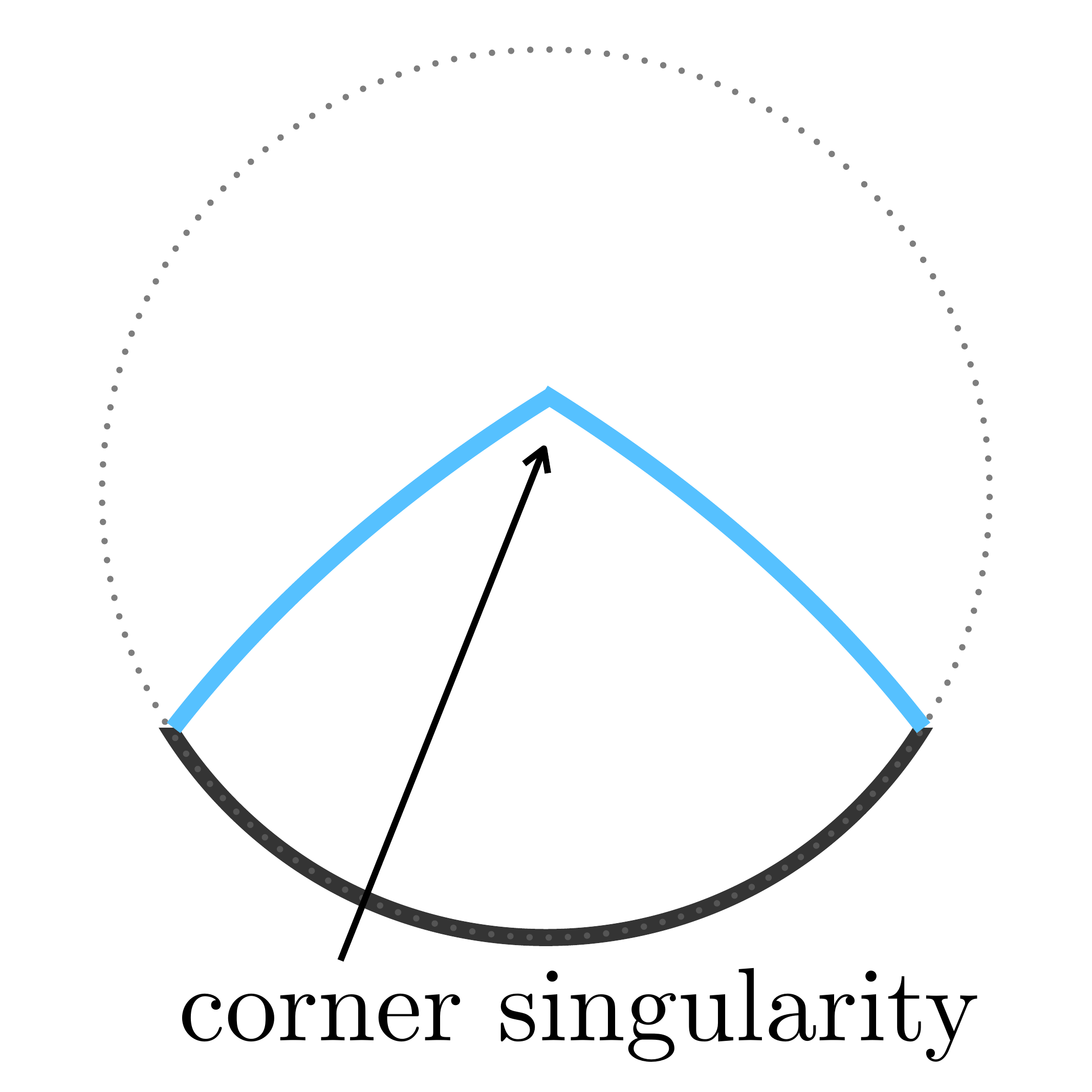}
 \end{center}
 \caption{Positive tension brane with a corner defect on it.}
 \label{fig:branedefect_positive}
\end{figure}

One may wonder if the we can consider a similar geometry for positive tension cases shown in Figure \ref{fig:branedefect_positive} to realize a dual of a two-point function of bulk primaries. We propose that this solution can be excluded in the following way.
In finding a solution to the action (\ref{eq:hayward}),
we first seek a solution with $T_{\ca{C}} = \pi$.
In the case where we cannot find such a solution,
we second seek a solution with the dynamical $T_{\ca{C}}$.
This process can be justified by the interpretation of the defect on the brane as the conical singularity close to the brane.
In fact, the conical defect attaches to the brane in the negative tension case, but this is not the case for the positive tension.
Consequently, while we need the dynamical $T_{\ca{C}}$ in the negative tension case,
this is not necessary in the positive tension case.
As a result, configurations like \ref{fig:branedefect_positive} are excluded.

It is worth noting the relation between the setups in \cite{Miyaji2022, Biswas2022} and ours. 
In \cite{Miyaji2022, Biswas2022}, the authors proposed a holographic dual of an open string with distinct boundaries by introducing a corner defect on the end-of-the-world brane. 
One main difference comes from this corner defect.
In their case, the defect tension is a parameter which can be freely chosen. This parameter fixes the ground state of the open string spectrum.
In other words, $T_{\ca{C}}$ can be regarded as a parameter which specifies a theory.
Contrary to this, what we would like to do is to realize the negative tension setup in an already-given theory, and therefore our defect tension $T_{\ca{C}}$ should be determined from $T$ and $m$, and it just represents the hidden conical singularity in a non-trivial way.

It would be an interesting future direction to consider the braneworld interpretation of brane-localized defects.

%%%%%%%%%%%%%%%%%%%%%%%%%%%%%%%%%%%%%%%%%%%%%%%%%%%%%%%%%%%%%%%%%%%%%%%%%%%%%%%%%%%%%%%%%%%%%%
%%%%%%%%%%%%%%%%%%%%%%%%%%%%%%%%%%%%%%%%%%%%%%%%%%%%%%%%%%%%%%%%%%%%%%%%%%%%%%%%%%%%%%%%%%%%%%
\section{Discussion}
%%%%%%%%%%%%%%%%%%%%%%%%%%%%%%%%%%%%%%%%%%%%%%%%%%%%%%%%%%%%%%%%%%%%%%%%%%%%%%%%%%%%%%%%%%%%%%
%%%%%%%%%%%%%%%%%%%%%%%%%%%%%%%%%%%%%%%%%%%%%%%%%%%%%%%%%%%%%%%%%%%%%%%%%%%%%%%%%%%%%%%%%%%%%%
We propose some remaining questions and interesting future works at the end of this paper. 

\paragraph{Other consistency conditions}~\par

In this article, we focus on gravity with a single end-of-the-world brane and a single massive particle,
which requires the consistency condition to avoid the self-intersection.
The interesting point is that this consistency condition can also be obtained from the conformal bootstrap.
One straightforward future direction is to consider more complicated setups.
It is known that there are other non-trivial conformal bootstrap equations in BCFT \cite{Lewellen1992}.
These bootstrap equations may give additional conditions for brane profiles or the interaction between branes and massive particles.
It would be interesting to see whether the bottom-up model still nicely works as semiclassical gravity by further analysis of other bootstrap equations.
A specific interesting future direction is to consider the consistency conditions in the setup where two branes can intersect with each other \cite{Miyaji2022, Biswas2022}.

\paragraph{Quantum Regge trajectory from gravity}~\par
In this article, we show that the spectrum of the binding energy between a brane and a massive particle follows the quantum  Regge trajectory.
It would be interesting to reproduce this spectrum from the gravity side.
One may obtain some hints from an interesting result in \cite{Maxfield2022}.
It is known that the spectrum of a two-particle state with a large spin shows the quantum Regge trajectory \cite{Kusuki2019a, Collier2019},
and it is shown in \cite{Maxfield2022} that this trajectory can be reproduced from the spinning strings coupled to gravity.

\paragraph{Refined holographic R\'enyi entropy}~\par

In this article, we present a computation of the R\'enyi entropy on the gravity side.
We propose that the proper cutoff surface is given by the end-of-the-world brane that arises from the entangling surface.
It would be interesting to give further checks for this cutoff proposal by evaluating the constant part in more complicated cases or by considering other entanglement measures.

\paragraph{Spinning particle}~\par

Most works about excited geometries in AdS/BCFT (even in AdS/CFT)
are restricted to setups with only scalar primaries.
With twisted identification presented in this paper, one can investigate, for example, holographic entanglement entropy, in geometries with spinning particles.
As one can see for a one-point function on a disk, the generalization to non-scalar primary drastically changes some nature of physics.
Therefore, it would be interesting to consider this non-scalar generalization in various setups.

\paragraph{Negative tension}~\par

We have shown that the negative tension case can be dealt with in the same framework as the original one.
This result is consistent with the non-sensitivity to the boundary entropy shown from the conformal bootstrap.
Nevertheless, there are still unclear aspects since we only consider the simplest case.
It would be nice to give a further investigation in this direction.
In particular, it would be interesting to consider the brane-world \cite{Karch:2000ct} interpretation of the defect localized on the brane. 

\paragraph{Wightman functions with branes and defects}~\par
Throughout this article, we have presented several gravity duals with branes and conical defects. It would be intriguing to study the Wightman functions in these backgrounds as a generalization of \cite{Ageev:2015qbz,Arefeva:2016nic,Berenstein:2022ico}. These corresponds to heavy-heavy-light-light correlation functions on the BCFT side, and will give more information about the AdS/BCFT holography.

\paragraph{Higher dimensional generalization}~\par

In this article, we highly use the specialty of 2D BCFTs.
For example, it is (rather) easy to classify boundary conditions by using the state/boundary correspondence in 2D BCFTs.
This is not the case in higher dimensions.
For many reasons including the above one, it is not straightforward to generalize our results to higher dimensions,
nevertheless, it is very important and interesting to give further consideration in this direction.

%%%%%%%%%%%%%%%%%%%%%%%%%%%%%%%%%%%%%%%%%%%%%%%%%%%%
\section*{Acknowledgments}
%%%%%%%%%%%%%%%%%%%%%%%%%%%%%%%%%%%%%%%%%%%%%%%%%%%%
We are grateful to Nathan Benjamin, Cyuan-Han Chang, Scott Collier, Xi Dong, Hao Geng,  Andreas Karch, Marco Meineri, Takato Mori, Hidetoshi Omiya, Sridip Pal, Shan-Ming Ruan, David Simmons-Duffin, Tadashi Takayanagi, Maria Tikhanovskaya, Zhencheng Wang, Masataka Watanabe and Ying Zhao for useful discussions.
YK is supported by the Brinson Prize Fellowship at Caltech and the U.S. Department of Energy, Office of Science, Office of High Energy Physics, under Award Number DE-SC0011632.
ZW is supported by Grant-in-Aid for JSPS Fellows No. 20J23116, in part by the National Science Foundation under Grant No. NSF PHY-1748958 and by the Heising-Simons Foundation.

\appendix

%%%%%%%%%%%%%%%%%%%%%%%%%%%%%%%%%%%%%%%%%%%%%%%%%%%%%%%%%%%%%%%%%%%%%%%%%%%%%%%%%%%%%%%%%%%%%%
%%%%%%%%%%%%%%%%%%%%%%%%%%%%%%%%%%%%%%%%%%%%%%%%%%%%%%%%%%%%%%%%%%%%%%%%%%%%%%%%%%%%%%%%%%%%%%
\section{Constructing a Spinning Defect from Global \texorpdfstring{AdS$_3$}{AdS3}}\label{app:spinning}
%%%%%%%%%%%%%%%%%%%%%%%%%%%%%%%%%%%%%%%%%%%%%%%%%%%%%%%%%%%%%%%%%%%%%%%%%%%%%%%%%%%%%%%%%%%%%%
%%%%%%%%%%%%%%%%%%%%%%%%%%%%%%%%%%%%%%%%%%%%%%%%%%%%%%%%%%%%%%%%%%%%%%%%%%%%%%%%%%%%%%%%%%%%%%

In this appendix, we present a method to realize a spinning defect by performing a cut-and-paste to global AdS$_3$. We will first present the intuitive interpretation of why this construction is considered to be true. Then we will quantitatively check our expectation by comparing it with the standard metric obtained from an analytic continuation of the Kerr-BTZ metric. Note that such a construction has been discussed in \cite{Carlip1994}. 

First of all, let us review how a standard conical defect corresponding to a scalar primary is constructed. Starting from a global AdS$_3$ whose metric is given by
\begin{align}
    ds^2 = -(r^2+1)dt^2 + \frac{dr^2}{r^2+1} + r^2 d\theta^2,
\end{align}
with $\theta\in[0,2\pi)$. By keeping the same metric and changing the periodicity of $\theta$ into $2\chi\pi$, one can realize a conical defect with deficit angle $2\pi(1-\chi)$ at $r=0$. Here, when translating the current configuration into a 2D CFT setup, the conformal dimension $h_i$ of the primary operator which creates the conical defect is related to $\chi$ as 
\begin{align}
    \frac{24h_i}{c} = 1-\chi^2. 
\end{align}
Let us then think about the configuration which corresponds to a two-point function of a non-scalar primary. Inspired by the factorization structure in CFT$_2$, we may work with the lightcone coordinate $\theta-t$ and $\theta+t$. 

\paragraph{A conjecture on how to construct a spinning defect geometry}~\par
By intuition, it seems that imposing identification conditions separately on the chiral coordinate and the anti-chiral coordinate will reproduce a geometry with a spinning defect. Therefore, let us impose an identification condition separately: 
\begin{align}
    (\theta-t,\theta+t) \sim (\theta-t+2\chi_-\pi, \theta+t+2\chi_+\pi).
\end{align}
In other words, we can say this is a twisted identification with
\begin{align}
    (\theta, t) \sim (\theta+(\chi_++\chi_-)\pi, t+(\chi_+-\chi_-)\pi) = (\theta, t) + ((\chi_++\chi_-)\pi, (\chi_+-\chi_-)\pi).
\end{align}
Also by intuition, the geometry created in this way corresponds to a CFT two-point function of a primary with conformal dimension $(h_+,h_-) \equiv (h_i, \bar{h}_i)$ satisfying
\begin{align}\label{eq:handchi_spin}
    \frac{24h_+}{c} = 1-\chi_+, \quad
    \frac{24h_-}{c} = 1-\chi_-.
\end{align}
Our goal is to test this conjecture. 

\paragraph{Geometry made from the twisted identification}~\par
Let us then perform the twisted identification given above to the global AdS$_3$ and figure out its structure. First of all, let us change our coordinate to a new coordinate $(\theta',t')$ such that the identified points are on the same time slice $t'={\rm const.}$. Obviously, we can achieve this by performing the following Lorentz boost: 
\begin{align}
    \theta' &= \cosh \eta~ \theta + \sinh \eta ~ t,  \\
    t' &= \sinh \eta ~\theta + \cosh \eta ~ t, 
\end{align}
where 
\begin{align}
    \cosh \eta &= \frac{\chi_+ + \chi_-}{2\sqrt{\chi_+\chi_-}} , \\
    \sinh \eta &= \frac{\chi_+ - \chi_-}{2\sqrt{\chi_+\chi_-}}. 
\end{align}
The inverse transformation is given by
\begin{align}
    \theta &= \cosh \eta~ \theta' - \sinh \eta ~ t' ,  \\
    t &= -\sinh \eta ~\theta' + \cosh \eta ~ t'.  
\end{align}
The periodicity then turns out to be 
\begin{align}
    (\theta', t') \sim (\theta'+2\sqrt{\chi_+\chi_-}\pi, t').
\end{align}
Then the metric turns out to be 
\begin{align}
    ds^2 &= -(r^2+1)dt^2 + \frac{dr^2}{r^2+1} + r^2 d\theta^2 \nonumber \\
    &= -(r^2+1)dt^2 + \frac{dr^2}{r^2+1} + (r^2+1) d\theta^2 - d\theta^2 \nonumber \\
    &= -(r^2+1)dt'^2 + \frac{dr^2}{r^2+1} + (r^2+1) d\theta'^2 - d\theta^2 \nonumber \\
    &=-(r^2+1)dt'^2 + \frac{dr^2}{r^2+1} + (r^2+1) d\theta'^2 - (\cosh \eta~ d\theta' - \sinh \eta ~ dt')^2 \nonumber\\
    &=-(r^2+1+\sinh^2\eta)dt'^2 + \frac{dr^2}{r^2+1} + (r^2+1-\cosh^2\eta) d\theta'^2 + 2\cosh\eta \sinh\eta ~d\theta'dt' \nonumber\\
    &= -(r^2+\cosh^2\eta)dt'^2 + \frac{dr^2}{r^2+1} + (r^2-\sinh^2\eta) d\theta'^2 + 2\cosh\eta \sinh\eta~ d\theta'dt'. 
\end{align}
As the next step, we would like to change the periodicity of the angle direction to $2\pi$ to match standard conventions for the BTZ metric. This can be achieved by 
\begin{align}
    \theta'' = \frac{1}{\sqrt{\chi_+\chi_-}} \theta', \nonumber\\
    t'' = \frac{1}{\sqrt{\chi_+\chi_-}} t'.
\end{align}
As a result, we obtain
\begin{align}\label{eq:myBTZ}
    ds^2 &= -(r^2+\cosh^2\eta)\chi_+\chi_-dt''^2 + \frac{dr^2}{r^2+1} + (r^2-\sinh^2\eta) \chi_+\chi_- d\theta''^2 + 2\cosh\eta \sinh\eta \chi_+\chi_- d\theta''dt'' \nonumber \\
    &= -\left(r^2+\frac{(\chi_++\chi_-)^2}{4\chi_+\chi_-}\right)\chi_+\chi_-dt''^2 + \frac{dr^2}{r^2+1} + \left(r^2-\frac{(\chi_+-\chi_-)^2}{4\chi_+\chi_-}\right) \chi_+\chi_- d\theta''^2 + \frac{\chi_+^2 - \chi_-^2}{2} d\theta''dt'' \nonumber. \\
\end{align}
On the other hand, a standard BTZ metric is given by \cite{Banados1992}
\begin{align}\label{eq:BTZBTZ}
    ds^2 &= -\left(-M+\rho^2+\frac{J^2}{4\rho^2}\right) dt''^2 + \frac{d\rho^2}{-M+\rho^2+\frac{J^2}{4\rho^2}} + \rho^2 \left(-\frac{J}{2\rho^2} dt'' + d\theta''\right)^2  \nonumber\\
    &= -\left(-M+\rho^2\right) dt''^2 + \frac{d\rho^2}{-M+\rho^2+\frac{J^2}{4\rho^2}} + \rho^2 d\theta''^2 - J  d\theta'' dt'',
\end{align}
where $M$ and $J$ are the mass\footnote{The mass defined here is a little bit different from that we used in the main text. The difference should be obvious from the context.} and the angular momentum of the BTZ black hole respectively. 
By intuition, here $\theta''$ and $t''$ in \eqref{eq:myBTZ} and \eqref{eq:BTZBTZ} should be identified, and that is the reason why we wrote them with the same notation from the first place. 

Let us see how the angular momentum and the mass turn out to be with this identification. By comparing the coefficient of the $d\theta'' dt'' $ term, we have 
\begin{align}
    J = -\frac{\chi_+^2 - \chi_-^2}{2} . 
\end{align}
By comparing the coefficient of the $d\theta''$ term, we have 
\begin{align}
    \rho^2 = \left(r^2-\frac{(\chi_+-\chi_-)^2}{4\chi_+\chi_-}\right) \chi_+\chi_-,
\end{align}
and hence 
\begin{align}
    2\rho d\rho = 2r dr \chi_+\chi_-. 
\end{align}
Therefore, 
\begin{align}
    \frac{dr^2}{r^2+1} &= \frac{\frac{\rho^2}{\frac{\rho^2}{\chi_+ \chi_-}+ \frac{(\chi_+-\chi_-)^2}{4\chi_+\chi_-}}\frac{1}{\chi_+^2\chi_-^2}d\rho^2}{\frac{\rho^2}{\chi_+ \chi_-} + \frac{(\chi_+-\chi_-)^2}{4\chi_+\chi_-}+1} \nonumber\\
    &= \frac{d\rho^2}{\rho^2 + \frac{\chi_+^2+\chi_-^2}{2} + \frac{(\chi_+^2 - \chi_-^2)^2}{16\rho^2}}. 
\end{align}
By comparing the coefficient of this term, we can see that the angular momentum $J$ given above is consistent, and 
\begin{align}
    M = -\frac{\chi_+^2+\chi_-^2}{2} . 
\end{align}
We can also explicitly check that the coefficient of $dt''$ matches under the identifications given above. 

As a result, we have checked that the geometry created from the twist identification of the global AdS$_3$ is a spinning defect geometry with mass and angular momentum 
\begin{align}\label{eq:BTZ_MandJ}
    M = -\frac{\chi_+^2 + \chi_-^2}{2}, \quad J = -\frac{\chi_+^2 - \chi_-^2}{2}.  
\end{align}
The spectrum of such defects is discussed in \cite{Miskovic2009}. 

\paragraph{The CFT dual} ~\par

We have succeeded to see that a spinning defect can be obtained by performing a twist identification in global AdS$_3$. The next step is to check if our expectation about the relation \eqref{eq:handchi_spin} between the AdS side and the CFT side is correct or not. 

By plugging \eqref{eq:handchi_spin} into \eqref{eq:BTZ_MandJ}, we have 
\begin{align}
    J &= -\frac{\chi_+^2 - \chi_-^2}{2} = \frac{12}{c} (h_+-h_-) = 8G_N (h_+ - h_-).  \nonumber\\
    M &= -\frac{\chi_+^2+\chi_-^2}{2} = \frac{12}{c} (h_+ + h_-) -1 = 8G_N (h_+ + h_-) -1. 
\end{align}
This is exactly what we expect for non-scalar primaries with conformal dimension $(h_+,h_-)$.

\clearpage
\bibliographystyle{JHEP}
\bibliography{main}

\end{document}